\documentclass[%
 reprint,
superscriptaddress,
nofootinbib,
 amsmath,amssymb,
 aps,
]{revtex4-1}

\usepackage{nicefrac}
\usepackage[euler]{textgreek}
\usepackage{braket}
\usepackage{graphicx}
\usepackage{float}
\usepackage{dcolumn}
\usepackage{bm}
\usepackage[colorlinks,citecolor=blue,urlcolor=blue,linkcolor=blue,bookmarks=false,hypertexnames=true]{hyperref}
\usepackage{capt-of}
\usepackage{caption}
\usepackage{ulem}
\usepackage{blkarray}
\usepackage{xcolor}
\usepackage{mathtools}
\usepackage{braket}
\usepackage{soul}
\usepackage{amsmath}
\usepackage{relsize}
\usepackage{graphicx}
\usepackage{dcolumn}
\usepackage{bm}
\usepackage{hyperref}
\DeclareCaptionLabelFormat{adja-page}{\hrulefill\\#1 #2 \emph{(previous page)}}

\usepackage{color}
\usepackage{xcolor}
\usepackage{graphicx}
\usepackage{etoolbox}
\makeatletter
\patchcmd{\frontmatter@abstract@produce}
  {\vskip200\p@\@plus1fil
   \penalty-200\relax
   \vskip-200\p@\@plus-1fil}
  {}
  {}
  {}
\makeatother

\begin{document}

\title{Stochastic modeling of x-ray superfluorescence}

\author{Stasis Chuchurka}
\email{stasis.chuchurka@desy.de}
\affiliation{
Deutsches Elektronen-Synchrotron DESY, Hamburg 22603, Germany}
\affiliation{Department of Physics, Universit{\"a}t Hamburg, Hamburg 22761, Germany
}

\author{Andrei Benediktovitch}

\affiliation{Center for Free-Electron Laser Science (CFEL), Deutsches Elektronen-Synchrotron DESY, Hamburg 22607, Germany}

\author{Špela Krušič}
\affiliation{Jožef Stefan Institute, Ljubljana 1000, Slovenia}

\author{Aliaksei Halavanau}
\affiliation{Accelerator Research Division, SLAC National Accelerator Laboratory, Menlo Park, CA 94025}

\author{Nina Rohringer}
\email{nina.rohringer@desy.de}
\affiliation{Center for Free-Electron Laser Science (CFEL), Deutsches Elektronen-Synchrotron DESY, Hamburg 22607, Germany}
\affiliation{Department of Physics, Universit{\"a}t Hamburg, Hamburg 22761, Germany
}

\begin{abstract}
An approach to modeling the dynamics of x-ray amplified spontaneous emission and superfluorescence---the phenomenon of collective x-ray emission initiated by intense pulses of X-ray Free Electron Lasers---is developed based on stochastic partial differential equations. The equations are derived from first principles, and the relevant approximations, derivation steps, and extensions specific to stimulated x-ray emission are presented. The resulting equations take the form of three-dimensional generalized Maxwell-Bloch equations augmented with noise terms for both field and atomic variables. The derived noise terms possess specific correlation properties that enable the correct reconstruction of spontaneous emission. Consequently, the developed theoretical formalism is universally suitable for describing all stages of stimulated x-ray emission: spontaneous emission, amplified spontaneous emission, and superfluorescence. We present numerical examples that illustrate various properties of the emitted field, including spatio-temporal coherence, spectral-angular and polarization characteristics. We anticipate that the proposed theoretical framework will establish a robust foundation for interpreting measurements in stimulated x-ray emission spectroscopy, modeling x-ray laser oscillators, and describing other experiments leveraging x-ray superfluorescence.
\end{abstract}

\maketitle

\section{\label{sec: Introduction} Introduction}
X-rays are naturally suited for studying the dynamical structure of matter with atomic resolution and on sub-femtosecond timescales. X-ray Free Electron Lasers (XFELs)~\cite{2009'Emma_first-LCLS, 2012'Ishikawa_first-SACLA, 2017'Kang_first-PALFEL, 2017'Milne_first-swissfel, 2018'Weise_first-EuXFEL} create a paradigm shift, opening the realm of exploring high-intensity, nonlinear x-ray--matter interaction phenomena. The bright and femtosecond-short XFEL pulses can drive matter into previously unexplored, highly excited states, enabling unique insights into its structure and dynamics~\cite{2016'Bostedt, 2017'Bergmann_applications-of-XFELs, 2020'Marangos, 2021'Bergmann, 2020'Jaeschke, 2018'Young}.

For instance, focused XFEL beams can prepare atoms in a state of sizeable population inversion of core-valence transitions through rapid inner-shell photoionization. In the optical domain, a medium that is kept in a population-inverted state and placed in a resonator forms a classical laser oscillator. In the x-ray domain, sustaining a steady state of population inversion is hampered by fast decay processes on the femtosecond timescale. We consider the case of a transient population inversion produced by a short x-ray pulse traveling through a pencil-shaped medium. The x-ray emission process starts from isotropic, spontaneous x-ray fluorescence, which, upon propagating through the excited medium, is exponentially amplified until saturation, resulting in short, directed x-ray emission bursts. We later refer to the exponential amplification regime as the amplified spontaneous emission (ASE) regime and the saturation as the superfluorescence (SF) regime.

Soft x-ray SF was first realized in Ne gas~\cite{2012'Rohringer, 2013'Weninger}, with the observed emission also referred to as atomic x-ray lasing. Subsequent experiments demonstrated hard x-ray SF for solid targets~\cite{2015'Yoneda, 2022'Zhang} and liquid jets~\cite{2018'Kroll, 2020'Kroll}. Several applications for these types of x-ray pulses have been proposed. The directivity and high intensity of SF pulses facilitate high signal-to-noise ratio measurements, beneficial in x-ray spectroscopy. As experimentally demonstrated in~\cite{2018'Kroll}, chemical shifts are preserved in stimulated x-ray emission spectroscopy (sXES). Furthermore, it has been shown that weaker x-ray emission lines can be seeded and selected from other lines~\cite{2020'Kroll}. The development of these sXES techniques is one of the future directions at XFELs~\cite{2021'Bergmann}. Furthermore, x-ray SF may be used as a source of x-ray radiation with unique characteristics. In~\cite{2022'Zhang}, it was demonstrated that employing a SASE XFEL pump pulse can result in double-pulse x-ray SF. Further improvement of this technique may create x-ray sources needed for coherent nonlinear spectroscopy techniques~\cite{2017'Kowalewski}. In Ref.~\cite{2020'Halavanau}, a lasing medium, operating in hard x-ray ASE or SF regimes, is considered in a Bragg cavity, with the ultimate goal of forming an X-ray Laser Oscillator (XLO) resulting in spatially and temporally coherent x-ray pulses, with properties comparable to planned cavity-based XFEL pulses~\cite{2006'Huang, 2020'Marcus, 2008'Kim, 2011'Lindberg}.

The interpretation of sXES data and modeling of x-ray SF-based sources strongly benefit from predictive, quantitative modeling. In general, the phenomenon can be considered as a particular case of superradiance, which historically attracted significant interest~\cite{1954'Dicke, 1982'Gross, 2018'Trifonov}. A full quantum description of the interaction between the continuum of electromagnetic field modes and an ensemble of few-level emitters can be performed for certain particular cases, assuming the permutational invariance of emitters~\cite{2016'Gegg, 2018'Shammah, 2022'Sukharnikov} or restricting the evolution to early times~\cite{2021'Robicheaux, 2022'Sierra}. However, these methods cannot be directly applied to our setting since we are interested in systems containing $\sim10^{12}$ emitters. Therefore, we opt for a coarse-grained description of the problem, approximating the ensemble of atoms as a continuous medium. Our approach encompasses, in the general case, the dynamics of pumping and building up the transient population inversion, the initial stage of spontaneous emission, subsequent propagation and diffraction of the amplified emitted field, and the dynamics in the nonlinear saturation regime. If quantum properties of the electromagnetic field can be neglected, the description can be done with the help of optical Maxwell-Bloch (MB) equations~\cite{2019'Boyd}. However, in the case of SF, no atomic coherences nor emitted fields are initially present, hence homogeneous MB equations lack the source of spontaneous emission. Combining the MB equations with quantum effects triggering SF is necessary.

A rigorous description of both quantum and classical effects, e.g. diffraction, is possible in the ASE regime. In this case, the emitted fields are not strong enough to cause a change in the population inversion, and the equations for field and atomic operators become linear. Under these conditions, analytical expressions for emitted field properties can be derived for various shapes of the inverted medium~\cite{1985'Prasad, 1990'Hazak, 2012'Manassah}. Once the emitted field becomes strong, the ASE regime transforms into the SF regime where nonlinear effects play an important role, and the quantum fluctuations have negligible contributions. In the case of instantaneous excitation of atoms, the influence of quantum fluctuations can be represented by a suitable distribution of initial conditions for the MB equations. In two-level systems, the distribution of initial conditions can be mapped onto the distribution of tipping Bloch-vectors from the pole of the Bloch sphere~\cite{1982'Gross, 1979'Vrehen}. The numerical modeling of SF including diffraction effects is possible in paraxial approximation~\cite{1982'Gross, 1983'Watson} as well as within rigorous finite-difference time-domain methods~\cite{2009'Andreasen, 2012'Pusch}. However, in the x-ray domain, the rapid depopulation of the core-excited states on fs timescale due to the Auger-Meitner and radiative decays limits the approximation of instantaneous excitation. In this case, pumping, decay, and SF take place on the same time scale. In addition, different regions of the medium may experience evolution in different regimes---e.g., the central part may experience saturation, while the edges may be still within ASE. Hence, a formalism that enables a uniformly-suitable description of both quantum spontaneous emission and semi-classical MB-like behavior is needed.

It is possible to modify semi-classical equations in a phenomenologic way to include quantum effects responsible for spontaneous emission, by for example augmenting the MB equations with noise terms in the field equations~\cite{2004'Ziolkowski}, or in the atomic equations~\cite{2000'Larroche}, by including stochastic relaxation terms in the atomic equations and performing rescaling of the electric field---so-called Ehrenfest+R method~\cite{2019'Subotnik-EhrenfestR,2019'Subotnik_comparison}, and other ways~\cite{2018'Subotnik_qced, 2023'Park2023_noise-modified}. However, since those methods are not derived from the first principles, they possess certain limitations. Among those methods, the approach based on augmenting the MB equations with phenomenological noise terms~\cite{2000'Larroche} is widespread and has been applied for a series of applications~\cite{2014'Weninger,2020'Lyu,2012'Oliva,2015'Oliva,2018'Kroll,2020'Kuan,2022'Zhang}. This approach has the same computational complexity as MB equations and describes well the nonlinear dynamics in the saturation stage, however, has deficiencies in the description of the initial spontaneous-emission-dominated stage. Namely, the resulting temporal profile of the spontaneous emission is not reproduced correctly~\cite{2018'Krusic,2019'Benediktovitch}. A correct description of the spontaneous emission and cross-over to MB equations can be realized based on solving equations for the correlation function of the field and atomic coherences~\cite{2019'Benediktovitch}. This approach has been applied to several systems~\cite{2019'Mercadier,2020'Krusic}, but is computationally costly, since two-point quantities need to be computed. Moreover, extending this approach beyond two-level systems is challenging since the factorization of higher-order correlation functions into one- and two-point correlation functions, which is crucial to obtain a closed system of equations, becomes problematic even for the three-level systems.

In this paper, we present an approach that is general enough to describe spontaneous x-ray emission, ASE, and SF under realistic conditions and is free from uncontrollable approximations. We build on the formalism presented in paper~\cite{chuchurka2023stochastic}, and apply it to the case of lasing in copper atoms. We consider a typical XFEL pump pulse and parameters of the medium that result in pencil-shaped geometry. In this case, we can apply the paraxial approximation and---due to a short pump-pulse duration as well as rapid level decay compared to the propagation time---neglect the back-propagating wave and thus take advantage of using a co-moving frame by the concept of retarded time. Under these simplifications, we obtain equations in the structure similar to MB augmented with noise terms. The derived noise terms possess non-trivial correlation properties and can correctly reproduce the spontaneous emission. 

{In practical applications, when sampled in a Monte Carlo fashion, the proposed formalism can often result in diverging statistical realizations, a characteristic shared with similar phase-space methods \cite{2014'Drummond_book, 2006'Deuar_stochastic-gauges, 2005'Deuar_PhD}. In this article, we introduce an empirical modification designed to mitigate this divergent behavior. A more rigorous strategy addressing this issue will be explored and detailed in subsequent publications.}

The paper is organized as follows: In Sec.~\ref{sec: problem statement}, we formulate the master equation for Cu-K$\alpha_1$ lasing in a pencil-shape medium. Specific details about the pumping, decay, and decoherence processes can be found in Appendixes~\ref{sec: Appendix: pump and decay} and~\ref{sec: Appendix: Photoionization cross-sections}. In Sec. \ref{sec: SDE}, the master equation is converted into a system of stochastic differential equations. In Appendix~\ref{sec: Numerical}, the numerical scheme for solving these equations is presented. Finally, in Sec.~\ref{sec: Examples}, we give an example of numerical modeling and discuss the relationship between the output of the stochastic equations and the physical observables of interest.

{\section{\label{sec: problem statement}Problem statement}

\subsection{Resonant interactions with the light}

We consider an ensemble of many-level atoms in free space interacting through the quantized electromagnetic field. Each atom bears its own index $a$, to differ from the others. Its inner structure is characterized by a set of levels $\{\ket{p}\}$ and energies $\hbar\omega_p$. The free Hamiltonian of atom $a$ has the form
\begin{equation}
\label{eq: atomic Hamiltonian}
\hat{H}_a=\sum_p\hbar\omega_{a,p}\hat{\sigma}_{a,pp},
\end{equation}
Here, we introduce operators $\hat{\sigma}_{a,pq}=\ket{p}_a\!\bra{q}_a$ that measure the occupations and transitions between states of a particular atom $a$.

Prior to being excited by, for example, an XFEL beam, the atoms are in their ground state. Being ionized, the atoms start interacting with each other through the quantized electromagnetic field, resonant to the open transitions. The Hamiltonian of the field reads as
\begin{equation}
    \label{eq: field Hamiltonian}
   \hat{H}_f = \sum_{\textbf{k},s}\hbar\omega_k \hat{a}^\dag_{\textbf{k},s}\hat{a}_{\textbf{k},s}.
\end{equation}
Each mode of the field is characterized by a wave vector $\textbf{k}$, frequency $\omega_k=|\textbf{k}|c$, and polarization vector $\textbf{e}_{s}$. The pumping by a focused XFEL beam typically results in a pencil-shaped geometry of the excited medium. According to Ref.~\cite{1998'Varga}, for Gaussian beams with beam waist $w_0 > 100 \lambda$ ($\lambda$ is radiation wavelength), the difference between the solutions of full Maxwell equations and the paraxial scalar wave equations is less than a few percent. Since the XFEL focus size, even for the best x-ray focusing optics, is much larger than the wavelength~\cite{2018'Bajt,2018'Matsuyama}, we can use the paraxial approximation. In this case, the field propagating in the medium includes only the paraxial modes whose wave vectors $\textbf{k}$ are close to the central carrier wave vector $\textbf{k}_0=\omega_0/c$. Its propagation direction is denoted as $z$. The polarization vectors $\textbf{e}_{s}$ remain independent of the wave vectors $\textbf{k}$ and are orthogonal to the $z$ axis, forming the basis for a two-dimensional space. In this article, we employ right- and left-hand circular polarized Jones vectors as the chosen polarization basis (see Ref.~\cite{doi:10.1080/09500340408232511}):
\begin{equation*}
\begin{aligned}
    \textbf{e}_{-1} &= (\textbf{e}_x-i\textbf{e}_y)/\sqrt{2},\\
    \textbf{e}_{+1} &= (\textbf{e}_x+i\textbf{e}_y)/\sqrt{2}.
    \end{aligned}
\end{equation*}

In addition to the field propagating along the $z$ axis, the atoms exhibit isotropic spontaneous emission. This phenomenon cannot be accurately analyzed using the paraxial approximation. Given the negligible interaction of this emission with the medium, we exclude it from the field variable and consider it solely in the context of the lifetimes of the excited states. See Sec.~\ref{sec: Inclusion of the pump and decay processes} for more details.

The light is assumed to be resonant with the two manifolds of atomic levels: upper levels $\{\ket{u}\}$, and lower levels $\{\ket{l}\}$, whose transition energies $\omega_{uu'}=\omega_u-\omega_{u'}$ and $\omega_{ll'}=\omega_l-\omega_{l'}$ are assumed to be much smaller than the carrier frequency $\omega_0$. We reserve the indices $u$ and $l$ for the upper and lower states, respectively. For the numerical example, we will consider the level scheme corresponding to the $K{\alpha_1}$ transition of Cu atoms and stimulated emission following $1s$ ionization. 
SF on this transition was observed in~\cite{2015'Yoneda}. The Cu-$K{\alpha_1}$ transition is a candidate for the first implementation of the x-ray laser oscillator concept~\cite{2020'Halavanau}. To address the polarization properties of the emitted field, we have to explicitly treat the degenerate sublevels with different magnetic numbers. The manifolds of upper and lower levels have the following explicit form
\begin{equation}
\label{eq: e and g explicit for Cu}
\begin{aligned}
\{u\}&=\left\{1s_{\frac{1}{2},m=-\frac{1}{2}}, 1s_{\frac{1}{2},m=\frac{1}{2}}\right\},\\
\{l\}&=\left\{2p_{\frac{3}{2},m=-\frac{3}{2}}, 2p_{\frac{3}{2},m=-\frac{1}{2}}, 2p_{\frac{3}{2},m=\frac{1}{2}}, 2p_{\frac{3}{2},m=\frac{3}{2}}\right\}.
\end{aligned}
\end{equation}
The considered level scheme is sketched in Fig.~\ref{cuScheme}. 
\begin{figure}
	\centering
	\includegraphics[width=\linewidth]{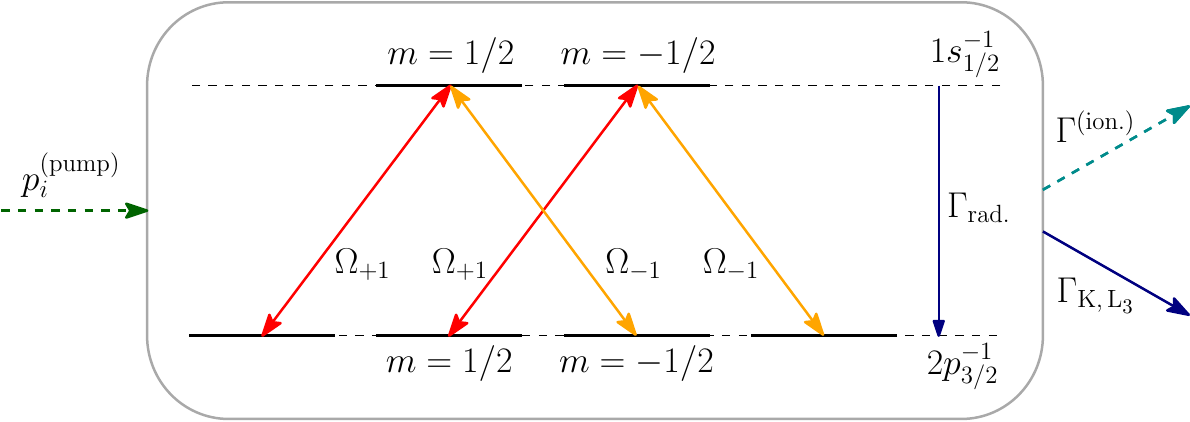}
	\caption[Schematic representation of the K$\alpha_1$ system in copper.]{Illustration of the K$\alpha_1$ system within a copper atom. Upper states $1s_{1/2}^{-1}$ experience radiative decay to lower states $2p_{3/2}^{-1}$ at a spontaneous rate $\Gamma_{\text{rad.}}$, and are coupled by two radiation modes ($\Omega_{\pm 1}$). All ionic states are generated through nonresonant photoionization from the ground state ($p_i^{(\mathrm{pump})}$). These ionic states can subsequently decay either spontaneously ($\Gamma_{\mathrm{K}},\Gamma_{\mathrm{L_3}}$) or via photoionization triggered by the pump and emitted fields ($\Gamma^{(\mathrm{ion.})}$).}
	\label{cuScheme}
\end{figure} 

The dynamics of the atomic populations is supposed to be incomparably slower than the oscillations of the field; therefore, we neglect all non-resonant interactions. Based on these assumptions, we write the following interaction Hamiltonian:

\begin{equation}
\label{eq: interaction Hamiltonian}
   \hat{V}=-ig_0\hbar\sum_{a,u,l}\textbf{d}_{ul}\hat{\sigma}_{a,ul}\sum_{\textbf{k},s} \hat{a}_{\textbf{k},s}\textbf{e}_{s}e^{i\textbf{k}\cdot\textbf{r}_a}+\text{H.c.},
\end{equation}
where the indices $u$ and $l$ represent the upper and lower states, $g_0=\sqrt{\omega_0/[2V\hbar\varepsilon_0]}$, $V$ is the quantization volume, $\textbf{d}_{pq}$ are the matrix elements of the dipole moment operators, and $\textbf{r}_a$ is the coordinates of atom $a$. The size of the atoms is typically assumed to be small in comparison with the wavelength of the electromagnetic field in the system, allowing the application of the dipole approximation. In the case of K$\alpha$ transitions, the wavelength is comparable to the atom size; however, it is still much larger than the overlap between atomic orbitals involved in K$\alpha$ transitions. In this case, the dipole approximation could be used as well.}

We decompose $\textbf{d}_{ul}$ into the product of the reduced dipole moment $d_0$~\cite{landau1977} and dimensionless coefficients $T_{ul,s}$ and $T_{lu,s}$:
\begin{equation}
\label{eq: d via T}
\textbf{d}_{ul}\cdot{\textbf{e}}_s    = d_0 T_{ul,s}, \quad 
\textbf{d}_{lu}\cdot{\textbf{e}}_s^*  = d_0 T_{lu,s}.
\end{equation}
The reduced dipole moment $d_0$ defines the strength of the transition, whereas the coefficients $T_{lu,s}=T_{ul,s}^*$ store the directional information and are proportional to Clebsch–Gordan coefficients. They can be calculated based on the Wigner-Eckart theorem (more in Appendix~\ref{app: dipole moment calculations}):
\begin{widetext}
\begin{subequations}
\label{eqs: cu-Tijs}
\begin{equation}
 \begin{tabular}{c}\\
 $\Big\{T_{ul,s=1}\Big\}\,\,=\,$ 
 \end{tabular}\quad \begin{blockarray}{ccccc}
 \color{darkgray}{2p_{\frac{3}{2},\,m=-\frac{3}{2}}} & \color{darkgray}{2p_{\frac{3}{2},\,m=-\frac{1}{2}}} &  \color{darkgray}{2p_{\frac{3}{2},\,m=\frac{1}{2}}}& \color{darkgray}{2p_{\frac{3}{2},\,m=\frac{3}{2}}}& \\
\begin{block}{(cccc)c}
  0 & 0 & 1/3 & 0 & \quad \color{darkgray}{1s_{\frac{1}{2},\,m=-\frac{1}{2}}} \\
  0 & 0 & 0 & 1/\sqrt{3} & \quad \color{darkgray}{1s_{\frac{1}{2},\,m=\frac{1}{2}}}\\
\end{block}
\end{blockarray}
\end{equation}
\begin{equation}
 \begin{tabular}{c}\\
 $\Big\{T_{lu,s=-1}\Big\}\,\,=\,$ 
 \end{tabular}\quad \begin{blockarray}{ccccc}
 \color{darkgray}{2p_{\frac{3}{2},\,m=-\frac{3}{2}}} & \color{darkgray}{2p_{\frac{3}{2},\,m=-\frac{1}{2}}} &  \color{darkgray}{2p_{\frac{3}{2},\,m=\frac{1}{2}}}& \color{darkgray}{2p_{\frac{3}{2},\,m=\frac{3}{2}}}& \\
\begin{block}{(cccc)c}
  1/\sqrt{3} & 0 & 0 & 0 & \quad \color{darkgray}{1s_{\frac{1}{2},\,m=-\frac{1}{2}}} \\
  0 & 1/3 & 0 & 0 & \quad \color{darkgray}{1s_{\frac{1}{2},\,m=\frac{1}{2}}}\\
\end{block}
\end{blockarray}
\end{equation}
\end{subequations}
\end{widetext}
where the index $s$ describes the polarization of the emitted field and takes the values $-1$ or $1$, corresponding to circular polarizations of the field traveling along the sample. The remaining coefficients $T_{lu,s}$ can be derived by conjugation, namely, $T_{lu,s}=T_{ul,s}^*$. The transitions corresponding to non-zero $T_{ul,s}$ and  $T_{lu,s}$ are depicted in Fig.~\ref{cuScheme}. The analysis of possible transitions shows that the considered $K{\alpha_1}$ level scheme is equivalent to two $\Lambda$ systems, composed of levels $\left\{2p_{\frac{3}{2},m=-\frac{3}{2}}, 1s_{\frac{1}{2},m=-\frac{1}{2}}, 2p_{\frac{3}{2},m=\frac{1}{2}}\right\}$ and $\left\{2p_{\frac{3}{2},m=\frac{3}{2}}, 1s_{\frac{1}{2},m=\frac{1}{2}}, 2p_{\frac{3}{2},m=-\frac{1}{2}}\right\}$. Each of the $\Lambda$ systems interacts with fields of both polarizations; as a result, in the general case, neither field polarization modes nor the $\Lambda$ systems can be decoupled from one another.


{Finally, we note that by assuming $g_0$ is independent of $\omega$, we disregard dipole-dipole interactions that can lead to decoherence between neighboring atoms. The effect of dipole-dipole interactions is local and solely determined by the density of the atoms, while the collective behavior of superfluorescence is mainly influenced by the total number of atoms. A proper geometry of the system can minimize the loss of coherence. Consequently, neglecting dipole-dipole interaction is well-justified for large, elongated systems.}
\subsection{Inclusion of the pump and decay processes}
\label{sec: Inclusion of the pump and decay processes}

Superfluorescence in Cu is initiated by an intense and focused pump pulse with an x-ray photon energy above the $1s$ ionization threshold. As a result, the Cu atoms are transferred from the neutral ground state to the core-ionized state, predominantly leaving the Cu atom in the $1s_{1/2}^{-1}$ state. This state can decay radiatively to the $2p^{-1}$ manifold of states or undergo other radiative processes, as well as Auger-Meitner decay. Equation~(\ref{eq: interaction Hamiltonian}) with levels from Eq.~(\ref{eq: e and g explicit for Cu}) describes the evolution of a small subsystem of atomic levels conditioned by the interaction with the resonance and paraxial fields. Processes such as photoionization, Auger-Meitner decay, fluorescence, electron-impact ionization, shake-off, and others that follow the irradiation by an XFEL pulse~\cite{2016'Jurek} need to be incorporated. Since the paraxial fields do not include all spontaneous emission, it is necessary to consider its impact at the level of lifetimes of the excited states.

{In addition to the states listed in Eq.~(\ref{eq: e and g explicit for Cu}), we also analyze the population of the neutral ground state, which is required for describing the pumping via photoionization. To describe the absorption of the pump pulse, we will consider the cumulative population of singly-ionized states $\rho^{\text{(aux.)}}(\textbf{r},\tau)$ that are not explicitly mentioned in Eq.~(\ref{eq: e and g explicit for Cu}) (see Appendix~\ref{sec: Appendix: pump and decay} for more details). The inclusion of pump, decay, and decoherence is typically performed in Markov approximation with the help of a master equation~\cite{Meystre2007, 2019'Benediktovitch}. Assuming a separate independent reservoir for each atom, the master equation is modified as follows:}
\begin{subequations}
{
\label{eq: master equation}
\begin{equation}
\begin{split}
\frac{d\hat{\rho}\!\left(t\right)}{dt}\Big|_{\text{incoh.}}&=\hat{\mathcal{L}}_{\text{incoh.}}[\hat{\rho}(t)]\\&=\sum_{i}p_{i}^{(\text{pump})}(\textbf{r}_a,t)\hat{\sigma}_{a,i0}\hat{\rho}(t)\hat{\sigma}_{a,0i}\\&+\Gamma_{\text{rad.}}\sum_{ik}G_{ik}^{(\text{rad.})}\hat{\sigma}_{a,ik}\hat{\rho}(t)\hat{\sigma}_{a,ki}\\&-\frac{1}{2}\sum_i\Gamma_i(\textbf{r}_a,t)\left(\hat{\rho}\!\left(t\right)\hat{\sigma}_{a,ii}+\hat{\sigma}_{a,ii}\hat{\rho}\!\left(t\right)\right).
\end{split}
\end{equation}}
Here, $\Gamma_{i}(\textbf{r},t)$ represents the inverse lifetime of the state $\ket{i}$. {The non-stationary pump field causes secondary ionization, subsequently making the lifetimes non-stationary as well. Without the time-dependent contributions, $\Gamma_{u} = 2.24\text{ fs}^{-1}$ and $\Gamma_{l} = 0.96\text{ fs}^{-1}$, both of which are comparable to the duration of the pump pulse.} $p_{i}^{(\text{pump})}(\textbf{r},t)$ represents the transition rates from the neutral ground state $\ket{0}$ due to photoionization, $G_{ik}^{(\text{rad.})}$ describes spontaneous radiative transitions between levels listed in Eq.~(\ref{eq: e and g explicit for Cu}), and $\Gamma_{\text{rad.}}$ is the spontaneous radiation emission rate calculated based on $d_0$ and given by $\Gamma_{\text{rad.}}=\omega_0^3 d_0^2/[3 \pi \varepsilon_0 \hbar c^3]$. The explicit form of these coefficients as well as further details on the implementation of incoherent processes are discussed in Appendix~\ref{sec: Appendix: pump and decay}.

{Finally, we consider the absorption of the quantized electromagnetic field through non-resonant transitions. This can be described by the following additional terms in the master equation:

\begin{equation}
\begin{split}
\frac{d\hat{\rho}(t)}{dt}\Big|_{\text{absorp.}}=&\hat{\mathcal{L}}_{\text{absorp.}}[\hat{\rho}(t)]\\=&\frac{c}{2}\sum_s\int d\textbf{r}\Big(\left[\hat{A}_s(\textbf{r})\rho(t),A^\dag_s(\textbf{r})\right]\\&+\left[\hat{A}_s(\textbf{r}),\rho(t)\hat{A}^\dag_s(\textbf{r})\right]\Big)\mu_{s}(\textbf{r},t).
\end{split}
\end{equation}}
\end{subequations}

{Here, $\mu_s(\textbf{r},t)$ represents the absorption coefficients defined for each polarization $s$. These coefficients are assumed to be small compared to resonance absorption. To simplify the notation, we introduce the operator $\hat{\textbf{A}}(\textbf{r})=\sum_{\textbf{k},s} \hat{a}_{\textbf{k},s}\textbf{e}_{s}e^{i\textbf{k}\cdot\textbf{r}}\big/\sqrt{V}$, defined in coordinate space. It is important to note that $\mu_s(\textbf{r},t)$ varies with time to account for changes in the atomic states, which in turn affect the values of the cross-sections. The explicit form of the absorption coefficient $\mu_s(\textbf{r},t)$ can be found in Appendix \ref{sub: Absorption of the fields}.}

{\section{\label{sec: SDE} Stochastic differential equations}

\subsection{\label{sec: Stochastic equations}Stochastic variables}

Understanding the evolution of a macroscopic ensemble of atoms coupled to a quantized electromagnetic field is a complex and challenging topic. This complexity arises due to the exponential growth in the number of degrees of freedom associated with the underlying density matrix. In the study of superfluorescence in compact systems \cite{chuchurka2023stochastic}, the density matrix is represented as a factorized product of one-particle density matrices, with the dynamics of these individual one-particle density matrices described by Bloch equations. To account for collective many-body effects, additional noise terms are introduced. Each realization of these noise terms yields a distinct density matrix, and the average of different density matrices restores quantum effects, accurately reproducing the phenomenon of collective spontaneous decay in compact ensembles of atoms.

Simplifying the analysis of superfluorescence in compact systems involves tracing out the field degrees of freedom, leading to a parametrization that includes only atomic variables. However, in elongated systems, explicit consideration of the propagation of the field becomes necessary. The parametrization from Ref.~\cite{chuchurka2023stochastic} is extended to include the field variables:
\begin{subequations}
    \label{eq: stochastic ansatz for the density matrix}
\begin{equation}
\rho(t) = \Big\langle \prod_a \hat{\rho}_a(t)\prod_{\mathbf{k},s} \hat{\Lambda}(\alpha_{\mathbf{k},s}(t), \alpha^\dagger_{\mathbf{k},s}(t)) \Big\rangle.
\end{equation}
Here, each atom is characterized by a one-particle density matrix $\hat{\rho}_a$:
\begin{equation}
    \hat{\rho}_a = \sum_{p,q} \rho_{a,pq}(t) \hat{\sigma}_{a,pq}.
\end{equation} 
To incorporate the electromagnetic field, we draw inspiration from the concept of positive P representation (see Refs.~\cite{2014'Drummond_book, 2006'Deuar_stochastic-gauges} for more details). We expand the field in the basis of coherent states $\ket{\alpha_{\mathbf{k},s}(t)}$. In the density matrix formalism, the coherent states are combined into normalized projectors $\hat{\Lambda}(\alpha, \alpha^\dagger)$:
\begin{equation}
\label{eq: Lambda projector}
\hat{\Lambda}(\alpha, \alpha^\dagger) = \ket{\alpha}\bra{\alpha^{\dagger*}} \exp\left(-\alpha^\dagger \alpha + \frac{|\alpha|^2}{2} + \frac{|\alpha^\dagger|^2}{2}\right).
\end{equation}
\end{subequations}
The evolution of the one-particle density matrices $\rho_{a,pq}(t)$ and field mode amplitudes $\alpha_{\mathbf{k},s}(t)$ and $\alpha^\dag_{\mathbf{k},s}(t)$ is governed by stochastic differential equations, which will be introduced later. The presence of noise terms in these equations allows for the restoration of quantum many-body effects. Different realizations of the noise terms lead to different density matrices whose average is represented by the angle brackets in Eq.~(\ref{eq: stochastic ansatz for the density matrix}). While the constituent density matrices can be factorized, the resulting combination cannot be represented by a direct product.

The variables $\alpha_{\mathbf{k},s}(t)$ and $\alpha^\dag_{\mathbf{k},s}(t)$ represent the field in reciprocal space. To analyze the propagation effects, we combine these variables into slowly varying electric field amplitudes denoted as $\Omega^{(\pm)}_{s}(\textbf{r},t)$. In terms of Rabi frequency, these amplitudes have the following form:
\begin{subequations}
\label{eq: D via Omega}
\begin{align}
    \Omega^{(+)}_{s}(\textbf{r},\tau)&=\,id_0\sum_\textbf{k} g_0\alpha_{\textbf{k},s}(\tau+z/c)e^{i\textbf{k}\cdot\textbf{r}+i\omega_0\tau},\\ \Omega^{(-)}_{s}(\textbf{r},\tau)&=-id_0\sum_\textbf{k} g_0\alpha_{\textbf{k},s}^{\dag}(\tau+z/c)e^{-i\textbf{k}\cdot\textbf{r}-i\omega_0\tau}.\end{align}
\end{subequations}
Here, we have introduced the retarded time $\tau=t-z/c$, which conveniently incorporates the propagation effects. Additional information regarding the transition to the retarded time is provided in Appendix~\ref{app: retarded time}.

For K-$\alpha$ transitions in period-IV elements, the transient core-shell population inversion state created by the pump pulse relaxes on a femtosecond timescale due to the Auger-Meitner effect and x-ray fluorescence. As a result, the non-trivial dynamics of the atomic variables is mostly conditioned by the presence of the pump fields. For typical media of interest, such as solution jets and solid samples, the typical thickness of the medium is on the order of 100~\textmu m, while the pump length is on the order of $c$ $\times$ 10 fs, or about 3 \textmu m. In the context of the original coordinates, the moving pump fields cover a narrow diagonal strip in the $(z, t)$-plane. To avoid simulating the trivial dynamics outside of this strip, it is convenient to substitute the original time $t$ with the retarded time $\tau = t - z/c$. {In the $(z, \tau)$-plane, the pump and SF pulses remain stationary, and the region of relevant dynamics is compressed into a narrow horizontal strip.} Consequently, introducing retarded time helps save computational resources.

Finally, numerical simulations assume discretizing the dynamic variables on a grid. A finite grid step implies that we expect the variables to change slowly between neighboring nodes of the grid. When using the original time parameter $t$, the size of the coordinate grid step must be comparable to the time step due to the finite speed of light; otherwise, atoms within a single coordinate step may not evolve uniformly. However, when employing the retarded time parameter $\tau$, it conveniently accounts for propagation phenomena, and the coordinate step becomes constrained by other factors, such as amplification by an inverted medium.

By analogy to the field variables in Eq.~(\ref{eq: D via Omega}), we intend to redefine $\Gamma_{i}(\textbf{r},t)$, $p_{i}^{(\text{pump})}(\textbf{r},t)$, and $\mu_s(\textbf{r},t)$ involved in Eq.(\ref{eq: master equation}) in terms of the retarded time $\tau$. We perform the following substitution:
\begin{equation}
\label{eq: retarded time for the aux. variables}
\begin{matrix}
\Gamma_{i}(\textbf{r},\tau+z/c) &  & \Gamma_{i}(\textbf{r},\tau),\\
p_{i}^{\text{(pump)}}(\textbf{r},\tau+z/c) & \quad\to\quad\quad & p_{i}^{\text{(pump)}}(\textbf{r},\tau),\\
\mu_s(\textbf{r},\tau+z/c) &  & \mu_s(\textbf{r},\tau).
\end{matrix}    
\end{equation}
We will consistently use these redefined variables throughout the article.

In Appendix~\ref{app: collective variables}, we show that closely situated atoms can be grouped into collective variables. We divide the medium into small regions with a volume $\Delta V$ and define collective variables for each region. Given the assumption of a small $\Delta V$, the resulting variables can be treated as continuous:

\begin{subequations}
\label{eq: coarse grained approximation for the atoms}
\begin{align}
    \frac{1}{\Delta N}\sum_{a \in \Delta V} \rho_{a,u_1u_2}(\tau+z_a/c)\quad&\to \quad \rho_{u_1u_2}(\textbf{r},\tau),\\
 \frac{1}{\Delta N}\sum_{a \in \Delta V} \rho_{a,l_1l_2}(\tau+z_a/c)\quad&\to\quad  \rho_{l_1l_2}(\textbf{r},\tau).
\end{align}
Here, the indices $u_i$ and $l_i$ represent the upper and lower states, $\Delta N$ is the number of atoms included in the considered small region. For the coherences between the upper and lower states, we have to account for the frequently oscillating factor $e^{-i\omega_0 \tau}$:
\begin{align}
      \frac{1}{\Delta N}\sum_{a \in \Delta V} \rho_{a,ul}(\tau+z_a/c)e^{i\omega_0 \tau} \quad&\to\quad \rho_{ul}(\textbf{r},\tau),\\
 \frac{1}{\Delta N}\sum_{a \in \Delta V} \rho_{a,lu}(\tau+z_a/c)e^{-i\omega_0 \tau} \quad&\to\quad \rho_{lu}(\textbf{r},\tau).
\end{align}
\end{subequations}

In order to describe spontaneous emission, the equations for the atomic and field variables will incorporate elementary noise terms ${f_s}(\textbf{r},\tau)$, $f_s^\dag(\textbf{r},\tau)$, ${g_s}(\textbf{r},\tau)$, and $g_s^\dag(\textbf{r},\tau)$. The first pair is statistically independent of the second pair. ${f_s}(\textbf{r},\tau)$ and $f_s^\dag(\textbf{r},\tau)$ possess distinct correlation properties as follows:
\begin{subequations}
\label{eq: f_s noise correlations}
\begin{align}
\langle f_s(\textbf{r},t)f_{s'}(\textbf{r}',\tau')\rangle&=\langle f_{s}^\dag(\textbf{r},\tau)f_{s'}^\dag(\textbf{r}',\tau')\rangle=0,\\
\langle f_{s}(\textbf{r},\tau)f_{s'}^\dag(\textbf{r}',\tau')\rangle&= \delta_{ss'}\delta(z-z')\delta_\varepsilon(\tau-\tau')\delta_\varepsilon(\textbf{r}_\bot-\textbf{r}'_\bot).
\end{align}
\end{subequations}
Similar stochastic characteristics apply to both ${g_s}(\textbf{r},\tau)$ and $g_s^\dag(\textbf{r},\tau)$. The presence of the delta-function $\delta(z-z')$ simply reflects Ito's interpretation of integration along the $z$ axis (see Appendix~\ref{sec: Appendix: Equivalence of Ito and Stratonovich forms}). Furthermore, $\delta_\varepsilon(\tau-\tau')$ is a localized function serving a purpose similar to that of a delta-function. Its width is determined by $1/[c\Delta k_z]$, where $\Delta k_z$ represents the range of longitudinal wave vectors required for an accurate field representation. In a similar fashion, the width of the transverse correlator is determined by the range of relevant transverse modes required for an accurate representation of the paraxial fields. Consequently,  $\delta_\varepsilon(\textbf{r}_\bot-\textbf{r}'_\bot)$ is a bell-shaped function with a waist of $\sim\lambda_0/\sqrt{\Delta o}$, where $\Delta o$ represents the solid angle encompassing the paraxial modes. For more detailed information, refer to Appendices~\ref{app: retarded time} and~\ref{app: collective variables}.

\subsection{Stochastic Bloch equations}
\label{sec: Stochastic Bloch equations}

The detailed derivations of the equations presented in this section can be found in Appendices~\ref{sec: derivations}, \ref{app: retarded time}, and \ref{app: collective variables}. The Ito stochastic differential equations for ${\rho}_{pq}\!\left(\textbf{r},\tau\right)$ have the form of a semi-classical Bloch equations with additional noise terms. Their incoherent parts read as follows:
\begin{widetext}
\begin{subequations}
\label{eq: atomic equations}
\begin{multline}
\label{eq: pump decay decoherence for rho_ij}
 \!\!\!\!\!\frac{\partial}{\partial \tau}\rho_{pq}(\textbf{r},\tau)\Big|_{\text{incoh.}}=
    -(\Gamma_{p}(\textbf{r},\tau)+\Gamma_{q}(\textbf{r},\tau))\rho_{pq}(\textbf{r},\tau)/2
    \\+\delta_{pq}\left(p_{p}^{(\text{pump})}(\textbf{r},\tau)\rho^{(\mathrm{ground})}(\textbf{r},\tau)+\Gamma_{\text{rad.}}\sum_{k}G_{pk}^{(\text{rad.})}\rho_{kk}(\textbf{r},\tau)\right),\!\!\!\!\
\end{multline}
where $\rho^{(\mathrm{ground})}(\textbf{r},\tau)$ represents the population of the neutral ground state $\ket{0}$. The following terms capture the unitary evolution:
\begin{equation}
\label{eq: bloch equations}
\begin{aligned}
 \frac{\partial}{\partial \tau}\rho_{pq}(\textbf{r},\tau)\Big|_{\text{unitary}}=&-i\Delta\omega_{pq}{\rho}_{pq}\!\left(\textbf{r},\tau\right)+i\sum_{r,s}\Bigg[\Omega_s^{(+)}\!\left(\textbf{r},\tau\right)\Big(T_{p>r,s}{\rho}_{rq}\!\left(\textbf{r},\tau\right)-{\rho}_{pr}\!\left(\textbf{r},\tau\right)T_{r>q,s}\Big)\\&+\Omega_s^{(-)}\!\left(\textbf{r},\tau\right)\sum_{r}\Big(T_{p<r,s}{\rho}_{rq}\!\left(\textbf{r},\tau\right)-{\rho}_{pr}\!\left(\textbf{r},\tau\right)T_{r<q,s}\Big)\Bigg],
\end{aligned}
\end{equation}
where $p>q$ means that index $p$ corresponds to the subset of upper states $\{\ket{u}\}$ whereas index $q$ corresponds to the subset of lower states $\{\ket{l}\}$. Additionally, we have introduced the following energy differences:
\begin{align*}
\Delta \omega_{{uu'}}    = \omega_{{u}} - \omega_{{u'}},\quad             
\Delta \omega_{{ul}}     = \omega_{{u}} - \omega_{{l}} -\omega_0, \quad     
\Delta \omega_{{lu}}     = \omega_{{l}} - \omega_{{u}} +\omega_0,  \quad    
\Delta \omega_{{ll'}}    = \omega_{{l}} - \omega_{{l'}},
\end{align*} 
where the indices $u$ and $l$ represent the upper and lower states. In order to describe the spontaneous emission, we introduce the following stochastic terms:
\begin{equation}
\begin{aligned}
\label{eq: noise decomposition}
    \frac{\partial}{\partial \tau}\rho_{pq}(\textbf{r},\tau)\Big|_{\text{noise}}&=\sum_s\bigg(\sum_{r}{\rho}_{pr}(\textbf{r},\tau)T_{r>q,s}-{\rho}_{pq}(\textbf{r},\tau)\sum_{u,l}T_{ul,s}{\rho}_{lu}(\textbf{r},\tau)\bigg)g_s^\dag(\textbf{r},\tau)\\&+\sum_s\bigg(\sum_{r}T_{p<r,s}{\rho}_{rq}(\textbf{r},\tau)-{\rho}_{pq}(\textbf{r},\tau)\sum_{u,l}T_{lu,s}{\rho}_{ul}(\textbf{r},\tau)\bigg) f_s^\dag(\textbf{r},\tau),
\end{aligned}
\end{equation}
\end{subequations}
\end{widetext}
that involve $f_s^\dag(\textbf{r},\tau)$ and $g_s^\dag(\textbf{r},\tau)$ defined in Sec.~\ref{sec: Stochastic equations}. In addition to the previously mentioned approximations, we disregard contributions from Eq.~(\ref{eq: noise decomposition}) that exhibit a quadratic dependence on the atomic variables ${\rho}_{pq}(\textbf{r},\tau)$. These terms are proportional to the coherences $\rho_{lu}(\textbf{r},\tau)$ and $\rho_{ul}(\textbf{r},\tau)$, which are notably smaller when compared to the atomic populations during the pump stage. The coherences $\rho_{lu}(\textbf{r},\tau)$ and $\rho_{ul}(\textbf{r},\tau)$  gain significance only after substantial growth of the SF field. Given that noise terms play a critical role only in the initial stages when a strong SF field has not yet developed, it is justifiable to omit the quadratic terms.

\subsection{Stochastic wave equations for the field amplitudes}
\label{sec: Stochastic wave equations for the field amplitudes}

Similarly, the field variables are governed by traditional wave equations augmented by noise terms. These equations are linear, allowing for the decomposition of $\Omega_{s}^{(\pm)}(\textbf{r},\tau)$ into two components:
\begin{subequations}
\label{eq: paraxial Omega}
\begin{equation}
\label{field decomposition}
  \Omega_{s}^{(\pm)}(\textbf{r},\tau)= \Omega_{s,\,\text{det.}}^{(\pm)}(\textbf{r},\tau)+ \Omega_{s,\,\text{noise}}^{(\pm)}(\textbf{r},\tau),
\end{equation}
where $\Omega_{s,\,\text{det.}}^{(\pm)}(\textbf{r},\tau)$ are influenced by the initial conditions and deterministic parts, while the noise components $\Omega_{s,\,\text{noise}}^{(\pm)}(\textbf{r},\tau)$ are driven by the noise terms $f_s(\textbf{r},\tau)$ and $g_s(\textbf{r},\tau)$. The specific equations for these two components are given by:
\begin{multline}
\left[\frac{\partial}{\partial z}-\frac{i}{2 k_0}\left(\frac{\partial^2}{\partial x^2}+\frac{\partial^2}{\partial y^2}\right)+\frac{\mu_s(\textbf{r},\tau)}{2}\right]\begin{pmatrix}
\Omega_{s,\,\text{det.}}^{(+)}(\textbf{r},\tau)\\
\Omega_{s,\,\text{noise}}^{(+)}(\textbf{r},\tau)
\end{pmatrix}
  \\= i\frac{3}{8\pi}\lambda^2_0\Gamma_{\text{rad.}} \begin{pmatrix}
n(\textbf{r}) \sum_{u,\, l}T_{lus} \rho_{ul}(\textbf{r},\tau)\\
f_s(\textbf{r},\tau)
\end{pmatrix},
\end{multline}
\begin{multline}
\left[\frac{\partial}{\partial z}+\frac{i}{2 k_0}\left(\frac{\partial^2}{\partial x^2}+\frac{\partial^2}{\partial y^2}\right)+\frac{\mu_s(\textbf{r},\tau)}{2}\right]\begin{pmatrix}
\Omega_{s,\,\text{det.}}^{(-)}(\textbf{r},\tau)\\
\Omega_{s,\,\text{noise}}^{(-)}(\textbf{r},\tau)
\end{pmatrix} \\ =-i\frac{3}{8\pi}\lambda^2_0\Gamma_{\text{rad.}} \begin{pmatrix}
n(\textbf{r}) \sum_{u,\,l}\rho_{lu}(\textbf{r},\tau) T_{uls}\\
g_s(\textbf{r},\tau)
\end{pmatrix},
\end{multline}
\end{subequations}
where the indices $u$ and $l$ represent the upper and lower states, $n(\textbf{r})$ is the concentration of the atoms, and $\lambda_0$ is the wavelength of the carrier mode. From a qualitative standpoint, the atoms can be described as simultaneously carrying independent deterministic and stochastic dipole moments, corresponding to $\rho_{ul}(\textbf{r},\tau)$ and $f_s(\textbf{r},\tau)$, respectively. The deterministic dipoles give rise to the deterministic fields $\Omega_{s,\text{det.}}^{(\pm)}(\textbf{r},\tau)$, resembling solutions to traditional Maxwell equations. In contrast, the stochastic dipole moments generate the stochastic fields $\Omega_{s,\text{noise}}^{(\pm)}(\textbf{r},\tau)$.

It is crucial to emphasize that the right-hand side may encompass modes beyond the scope of the paraxial approximation. To address this issue, damping is introduced to the Laplace operator $\nicefrac{\partial^2}{\partial x^2}+\nicefrac{\partial^2}{\partial y^2}$ for non-paraxial modes. By using spectral methods, damping is implicitly implemented by considering a finite set of basis functions. Additionally, note that the integration along the $z$ axis should be carried out using Ito's interpretation.}

\subsection{The structure of the noise terms}
\label{subsec: Discussion of the structure of the equations and noise terms}

During the initial phase, when coherences or fields are absent, the deterministic terms in Eq.~(\ref{eq: bloch equations}) are zero. However, the noise contribution for the coherences becomes non-zero if the upper states are populated. Owing to the correlation properties detailed in Eq.~(\ref{eq: f_s noise correlations}), the noise terms in the equations governing atomic variables remain uncorrelated, just as the noise terms in the equations for field variables. Correlations solely manifest between the noise terms associated with field and atomic variables. This property allows for the accurate capture of the temporal profile of emitted radiation in the limit of pure spontaneous emission. For a more in-depth exploration of this aspect, please refer to Section~\ref{sec: Spontaneous emission within the stochastic methodology}.

{\subsection{\label{sec: expectation values}Expectation values}

From a set of realizations of the stochastic variables, various expectation values can be constructed. The atoms are characterized by the variables $\rho_{pq}\!\left(\textbf{r},\tau\right)$, which are directly linked to one-particle properties:
\begin{subequations}
    \label{eq:observables-1-particle atom}
    \begin{align}
        \!\!\!\!\!\!\text{Tr}\left[\hat{\sigma}_{a,uu'}\hat\rho(t)\right]&=\langle \rho_{a,u'u}(t)\rangle =\langle \rho_{u'u}(\textbf{r}_a,t-z_a/c)\rangle,\\
        \text{Tr}\left[\hat{\sigma}_{a,ll'}\hat\rho(t)\right]&=\langle \rho_{a,l'l}(t)\rangle =\langle \rho_{l'l}(\textbf{r}_a,t-z_a/c)\rangle.
    \end{align}
Constructing expectation values related to transitions between upper and lower states, it is essential to restore the phase:
\begin{align}
        \text{Tr}\left[\hat{\sigma}_{a,ul}\hat\rho(t)\right]&\nonumber=\langle \rho_{a,lu}(t)\rangle \\&=\langle \rho_{lu}(\textbf{r}_a,t-z_a/c)\rangle e^{i\omega_0(t-z_a/c)},\\
        \text{Tr}\left[\hat{\sigma}_{a,lu}\hat\rho(t)\right]&\nonumber=\langle \rho_{a,ul}(t)\rangle \\&=\langle \rho_{ul}(\textbf{r}_a,t-z_a/c)\rangle e^{-i\omega_0(t-z_a/c)}.
    \end{align}
Recall that the continuous variables $\rho_{pq}(\textbf{r},\tau)$ represent the collective atomic properties in the vicinity of coordinate $\textbf{r}$. To replace the discrete atomic variables $\rho_{a,pq}(t)$, which pertain to individual atoms, with their continuous analogs, we assume that the atomic variables exhibit sufficient smoothness. Further details can be found in Appendix \ref{app: collective variables}.

Let us provide an example of obtaining two-particle properties. The correlations between neighboring atoms can be measured by the product of their coherences as follows:
\begin{equation}
\begin{aligned}
\text{Tr}\left[\hat{\sigma}_{a,ul}\hat{\sigma}_{a',l'u'}\hat\rho(t)\right]&=\langle \rho_{a,lu}(t)\rho_{a',l'u'}(t)\rangle\\&
=\langle \rho_{lu}(\textbf{r}_a,t-z_a/c)\rho_{ul}(\textbf{r}_a,t-z_a/c)\rangle \\&\times e^{i\omega_0\Delta z/c}.
    \end{aligned}    
\end{equation}
\end{subequations}
Since the atoms are close to each other, we do not distinguish their coordinates when using the slowly varying continuous variables. The distance between the atoms, $\Delta z=z_a-z_{a'}$, is only involved in the frequently oscillating phase multiplier $e^{i\omega_0\Delta z/c}$.}

To analyze the properties of the emitted fields, we define the following first-order correlation functions:
\begin{equation}
\label{eq: Jsp-t1-t2_def}
J_{s}(\textbf{r},\tau_1,\tau_2)=\frac{\langle\Omega_{s}^{(+)}(\textbf{r},\tau_1)\Omega_{s}^{(-)}(\textbf{r},\tau_2)\rangle}{\frac{3}{8\pi}\lambda^2_0 \Gamma_{\text{rad.}}}.
\end{equation}
Thanks to the properly chosen multiplier, $J_{s}(\textbf{r},\tau,\tau)$ directly provides the photon flux:
\begin{equation}
\label{eq: Nsp-via-J}
I_s(\textbf{r},\tau) =\frac{d N_{s}^{(\text{ph.})}(\textbf{r},\tau)}{dt dS}= J_{s}(\textbf{r},\tau,\tau).
\end{equation}

\subsection{Spontaneous emission within the stochastic methodology}
\label{sec: Spontaneous emission within the stochastic methodology}

The noise terms in the equations of motion manifest most prominently in the case of spontaneous emission. The evolution due to spontaneous emission can be modeled by assuming a low atomic density, denoted as $n(\textbf{r})$, which reduces the chance of re-absorption. Practically, this limit is addressed by retaining terms linearly dependent on $n(\textbf{r})$ in Eq. (\ref{eq: Jsp-t1-t2_def}). It still requires integration of the equations for the atomic variables. We neglect deterministic parts of the fields $\Omega_{s, \text{det.}}^{(\pm)}(\textbf{r},\tau)$ in Eq.~(\ref{eq: atomic equations}) as they are proportional to $n(\textbf{r})$. Consequently, the equations become linear and can be straightforwardly integrated. Substituting the integrated expressions for the atomic coherences $\rho_{ul}(\textbf{r},\tau)$ and $\rho_{lu}(\textbf{r},\tau)$ into the field equations (\ref{eq: paraxial Omega}) and utilizing the correlation properties in Eq.~(\ref{eq: f_s noise correlations}), we obtain:
\begin{multline}
\label{eq: Jsp-t1-t2_num}
J_{s}({\textbf{r}},\tau_1,\tau_2)\approx
    \frac{3}{8\pi}\lambda^2_0\Gamma_{\text{rad.}}\,e^{-\gamma_{\text{dec.}} |\tau_1-\tau_2|}
    \int d\textbf{r}' n(\textbf{r}')\\\times\langle\rho_{s}^{(\text{up.})}(\textbf{r}',\min(\tau_1,\tau_2))\rangle|G_s(\textbf{r}-\textbf{r}')|^2 ,
\end{multline}
where we assume that the coherences decay with a rate $\gamma_{\text{dec.}}=\left[\Gamma_u+\Gamma_l\right]/2$. $G_s(\textbf{r})$~is the Green function for the propagation of the emitted field. $\rho_{s}^{(\text{up.})}(\textbf{r},\tau)$ and $\rho_{s}^{(\text{low.})}(\textbf{r},\tau)$ are defined as:
\begin{subequations}
\label{eq: E-and-G, num}
   \begin{align}
\rho_{s}^{(\text{up.})}(\textbf{r},\tau)&=\sum_{u,u',l}T_{{lu}s}\rho_{{uu'}}(\textbf{r},\tau)T_{{u'l}s},\\
\rho_{s}^{(\text{low.})}(\textbf{r},\tau)&=\sum_{l,l',u}T_{{ul}s}\rho_{{ll'}}(\textbf{r},\tau)T_{{l'u}s},
\end{align} 
\end{subequations}
where the indices $u$ and $l$ represent the upper and lower states. The difference between these two values, $\rho_{s}^{(\text{up.})}(\textbf{r},\tau) - \rho_{s}^{(\text{low.})}(\textbf{r},\tau)$, can be interpreted as an effective population inversion.

For a more comprehensive explanation of how the noise terms accurately replicate spontaneous emission, please refer to the details provided in Appendix \ref{sec: Illustr}.

\section{\label{sec: Examples} Numerical examples and discussion}

In this section, we provide a detailed demonstration of x-ray emission modeling, employing parameters closely aligned with the anticipated experimental conditions for the XLO project outlined in Ref.~\cite{2020'Halavanau}. To achieve a sizeable population inversion through rapid photoionization, we require a high pump-pulse energy and strong focusing. We assume an XFEL-pulse energy of 250~\textmu J, with the pump focused down to a 200~nm x 200~nm FWHM, and the x-ray photon energy set at 9~keV (above the Cu K-edge). The temporal profile of the XFEL pulse is conditioned by the self-amplified spontaneous emission (SASE) process and is thus composed of a large number of randomly generated spikes~\cite{2008'Saldin}. However, for this demonstration, we aim to disentangle the stochasticity inherent in the current simulation approach from the SASE stochasticity. To achieve this, we use a Gaussian temporal profile with an 11.7~fs FWHM.

As a medium that generates x-ray lasing, following~\cite{2020'Halavanau}, we consider a 270-\textmu m-thick jet of 8-molar solution of Cu(NO$_3$)$_2$ in water. Our calculations are performed on a 900 $\times$ 900 nm$^2$ spatial domain in the transverse direction, with 64 $\times$ 64 grid points, 40 grid points in the longitudinal direction, and 180 points for the 37-fs-long temporal moving window. Unless otherwise stated, all numerical results shown are based on these parameters.

{For technical details about the implementation of the numerical scheme, please refer to Appendix~\ref{sec: Numerical}. We discretized the equations using a split-step method, where the noise and deterministic parts of the fields are integrated by means of different schemes. The separation of the fields in Eq.~(\ref{field decomposition}) into two parts becomes apparent.}

\subsection{Run-away trajectories and diffusion gauges}
\label{Gauging run-away trajectories}

Before delving into the numerical results, it is essential to address a challenge inherent in approaches based on stochastic differential equations~\cite{2006'Deuar_stochastic-gauges, 2005'Deuar_PhD}. Apart from the stability requirement of the numerical scheme, the stochastic differential equations themselves should prevent unbounded, diverging solutions. {Given the exponential amplification involved in the phenomenon under analysis, it is crucial to clarify the following: diverging trajectories grow at a rate faster than any exponential function and reach infinity within a finite time interval.} In Ref.~\cite{chuchurka2023stochastic}, it was demonstrated that the freedom in constructing noise terms for superfluorescence in compact systems can be leveraged to suppress divergent behavior. This approach extends to the paraxial geometry. Here, we outline the main steps, with further details provided in Appendix~\ref{sec: Numerical}.

Diverging trajectories may arise when effective population inversion for any polarization $s$ is present:
\begin{equation}
\label{eq: condition for the correction}
    \text{Re}(\rho_{s}^{(\text{up.})}(\textbf{r},\tau))>\text{Re}(\rho_{s}^{(\text{low.})}(\textbf{r},\tau))\footnote{In the proposed formalism, the populations are complex, so we extract their real parts.}\!\!.
\end{equation}
For coordinates $\textbf{r}$ and retarded time $\tau$ satisfying the condition in Eq.~(\ref{eq: condition for the correction}), Eq.~(\ref{eq: paraxial Omega}) should be modified to suppress divergent behavior. This is achieved by replacing the density matrix elements in Eq.~(\ref{eq: paraxial Omega}) with their real parts:
\begin{subequations}
    \label{eq: gauge transformation}
\begin{align}
    \rho_{lu}(\textbf{r},\tau) \quad \to \quad \frac{1}{2}\left(\rho_{lu}(\textbf{r},\tau) +\rho_{ul}^*(\textbf{r},\tau) \right),\\
     \rho_{ul}(\textbf{r},\tau) \quad \to \quad \frac{1}{2}\left(\rho_{ul}(\textbf{r},\tau) +\rho_{lu}^*(\textbf{r},\tau) \right).
\end{align}
\end{subequations}
The consequences of the transformation in Eq.~(\ref{eq: gauge transformation}) for the evolution of the field are discussed in the text following Eq.~(\ref{eq: gain}). The justification for the transformation in Eq.~(\ref{eq: gauge transformation}) is the generalized Girsanov theorem or application of the stochastic drift gauge as described in Refs.~\cite{2006'Deuar_stochastic-gauges, 2005'Deuar_PhD}. In general, the stochastic drift gauge transformation in Eq.~(\ref{eq: gauge transformation}) should be accompanied by re-weighting the stochastic trajectories when computing expectation values in Sec. \ref{sec: expectation values}.

\begin{figure*}
	\centering
	\includegraphics[width=1\linewidth]{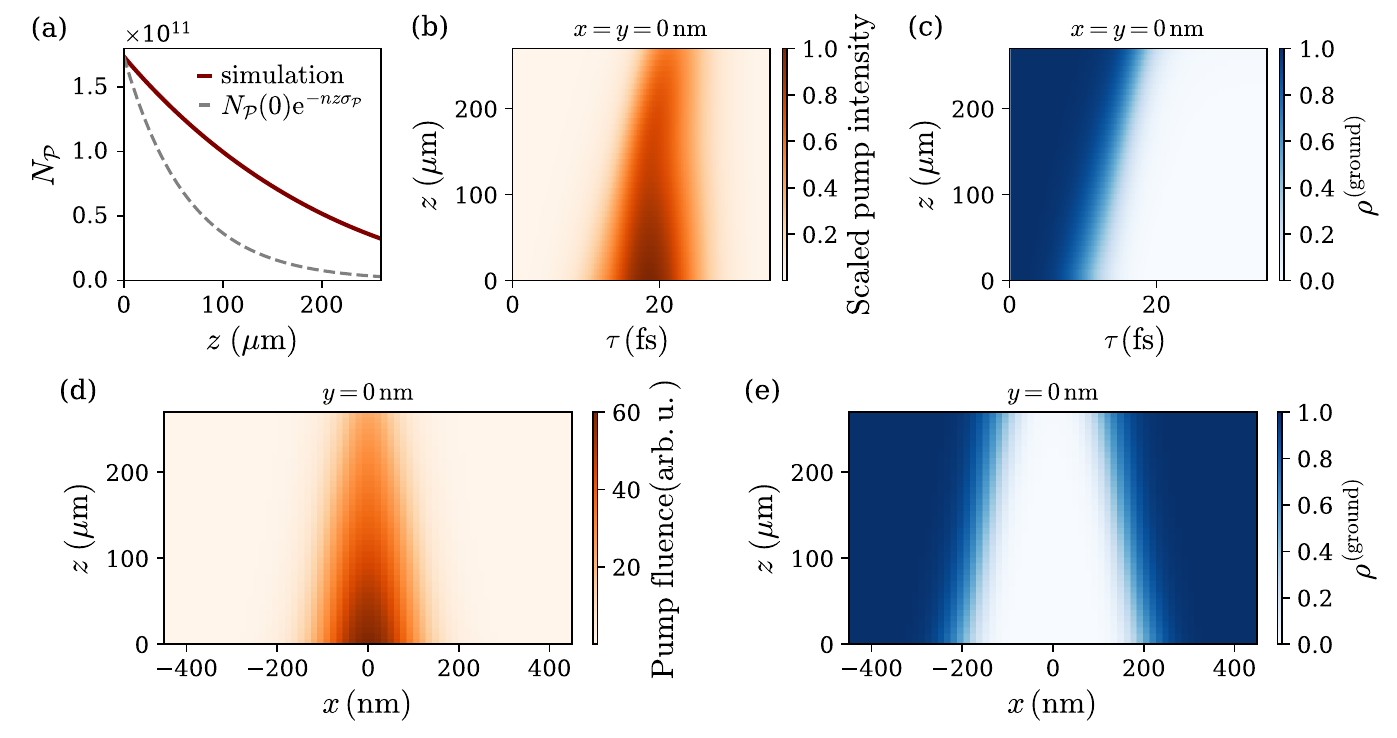}
	\caption{(a)~Number of pump photons as a function of propagation distance in the copper solution, the solid line denotes the numerical result of the simulation, and the dashed line denotes the number of photons obtained from the Beer-Lambert law; (b)~spatio-temporal evolution of the pump field intensity and (c)~initial neutral state population along the target axis; longitudinal sections at $y=0$ of (d)~pump photon fluence (number of pump photons per unit area) and (e)~population of the initial neutral state after the end of the pump pulse.}
	\label{fig: pump and ground state}
\end{figure*}

Unfortunately, for a large number of atoms, this weight coefficient might cause instabilities and worsen convergence. Therefore, \textit{we aim to neglect it in the current numerical implementation}. A more rigorous approach is the subject of further publications. To compensate for the absence of the weight coefficient, we reduce the need for gauging by minimizing the difference between atomic coherences $\rho_{ul}(\textbf{r},\tau)$ and $\rho_{lu}^*(\textbf{r},\tau)$. To minimize this difference, we take advantage of another degree of freedom in the representation of noise terms known as stochastic diffusion gauge analyzed in Refs.~\cite{2006'Deuar_stochastic-gauges, 2005'Deuar_PhD}. Namely, since there is no unique way to define noise terms satisfying correlation properties (\ref{eq: f_s noise correlations}), one can use this freedom to minimize the difference between atomic variables $\rho_{eg}(\textbf{r},\tau)$ and $\rho_{ge}^*(\textbf{r},\tau)$. Our goal is to minimize the average squared difference for each $s$, $\textbf{r}$, and $\tau$:
\begin{equation}
\label{eq: minimization problem}
\Big\langle\Big|\sum_{eg}T_{ges}\left(\rho_{eg}(\textbf{r},\tau)- \rho_{ge}^*(\textbf{r},\tau)\right)\Big|^2\Big\rangle,
\end{equation}
reducing the difference between the sources in the equations for $\Omega^{(+)}_{\text{det.}}(\textbf{r},\tau)$ and $\Omega^{(-)*}_{\text{det.}}(\textbf{r},\tau)$. The explicit form of the resulting noise terms used in the presented numerical simulations can be found in Appendix \ref{sec: Numerical}. In Section \ref{subsec: Spontaneous emission}, we demonstrate that the modified equations, as proposed in Appendix \ref{sec: Numerical} and this section, accurately reproduce spontaneous emission---the seeding stage of the amplification process.

\subsection{Pump propagation}
\label{subsec: Pump propagation}

The critical factor governing the dynamics of SF is the population inversion in the Cu ion. This inversion, in turn, is influenced by the dynamics of the pump pulse and the population of the neutral ground state. In Fig. \ref{fig: pump and ground state}, we illustrate the evolution of these quantities, computed using the expressions detailed in Appendix \ref{sec: Appendix: pump and decay}. 

Fig.~\ref{fig: pump and ground state} (a) presents the number of pump photons as a function of propagation distance. Notably, it displays a slower decline than anticipated by Beer's law, indicating substantial bleaching~\cite{2009'Nagler, 2014'Yoneda, 2015'Rackstraw}. In our formalism, this phenomenon is mainly driven by the reduction of the neutral ground state population, which possesses the largest absorption cross-section. As depicted in Fig.~\ref{fig: pump and ground state} (c), the ground state population diminishes to zero within the front part of the pump pulse, causing stronger absorption in this region compared to the tail of the pulse. This leads to pulse shortening~\cite{2023'Cardoch_modeling} and a shift of its peak to later times, as observed in Fig.~\ref{fig: pump and ground state} (b) and experimentally demonstrated in~\cite{2021'Inoue}. Additionally, the transverse profile of the pump pulse changes with propagation distance, as shown in Fig.~\ref{fig: pump and ground state} (d). Since bleaching is less pronounced for lower intensity pulses, the outer regions of the pulse experience stronger absorption than the central parts. As the pump pulse propagates, it decreases in energy, shrinks in transverse size, its duration shortens, and its peak shifts to a later time. These changes are reflected in the population of the ground state of the atoms, as illustrated in Figs. \ref{fig: pump and ground state} (c) and (e). The product of the pump flux and ground state population is the dominant contribution in Eq.~(\ref{eq: pump decay decoherence for rho_ij}), setting the stage for SF emission.

\begin{figure}[t]
	\centering
	\includegraphics[width=\linewidth]{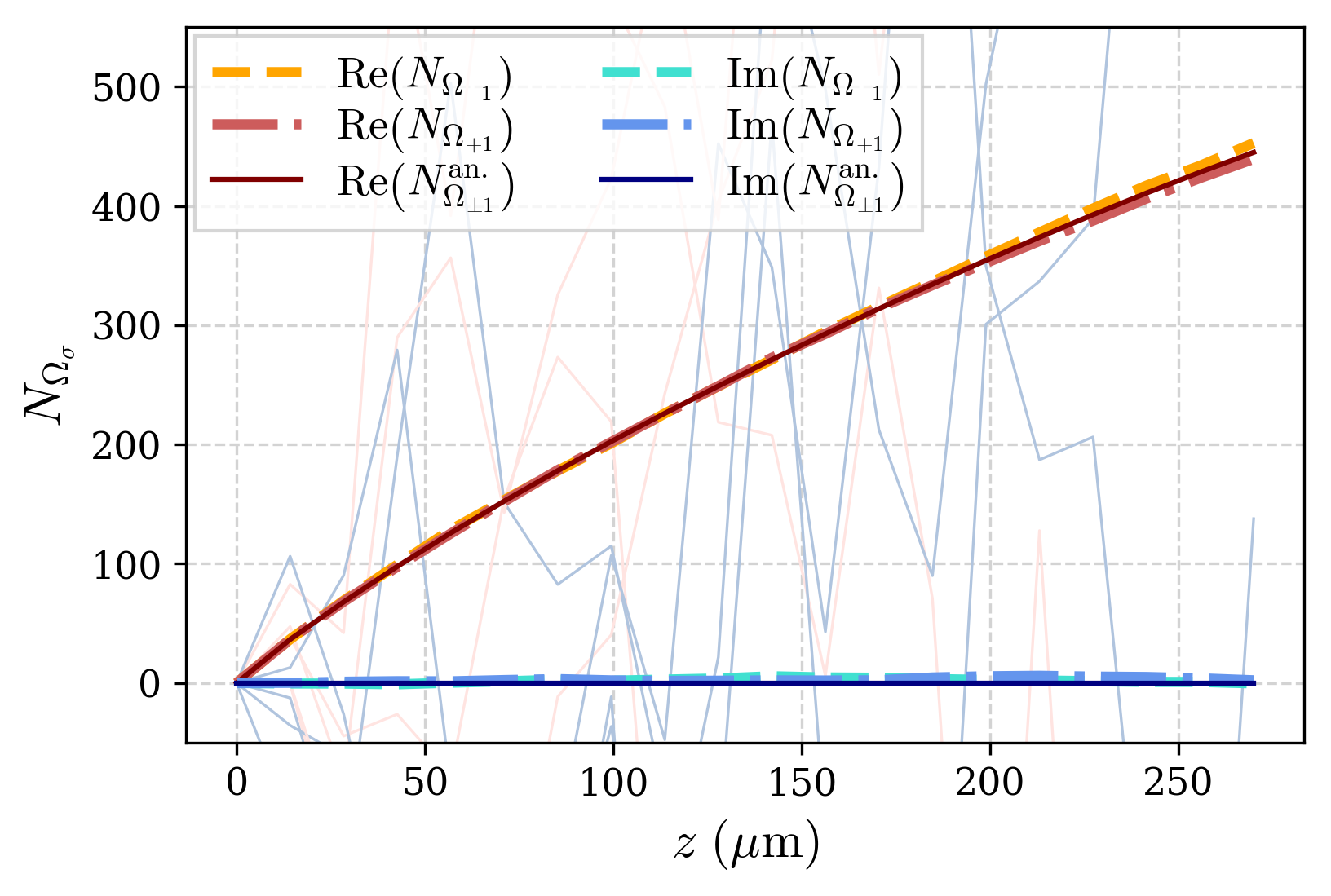}
	\caption[Numerically and analytically calculated average number of emitted photons as a function of propagation distance in Cu solution in the spontaneous emission approximation.]{Numerically ($N_{\Omega_{\sigma}}$) and analytically ($N_{\Omega_{\sigma}}^{\mathrm{an.}}$) calculated average numbers of emitted photons as functions of propagation distance in a Cu solution. Stimulated emission has been disregarded. Thin lines denote real and imaginary parts of the photon numbers for single realizations. The analytical solution is independent of the polarization $s=\pm1$.}
	\label{fig: spon_photons}
\end{figure}

\begin{figure}[t]
	\centering
	\includegraphics[width=0.95\linewidth]{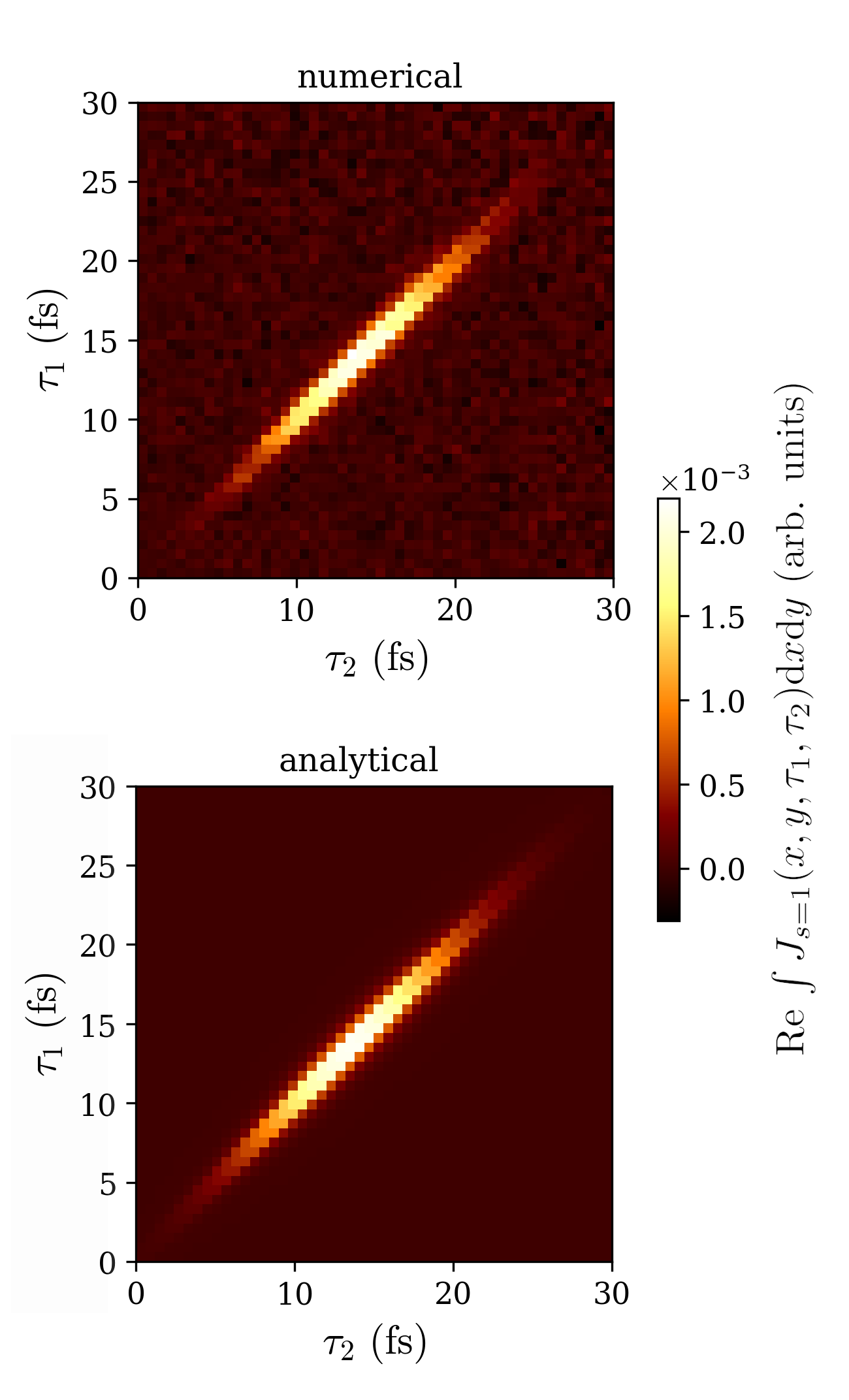}
	\caption{Comparison of the numerically and analytically calculated field correlation function (\ref{eq: Jsp-t1-t2_def}) at the target exit integrated over transverse spatial directions $x,y$ for $s=1$.
	}
	\label{fig: cu-spon_J}
\end{figure}

\subsection{Spontaneous emission}
\label{subsec: Spontaneous emission}

\begin{figure*}
	\centering
	\includegraphics[width=0.5\linewidth]{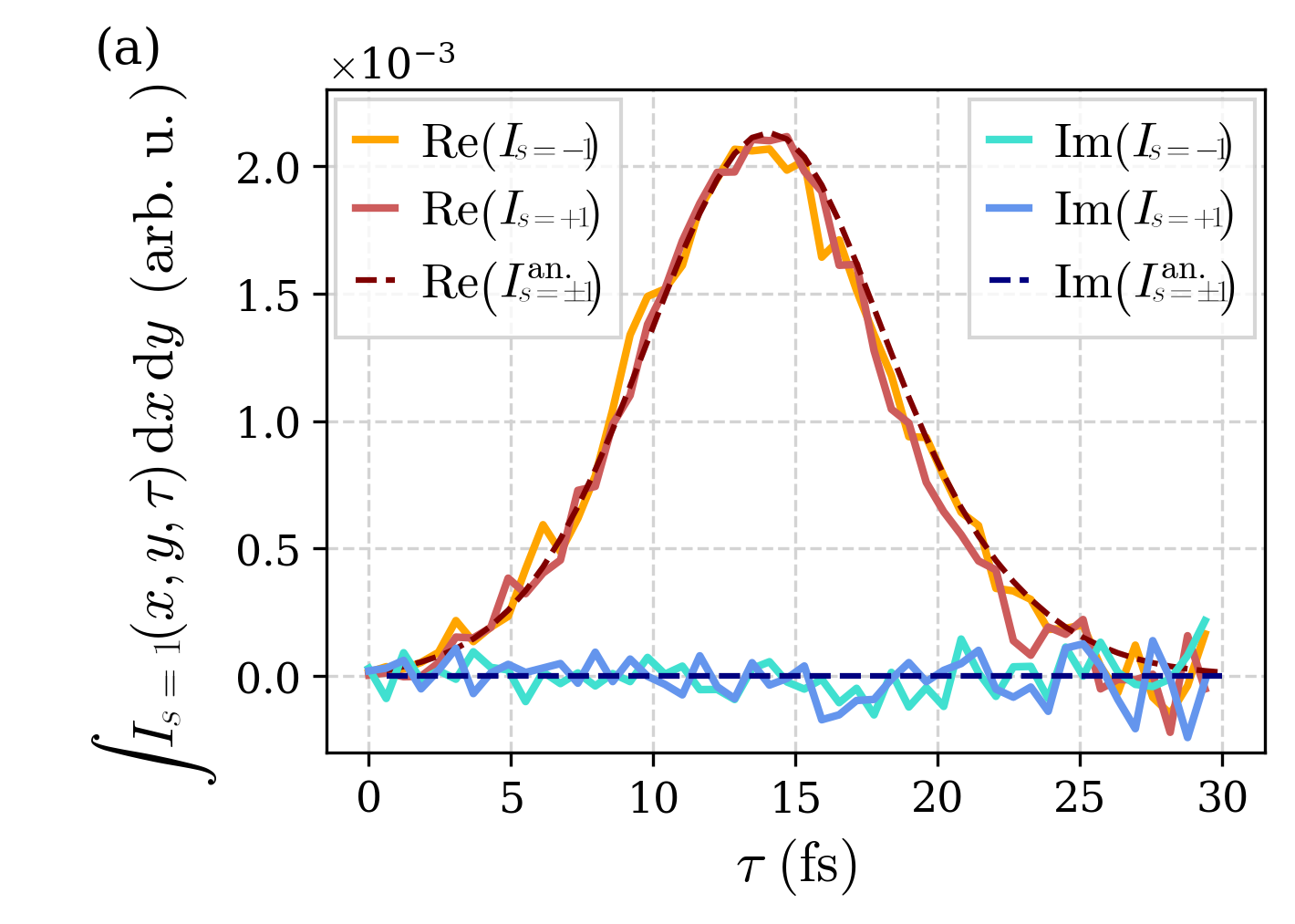}
	\includegraphics[width=0.43\linewidth]{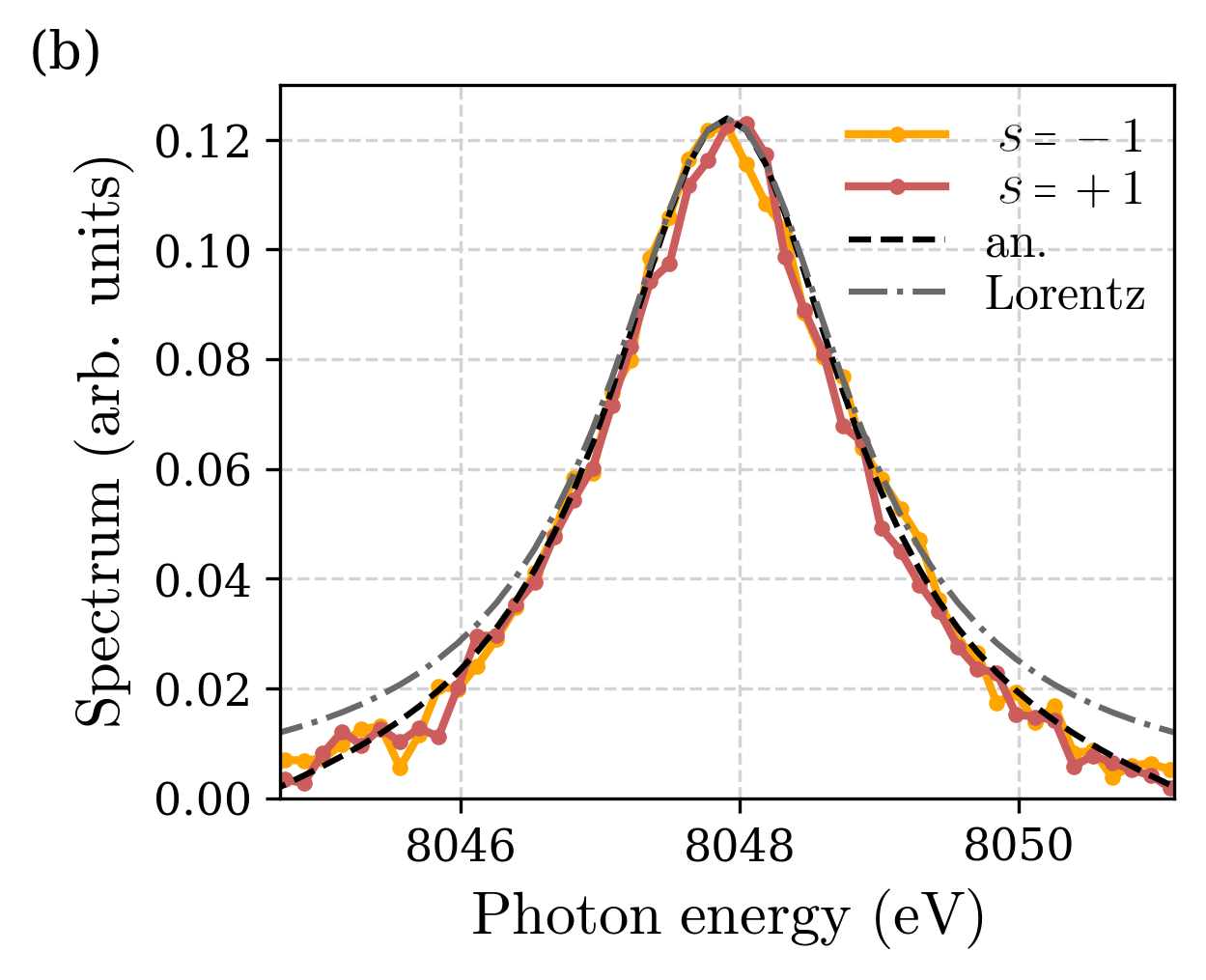}
	\caption[Temporal and spectral intensity profile of spontaneous radiation emitted by the target exit.]{(a)~Temporal profile of the averaged field intensity emitted at the target exit, integrated over transverse directions. (b)~Numerically calculated spectrum of the two radiation modes emitted at the target exit, along with the spectrum calculated from the analytical field correlation function (an.) and Lorentzian with the width corresponding to $2\gamma_{\text{dec.}}$ (Lorentz). The latter is scaled to the peak intensity of the analytical spectrum.
	}
	\label{fig: cu-spon_intensity}
\end{figure*}

Before delving into the analysis of SF simulation, it is advantageous to explore pure spontaneous emission, which serves as a valuable benchmark for our framework. To isolate spontaneous emission, we modify Eqs.~(\ref{eq: atomic equations}) and~(\ref{eq: paraxial Omega}) by eliminating the field variables in Eq.~(\ref{eq: bloch equations}). In other words, we exclude the stimulation responsible for amplification. Fig.~\ref{fig: spon_photons} compares the solutions of these modified equations to the photon number value derived from Eq.~(\ref{eq: Jsp-t1-t2_num}). Since the analysis of spontaneous emission properties necessitates averaging over a large number of stochastic realizations, and as this study does not focus on angular properties, a smaller grid has been used for Figs. \ref{fig: spon_photons}, \ref{fig: cu-spon_J}, and \ref{fig: cu-spon_intensity}. Specifically, we employ a 350 by 350 nm$^2$ spatial domain, with 6 by 6 grid points, with 20 grid points in the longitudinal direction, 50 points for the 30 fs temporal moving window, and average over $10^5$ stochastic realizations. To simplify the analysis, the absorption of the emitted field was omitted. The number of emitted photons varies significantly for different trajectories; moreover, it can take negative, as well as complex-valued values. This is expected from the structure of Eq.~(\ref{eq: paraxial Omega}). In the general case, the field variable representing the positive-frequency component $\Omega_s^{(+)}$ is not the complex conjugate of $\Omega_s^{(-)}$. This is an inherent property of the developed approach, directly related to the quantum-mechanical commutation relation of the field. The resulting doubling of the number of field variables, stemming from treating the variables corresponding to amplitudes of positive and negative frequency components as independent complex numbers, is typical for phase-space methods based on positive-P representation. A more in-depth discussion can be found in~\cite{2014'Drummond_book}. Single trajectories do not have a direct physical meaning and need to be averaged. As Fig.~\ref{fig: spon_photons} shows, after averaging over $10^5$ trajectories, the imaginary part of photon numbers vanishes, and the real part agrees with the analytically calculated values.

The dependence of the number of spontaneously emitted photons on the propagation distance $z$ is primarily defined by the behavior of the Green function in the paraxial approximation. For large distances $z$, the Green function can be approximated as:
\begin{align}
\label{eq: G num large z}
|G(\mathbf{r}-\mathbf{r}')| \approx \frac{1}{\lambda z}.
\end{align}
By substituting this asymptotic form of the Green function into Eqs.~(\ref{eq: Nsp-via-J}) and (\ref{eq: Jsp-t1-t2_num}), we obtain the following expression for the number of spontaneously emitted photons with polarization $s$ traversing cross-section $S$:
\begin{align}
\label{eq: Nph large z}
\frac{d N^{(\text{ph.})}_s}{d\tau} =
    \frac{3}{8\pi}
    \frac{S}{z^2}
    \Gamma_{\text{rad.}}
    \int  n(\textbf{r}) \langle\rho_{s}^{(\text{up.})}(\textbf{r},\tau)\rangle d\textbf{r}.
\end{align}
The number of emitted photons is proportional to the amount of excited atoms within the volume and the solid angle $S/z^2$ in which the spontaneous radiation is collected. The pre-factor agrees with quantum-mechanical calculations based on the Weisskopf-Wigner approximation~\cite{1997'Scully}, see also the discussion in~\cite{2019'Benediktovitch}.

For short propagation distances, the Green's function turns into a "broadened" delta-function:
\begin{align}
\label{eq: G num small z}
G(\textbf{r}-\textbf{r}')\approx \delta(\textbf{r}_\bot-\textbf{r}'_\bot),
\end{align}
meaning that the light travels almost parallel to the $z$ axis. The width of this delta-function is defined by the number of considered paraxial modes. Consequently, the number of spontaneously emitted photons of polarization $s$ traversing the cross-section of the sample is
\begin{align}
\label{eq: Nph small z}
\frac{d N^{(\text{ph.})}_s}{d\tau} =
    \frac{3}{8\pi}
   \Delta o\,
    \Gamma_{\text{rad.}}
    \int  n(\textbf{r}) \langle \rho_{s}^{(\text{up.})}(\textbf{r},\tau)\rangle d\textbf{r},
\end{align}
where $\Delta o$ is the solid angle spanned by the considered paraxial wave vectors.

The transition between the asymptotic behaviors of Eqs.~(\ref{eq: G num small z}) and (\ref{eq: G num large z}) determines the dependence of the number of spontaneously emitted photons on the propagation distance: for a small distance, the dependence is linear since the Green function in Eq.~(\ref{eq: G num small z}) is constant; for a larger propagation distance, the descending Green function results in the deceleration of the growth.

Fig.~\ref{fig: cu-spon_J} shows the correlation function of the field integrated\footnote{In the spontaneous emission regime, the contribution from each voxel is conditioned by the noise terms and is independent of other voxels. Consequently, the transversely-integrated quantities require fewer trajectories to obtain a given S/N level compared to the case of quantities at a specific transverse coordinate $x, y$.} over the simulation domain of the exit surface. The diagonal of the time correlation function determines the averaged temporal profile of the emitted intensity, while the width along the counter-diagonal is a measure of the temporal coherence. Averaging over stochastic realizations results in agreement between the numerically calculated values and calculations based on the analytical expression (\ref{eq: Jsp-t1-t2_num}). The temporal profile of the emitted radiation is presented in Fig.~\ref{fig: cu-spon_intensity} (a). Similarly to the case shown in Fig.~\ref{fig: spon_photons}, after averaging over trajectories, the imaginary part of the intensity vanishes, and the real part agrees with the analytical expressions obtained from (\ref{eq: Jsp-t1-t2_num}), following the population of the upper state. According to the Wiener–Khinchin theorem, the Fourier transform with respect to $\tau_1-\tau_2$ provides the spectrum of the emitted radiation. Fig.~\ref{fig: cu-spon_intensity} (b) shows the resulting spectrum, which agrees well with the analytical expression based on (\ref{eq: Jsp-t1-t2_num}) and is close to the Lorentzian profile with $\gamma_{\text{dec.}}$ HWHM. Also, as expected, Figs. (\ref{fig: spon_photons}) and (\ref{fig: cu-spon_intensity}) show that the behavior of spontaneous emission does not depend on polarization.

\begin{figure}[b]
	\centering
	\includegraphics[width=\linewidth]{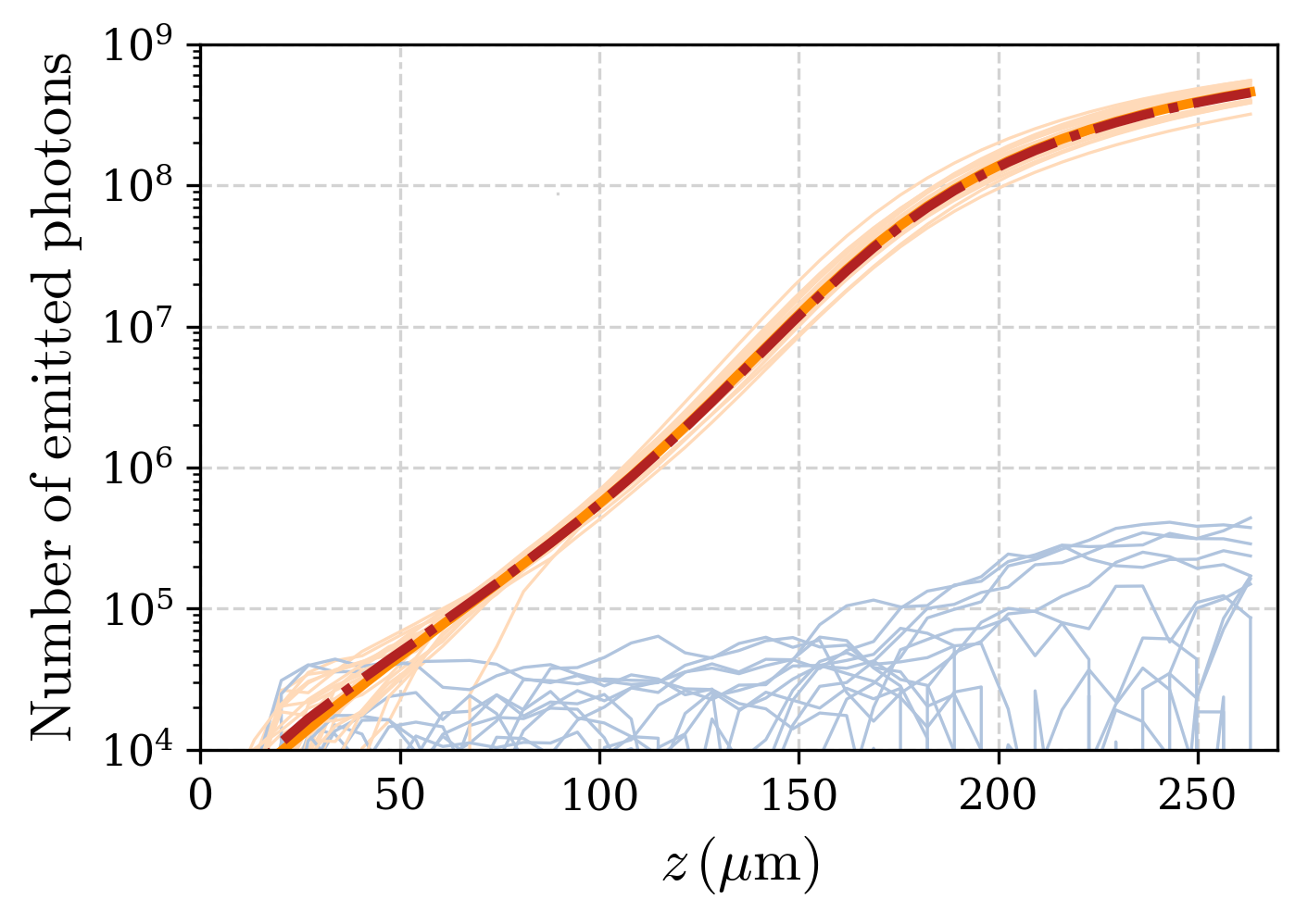}
	\caption{The total number of emitted photons as a function of propagation distance $z$. The thin lines represent individual trajectories, with orange and blue colors denoting the real and imaginary parts, respectively. The thick red line represents the mean value obtained from 300 trajectories. All lines correspond to polarization $s=1$.}
	\label{fig: Nph vs z}
\end{figure}

The ability to reproduce the field correlation function (\ref{eq: Jsp-t1-t2_num}) in the limit of spontaneous emission by means of the noise terms is a distinctive feature of the presented formalism, as opposed to the formalism in Ref.~\cite{2000'Larroche} based on phenomenological noise terms, which is not capable of reproducing the analytical results. The agreement in both spectral and temporal profiles is achieved by the non-trivial structure of the noise terms of the applied stochastic formalism.


\subsection{Field evolution and transverse properties}
\label{subsec: Field evolution and transverse properties}

The propagation of spontaneously emitted radiation, as discussed in Sec.~\ref{subsec: Spontaneous emission}, within the pumped medium, as discussed in Sec.~\ref{subsec: Pump propagation}, leads to the amplification of the radiation and subsequent saturation. Fig.~\ref{fig: Nph vs z} illustrates the number of emitted photons as a function of the propagation distance within the medium. The calculation of emitted photons is based on the photon flux in Eqs.~(\ref{eq: Jsp-t1-t2_def}) and (\ref{eq: Nsp-via-J}), integrated over time and the transverse simulation domain.

In the beginning of the medium, the evolution is conditioned by spontaneous emission, with the noise terms playing a dominant role. Here, similar to Fig.~\ref{fig: spon_photons}, individual trajectories exhibit a large scatter and comparable values of real and imaginary parts.
For larger propagation distances, an approximate exponential growth of the emitted photon number is observed. This regime is sometimes referred to as the linear gain regime or ASE~\cite{1972'Peters, 1998'Svelto, 2005'Kocharovsky}. In this regime, the emitted field and coherences are still small. Consequently, in the equations governing the evolution of the upper and lower states, terms proportional to the field-coherence product can be neglected. At larger distances, the evolution of upper and lower states is conditioned by pump and decay terms only. Under these conditions, Eqs.~(\ref{eq: atomic equations}) become linear in time- and space-dependent coefficients. Considering, for simplicity, the case of a large decoherence rate $\gamma_{\text{dec.}}$, the produced linear response of the coherences to the field results in amplification with the following gain coefficient $g_s(\textbf{r},\tau)$:
\begin{align}
\label{eq: gain}
    g_s(\textbf{r},\tau)=\frac{3}{8\pi}
    n(\textbf{r})\lambda^2
\frac{\Gamma_{\text{rad.}}}{\gamma_{\text{dec.}}}
    \left(\rho_s^{(\text{up.})}(\textbf{r},\tau)-\rho_s^{(\text{low.})}(\textbf{r},\tau)\right).
\end{align}
{Thanks to the modification of Eq.~(\ref{eq: paraxial Omega}) discussed in Sec. \ref{Gauging run-away trajectories}, the combinations $\Omega_s^{(\pm)}+\Omega_s^{(\mp)*}$ are amplified with the gain coefficient defined in Eq.~(\ref{eq: gain}), while the non-physical combinations $\Omega_s^{(\pm)}-\Omega_s^{(\mp)*}$ are not amplified. As a result, the spurious imaginary component of the photon number, computed exclusively from individual stochastic realizations, does not exhibit growth in the ASE regime, as demonstrated in Fig.~\ref{fig: Nph vs z}.}

The exponential growth of the photon number is eventually limited by nonlinear effects. When the produced field and the generated coherences become large enough, the corresponding terms in Eq.~(\ref{eq: atomic equations}) can no longer be neglected. As a result, the population inversion decreases, and saturation sets in. Depending on the parameters of the system, spatio-temporal ringing (due to Rabi oscillations in the population) of the emitted field intensity may be observed~\cite{2019'Benediktovitch}. As depicted in Fig.\,\ref{fig: Nph vs z}, the imaginary part of the photon number increases as the beam propagates. Nevertheless, it consistently remains negligible compared to the real part. In both ASE and saturation regimes, the noise terms are smaller than the regular terms in Eqs.~(\ref{eq: atomic equations}) and (\ref{eq: paraxial Omega}). As a consequence, the scatter of the real part of individual realizations of the number of photons exhibits, upon propagation, approximately the same width on the logarithm scale. This observation suggests that, within logarithmic accuracy, a few trajectories are sufficient to determine the mean photon number in the deep ASE and saturation regimes.   

\begin{figure*}
	\centering
	\includegraphics[width=0.85\linewidth]{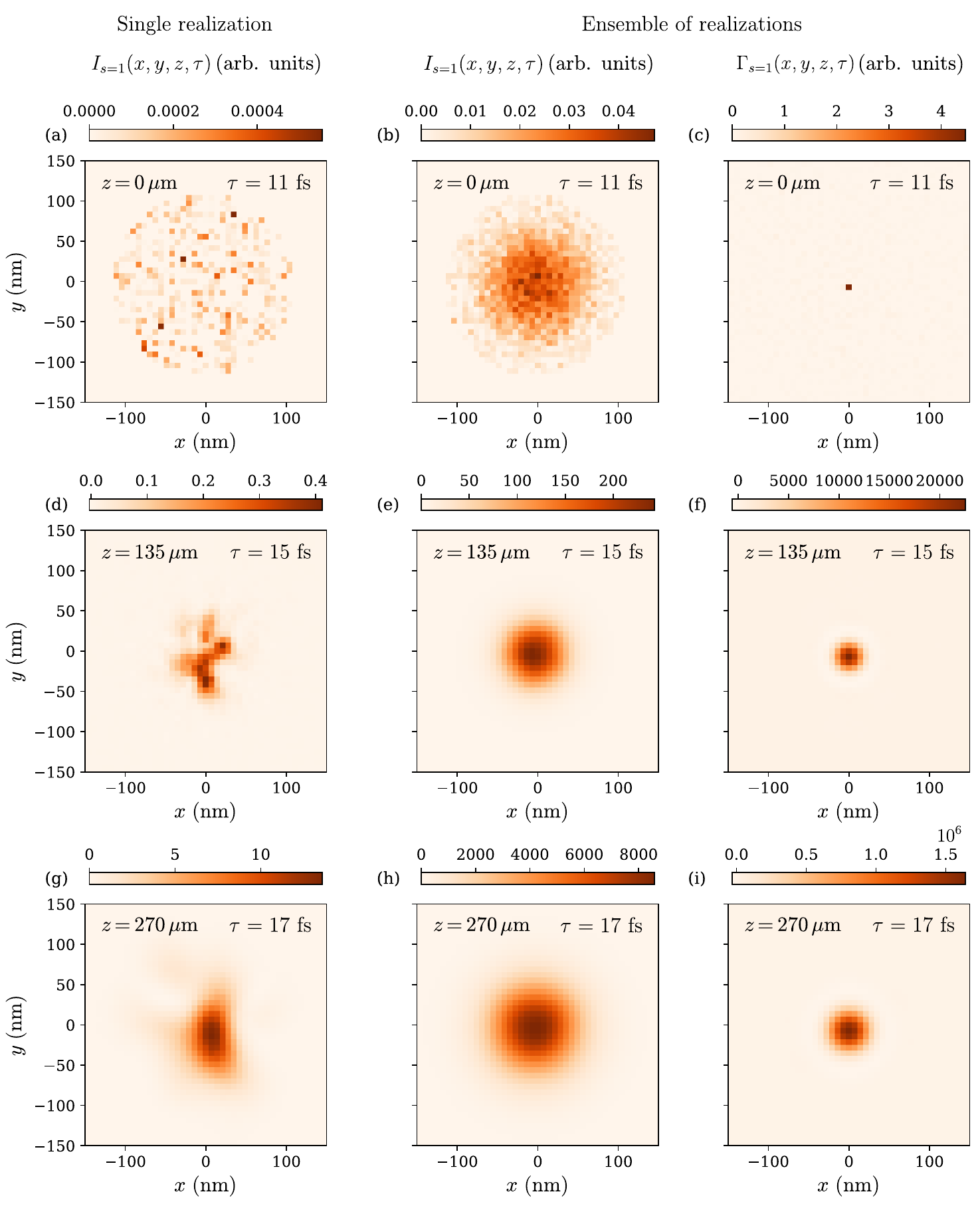}
	\caption{Transverse field distribution: (a), (d), (g) Intensity distribution of the emitted radiation for a single realization. (b), (e), (h) The same quantity calculated by averaging over an ensemble of realizations. (c), (f), (i) Field correlation function in the transverse direction calculated according to Eq.~(\ref{eq: transverse corr fucntion definition}). The rows of figures (a)~--~(c), (d)~--~(f), and (g)~--~(i) correspond to propagation distances of $z=0$~\textmu m~(SE~regime), $z=135$~\textmu m~(ASE~regime), and $270$~\textmu m~(SF~regime), with peak intensity observed at times $\tau_{\mathrm{peak}}=$ 11~fs, 15~fs, and 17~fs, respectively. For averaging, 20 000 numerical realizations are used for the SE regime and 1300 for the ASE/SF regimes. }
 	\label{fig: transverse field properties}
\end{figure*}

\begin{figure*}
	\centering
	\includegraphics[width=0.95\linewidth]{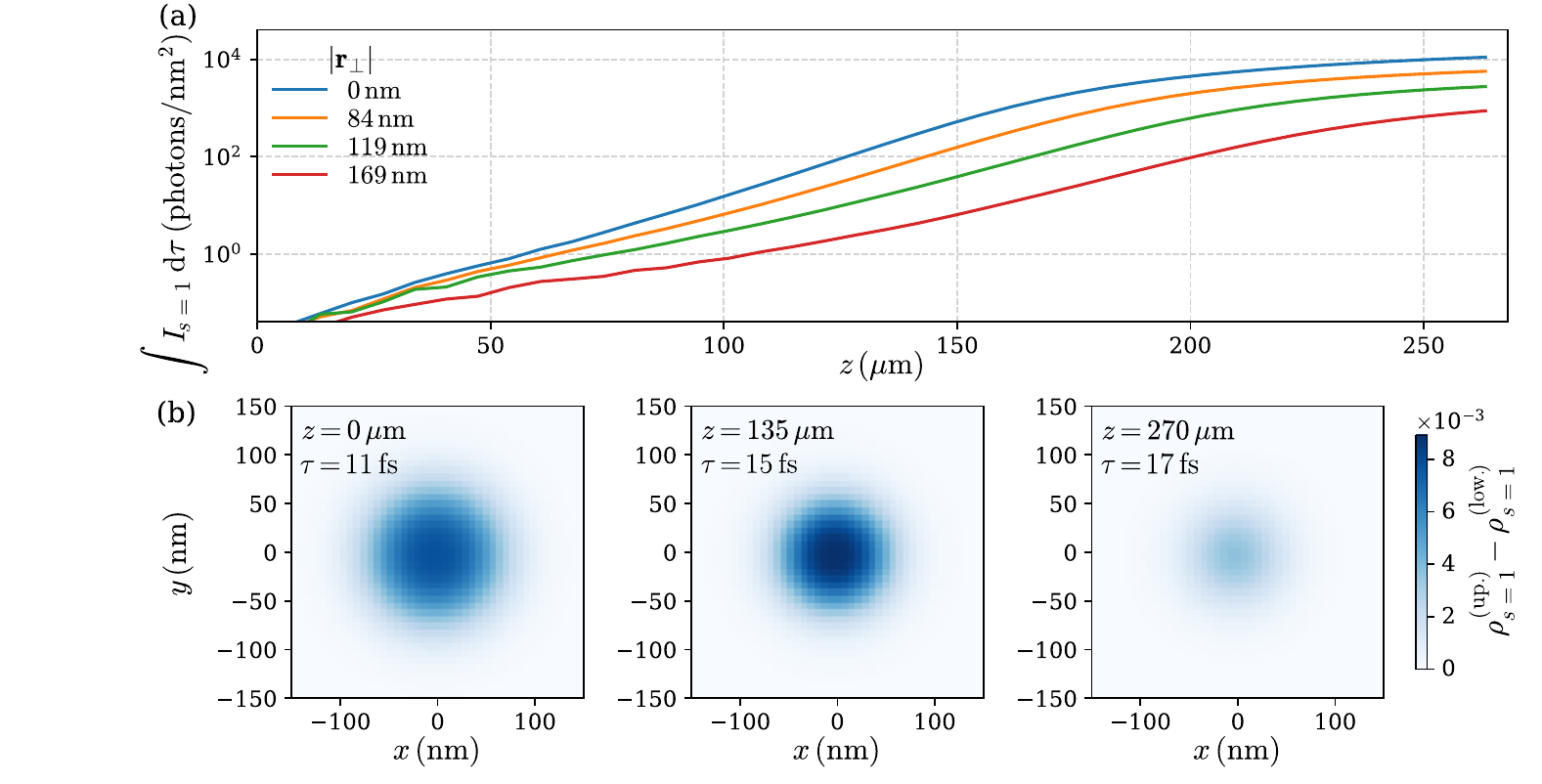}
	\caption{(a)~Emitted flux (time-integrated number of photons per unit area) as a function of propagation distance $z$ for several positions in the transverse direction: on the axis ($|\textbf{r}_{\bot}|=0$) and progressively further from the axis ($|\textbf{r}_{\bot}|=84$ nm,\,$119$\,nm,\,$169$\,nm); (b)~Transverse cuts of the effective population inversion for propagation distances $z=0$,\,$135$\,\textmu m,\,$270$\,\textmu m and times $\tau_{\mathrm{peak}}=$ 11~fs, 15~fs, and 17~fs, respectively. The number of numerical realizations is 1300.}
	\label{fig: transverse gain curves}
\end{figure*}

The knowledge of the transverse field distribution is essential for the applications of SF, as it is directly related to the angular distribution of the emitted intensity. In Ref.~\cite{2020'Kroll}, the larger angular spread of seeded-SF emission compared to the angular spread of the seed pulse enabled the detection of the seeded Mn K${\beta}$ signal. Another example for which the angular properties of the SF are crucial is the XLO: in this case, the angular divergence determines the efficiency of the in-coupling of the SF radiation into the crystal cavity.

Fig.~\ref{fig: transverse field properties} illustrates the transverse distribution of the field. Each row corresponds to a different propagation distance, representing qualitatively distinct regimes: spontaneous emission (SE), ASE, and saturation or superfluorescence (SF). Figs.~\ref{fig: transverse field properties} (a), (d), (g) show the intensity distribution for a single run of the numerical scheme (a single trajectory). In the case of SE represented by Fig.~\ref{fig: transverse field properties} (a), the spontaneously emitted intensity varies stochastically from pixel to pixel. Upon propagation, due to diffraction, neighboring pixels establish a correlation resulting in a speckle-like pattern. As Figs.~\ref{fig: transverse field properties} (d), (g) show, the size of the speckles grows upon propagation. As discussed for Figs. \ref{fig: spon_photons}, \ref{fig: Nph vs z}, a single realization does not, strictly speaking, have a direct physical meaning—--an ensemble of realizations is needed to determine the observable. In our case, one of the properties of interest is the transverse size of the emitted field, which can be deduced from the intensity profile shown in Figs.~\ref{fig: transverse field properties} (b), (e), (h). Averaging over several trajectories results in a smooth and axially-symmetric distribution. Another property of interest is the transverse coherence of the emitted radiation. A rough estimate of this property can be obtained based on the average size of the speckles. As an observable quantifying the transverse coherence, we can consider the transverse correlation function~\cite{1997'Scully}:
\begin{align}
\label{eq: transverse corr fucntion definition}
\Gamma_{s}(\textbf{r}_{\perp},z,\tau) = 
    \int 
        \langle\Omega_s^{(+)}(\textbf{r}',z,\tau)
        \Omega_s^{(-)}(\textbf{r}_{\perp}'+\textbf{r}_{\perp},z,\tau)\rangle d \textbf{r}_{\perp}' 
\end{align}
shown in Figs.~\ref{fig: transverse field properties} (c), (f), (i). As expected, the size of the transverse correlation function approximately agrees with the size of the speckles. Initially, in the SE regime Fig.~\ref{fig: transverse field properties} (c), the transverse coherence has the size of just one pixel and grows upon propagation. Since the transverse size of the correlation function is smaller than the transverse width of the intensity profile, the SF field is not fully transversely coherent, as also the speckle structure of single trajectories suggests. The ratio of the transverse width of the intensity profile to speckle size gives an estimate of the effective number of transverse modes.

The evolution of the transverse field profile is conditioned by the distribution of the population inversion as well as diffraction effects. Fig.~\ref{fig: transverse gain curves} (a) shows the photon flux as a function of the propagation distance at several transverse positions. As expected, the center of the beam exhibits the highest flux, which gradually decreases towards the edges, in accordance with the distribution of the population inversion shown in Fig.~\ref{fig: transverse gain curves} (b). The population inversion decreases as the beam propagates due to the combined effects of pump-pulse absorption and nonlinear (saturation) effects. In the ASE regime, a larger population inversion leads to larger amplification of the emitted radiation. As a result, the beam experiences gain guiding~\cite{1992'Salin} and decreases in size. However, as Fig.~\ref{fig: transverse gain curves} (a) shows, for the inner parts of the beam, the transition from ASE to saturation takes place at a shorter propagation distance than for the outer parts. As a result, after the saturation regime sets in for the on-axis part of the beam, the beam size increases—--as the middle column of Fig.~\ref{fig: transverse field properties} also illustrates.

\subsection{Spectral-angular properties}

The direct experimental measurement of the transverse and temporal profiles of the emitted x-ray field poses considerable difficulty. X-ray fields are typically measured in the far field, providing the angular distribution of the emission. The temporal properties of the x-ray pulse are often inferred from spectral analysis using a grating or crystal spectrometer. For example, in~\cite{2022'Zhang}, fringes in the spectrum were used to reconstruct the temporal separation between the peaks of the underlying field temporal profile. If a two-dimensional detector is used to measure the field after the analyzer crystal, the detector provides a spectral and angular distribution of the emission. To obtain this distribution from the presented formalism, we first perform the Fourier transform of the fields $\Omega_{s}^{(\pm)}(\textbf{r},\tau)$:
\begin{multline*}
    \bar{\Omega}_s^{(\pm)}(\theta_x,\theta_y,z,\omega)=\int \frac{dxdyd\tau}{(2\pi)^3}\Omega_s^{(\pm)}(\textbf{r},\tau)\\\times\exp\left[{\pm ik_0(x\theta_x+y\theta_y) \mp i\omega\tau}\right].
\end{multline*}
Similarly to Eqs. (\ref{eq: Jsp-t1-t2_def}) and (\ref{eq: Nsp-via-J}), we express the spectral and angular distribution $\bar{I}_s(\theta_x,\theta_y,z,\omega)$ as follows: 
\begin{equation}
\bar{I}_s(\theta_x,\theta_y,z,\omega)=\frac{\langle\bar{\Omega}_{s}^{(+)}(\theta_x,\theta_y,z,\omega)\bar{\Omega}_{s}^{(-)}(\theta_x,\theta_y,z,\omega)\rangle}{\frac{3}{8\pi}\lambda^2_0 \Gamma_{\text{rad.}}}.
\end{equation}

\begin{figure*}
    \centering
    \includegraphics[width=\linewidth]
    {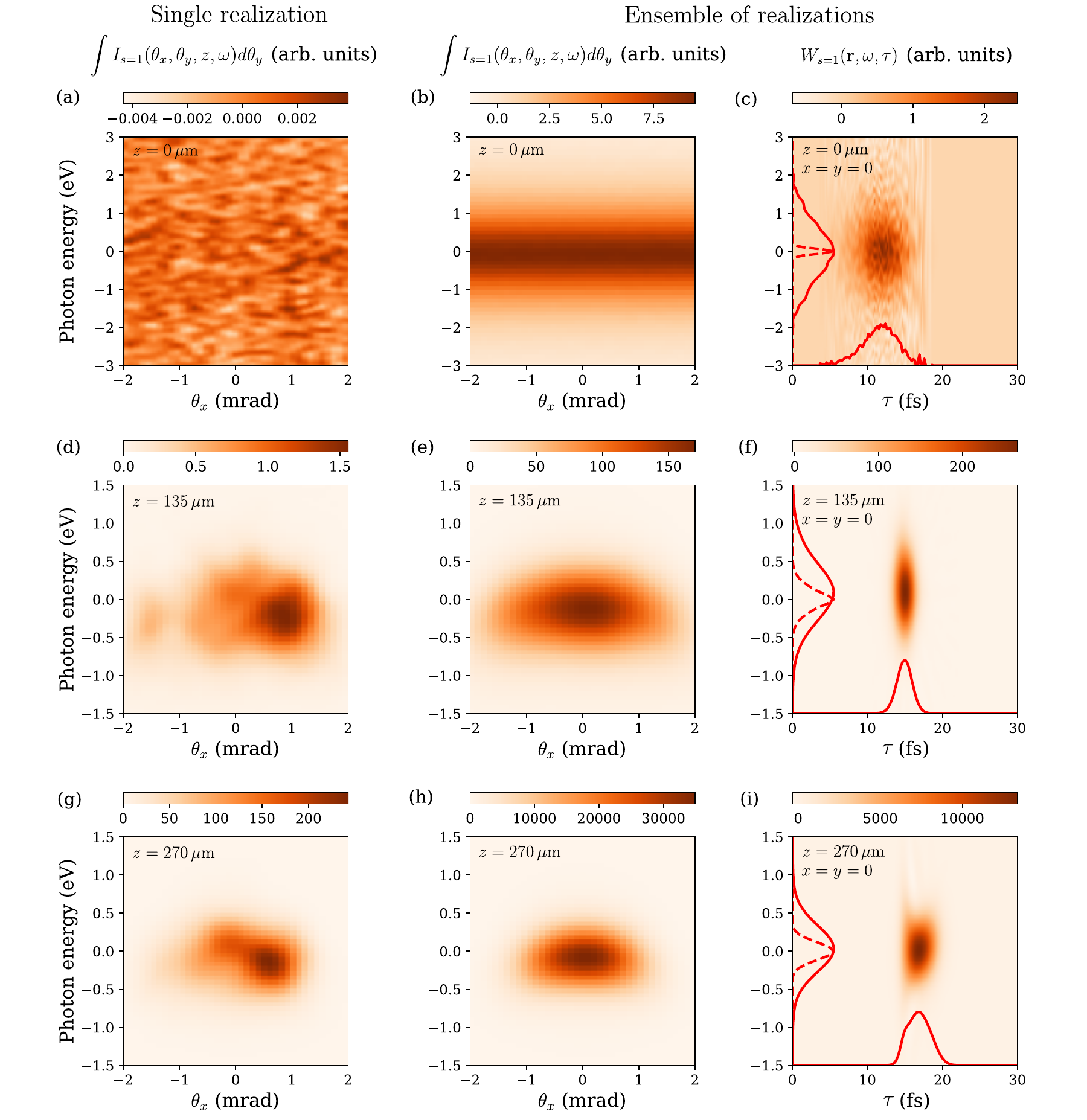}
    \caption{Spectral-angular properties of the emitted radiation: (a), (d), (g)  Spectral-angular distribution of the emitted radiation integrated over the $\theta_y$ direction for a single realization. (b), (e), (h) The same quantity, calculated by averaging over an ensemble of realizations. (c), (f), (i) Wigner distribution (\ref{eq: Wigner distribution}) at the center of the beam ($x=y=0$). The solid red lines represent projections of the Wigner function onto the time and energy axes, providing the temporal intensity profile and spectrum. The dashed red line illustrates how the spectrum would appear if calculated according to Eq.~(\ref{eqs: I tranform limited}). The series of figures (a) -- (c), (d) -- (f), and (g) -- (i) correspond to propagation distances of $z=0$ (SE regime), $z=135$\,\textmu m (ASE regime), and $z=270$\,\textmu m (SF regime). The number of numerical realizations is 1,000 for the SE regime and 100 for the ASE and SF regimes, respectively. }
    \label{fig: spectral-angular and Wigner plots}
\end{figure*}

Fig.~\ref{fig: spectral-angular and Wigner plots} (a) displays a typical spectral-angular distribution for a single realization in the SE regime. As expected, it exhibits isotropy in the angular direction and is highly stochastic. Fig.~\ref{fig: spectral-angular and Wigner plots} (d) corresponds to the ASE regime and reveals multiple spikes associated with the field modes emerging from spontaneous emission noise. In Fig.~\ref{fig: spectral-angular and Wigner plots} (g), we observe a similar distribution at a greater propagation distance in the SF regime. Here, the most intense modes are further amplified, while less intense modes diminish. This behavior is reminiscent of the well-known mode clearance phenomena in FEL physics~\cite{2016'Pellegrini}.

To confirm these observations at a single-trajectory level, spectral-angular intensity profiles are averaged over an ensemble of realizations, as shown in Figs.~\ref{fig: spectral-angular and Wigner plots} (b), (e), (h). Specifically, the angular distribution remains isotropic for the SE case [Fig.~\ref{fig: spectral-angular and Wigner plots}~(b)], becomes narrower in the ASE case [Fig.~\ref{fig: spectral-angular and Wigner plots}~(e)], and further narrows down in the SF regime [Fig.~\ref{fig: spectral-angular and Wigner plots}~(h)].

The temporal-spectral properties of the emitted field can be conveniently characterized in terms of the Wigner distribution:
\begin{multline}
\label{eq: Wigner distribution}
W_{s}(\textbf{r},\omega,\tau) = 
    \int \frac{d \tau'}{2\pi}
        \langle
            \Omega_s^{(+)}(\textbf{r}_{\perp},z,\tau+\tau'/2)
            \\\Omega_s^{(-)}(\textbf{r}_{\perp},z,\tau-\tau'/2)
        \rangle e^{i\omega \tau'}.
\end{multline}
The projection of the Wigner distribution on the time axis gives the averaged temporal intensity profile, and the projection on the frequency axis gives the spectral profile. Figs.~\ref{fig: spectral-angular and Wigner plots} (c), (f), (i) show the Wigner distribution for propagation distances corresponding to the SE, ASE, and SF regimes. The profile on the time-axis is influenced by the time dependence of the population inversion, while the profile on the frequency-axis is influenced by the decoherence rate of the transition. As a result, the spectral width of the produced radiation is broadened compared to the width determined by assuming a Fourier-limited pulse. The radiation is not transform-limited. If the radiation were fully coherent, and the field amplitude had a constant phase across the pulse, the connection between the spectrum 
\begin{equation*}
    I_s(\textbf{r},\omega) =  \frac{\int d\tau W_s(\textbf{r},\omega, \tau)}{{\frac{3}{4}\lambda^2_0 \Gamma_{\text{rad.}}}}
\end{equation*} and temporal intensity profiles 
\begin{equation*}
I_s(\textbf{r},\tau) = \frac{\int d\omega W_s(\textbf{r},\omega, \tau)}{{\frac{3}{8\pi}\lambda^2_0 \Gamma_{\text{rad.}}}}    
\end{equation*}
for transform-limited pulses would be given by:
\begin{align}
\label{eqs: I tranform limited}
I_s(\textbf{r},\omega) =
\left|
\int \frac{d\tau}{2\pi} e^{i\omega \tau} \sqrt{I_s(\textbf{r},\tau)}
\right|^2.
\end{align}
The spectral profile calculated according to Eq.~(\ref{eqs: I tranform limited}) is depicted as a dashed line in Figs.~\ref{fig: spectral-angular and Wigner plots} (c), (f), (i). During the propagation, the spectral profile becomes narrower due to the gain-narrowing effect of the ASE regime~\cite{1998'Svelto}, as shown in the comparison of Figs.~\ref{fig: spectral-angular and Wigner plots} (c) and (f). Deep in the SF regime, the emitted field eventually becomes so large that the induced dynamics, similar to Rabi oscillations, may cause additional broadening and splitting~\cite{2019'Mercadier}. In the presented example, this regime, however, is not pronounced.

{\subsection{Field polarization properties}

The scheme presented in Eq.~(\ref{eq: e and g explicit for Cu}) includes all the energy levels involved in the Cu-K${\alpha_{1}}$ emission. By considering the degeneracy of these states, we gain complete access to the polarization properties of the emitted radiation. For a quantitative analysis of polarization, we employ the Stokes parameters $S_i(\textbf{r},\tau)$, with $i = 0,1,2,3$ as introduced in Ref.~\cite{1998'Brosseau}. Specifically, to quantify the presence of circularly polarized components in the field, we focus on $S_3(\textbf{r},\tau)$, defined as:

\begin{figure*}[t!]
    
    \centering
    \includegraphics[width=0.95\linewidth]{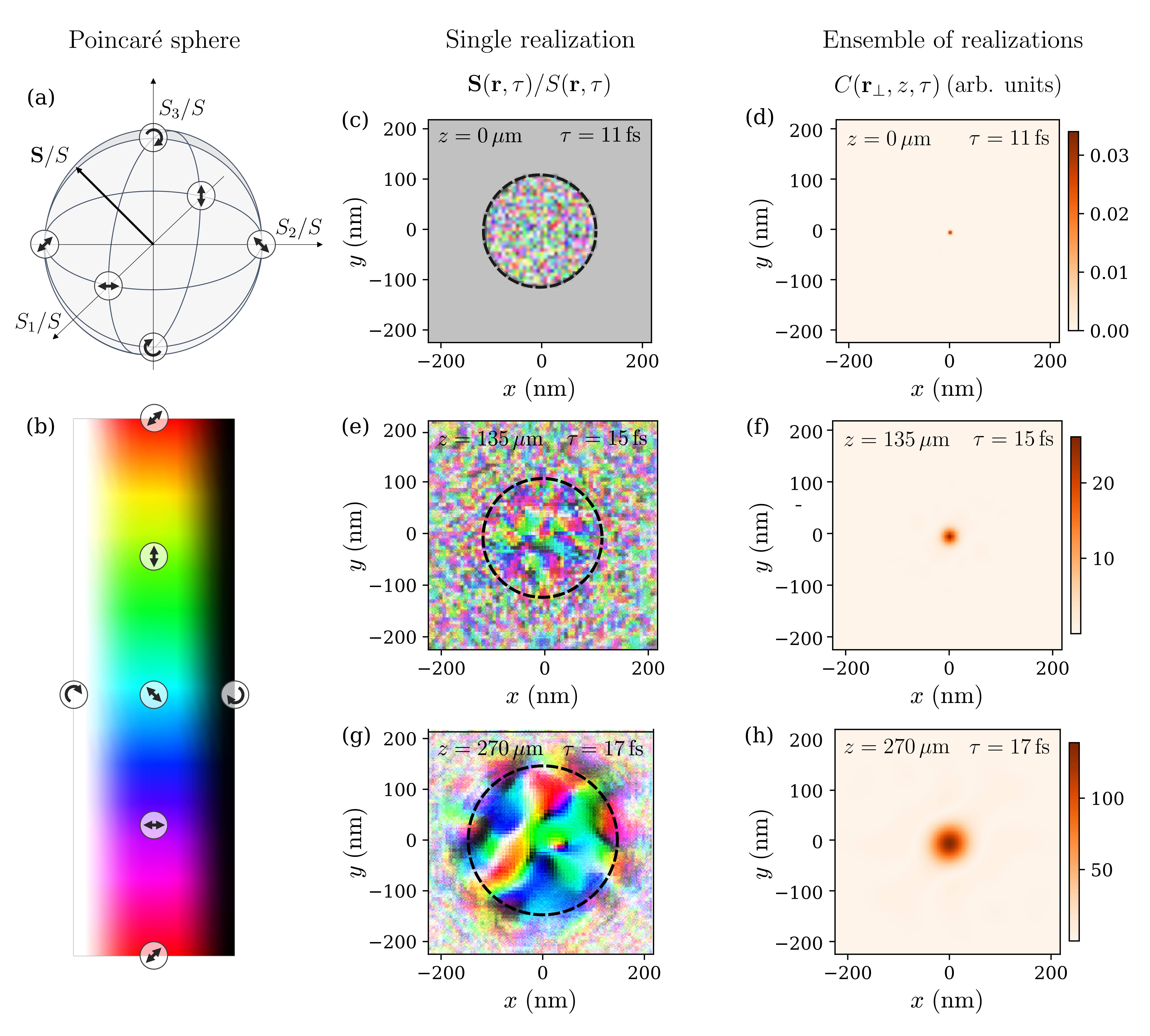}

    \caption{Field polarization of the emitted radiation. Left side: (a) Polarization states represented on the Poincar\'e sphere; (b) Color-coding of polarization states. Right side: (c), (e), (g) Transverse cross-section of polarization state for a single realization; (d), (f), (h) Correlation of the polarization state quantified according to Eq.~(\ref{eq: G4 C}). The black dashed line represents the area of noticeable pulse intensity. The row of figures (c) -- (d), (e) -- (f), and (g) -- (h) correspond to propagation distance of $z=0$ (SE regime),  $z=135$\,\textmu m (ASE regime), and $z=270$ \textmu m (SF regime) and time $\tau_{\mathrm{peak}}=$ 11~fs, 15~fs, and 17~fs respectively.} \label{fig: field polarization} 
\end{figure*}

\begin{subequations}
    \label{eq: G2ab via Stokes}
\begin{equation}
    S_3(\textbf{r},\tau) = \Omega_{-1}^{(+)}(\textbf{r},\tau)
        \Omega_{-1}^{(-)}(\textbf{r},\tau) - \Omega_{+1}^{(+)}(\textbf{r},\tau)
        \Omega_{+1}^{(-)}(\textbf{r},\tau).
\end{equation}
It is simply the difference in intensity between two distinct circular polarizations, reaching its maximum and minimum values when the field is solely represented by right- or left-hand polarization, respectively. Intermediate values of $S_3(\textbf{r},\tau)$ signify the presence of linearly polarized components that can be quantified by $S_1(\textbf{r},\tau)$ and $S_2(\textbf{r},\tau)$ defined as follows:
\begin{equation}
    S_1(\textbf{r},\tau) = \Omega_{+1}^{(+)}(\textbf{r},\tau)
        \Omega_{-1}^{(-)}(\textbf{r},\tau) + \Omega_{-1}^{(+)}(\textbf{r},\tau)
        \Omega_{+1}^{(-)}(\textbf{r},\tau),
\end{equation}
\begin{equation}
    S_2(\textbf{r},\tau) = i\left[\Omega_{+1}^{(+)}(\textbf{r},\tau)
        \Omega_{-1}^{(-)}(\textbf{r},\tau) - \Omega_{-1}^{(+)}(\textbf{r},\tau)
        \Omega_{+1}^{(-)}(\textbf{r},\tau)\right].
\end{equation}
Similarly to $S_3(\textbf{r},\tau)$, $S_1(\textbf{r},\tau)$ is the difference in intensities carried by horizontal and vertical polarizations while $S_2(\textbf{r},\tau)$ corresponds to diagonal polarizations. The introduced parameters $S_i(\textbf{r},\tau)$ form a vector $\textbf{S}(\textbf{r},\tau)$. Its length $S(\textbf{r},\tau)$ equals the last Stokes parameter $S_0(\textbf{r},\tau)$ defined as follows:
\begin{multline}
S_0(\textbf{r},\tau)=S(\textbf{r},\tau)=\Omega_{-1}^{(+)}(\textbf{r},\tau)
        \Omega_{-1}^{(-)}(\textbf{r},\tau) \\+ \Omega_{+1}^{(+)}(\textbf{r},\tau)
        \Omega_{+1}^{(-)}(\textbf{r},\tau),
\end{multline}
\end{subequations}
which is proportional to the total intensity of the field. To eliminate information about the intensities, which is irrelevant for this section, and to exclusively focus on the polarization state, normalizing the vector $\textbf{S}(\textbf{r},\tau)$ is a convenient step. For individual realizations, $\textbf{S}(\textbf{r},\tau)/S(\textbf{r},\tau)$ can be represented on the Poincaré sphere (see Fig. \ref{fig: field polarization} (a)). To visualize $\textbf{S}(\textbf{r},\tau)/S(\textbf{r},\tau)$, each point on the Poincaré sphere is associated with a distinct color, as shown in Fig. \ref{fig: field polarization} (b).

For a single realization, the emitted field is fully polarized; the polarization state changes randomly from speckle to speckle. This behavior can be observed in Figs. \ref{fig: field polarization} (c), (e), (g) as the appearance of colorful speckles. On average, the emitted field is unpolarized. This behavior agrees with the one-dimensional (1D) analysis presented in~\cite{1986'Crubellier}, where it was concluded that the single shots of the emitted SF radiation are fully polarized but with a random polarization direction. However, in our case, in contrast to the 1D case, there are several spatial modes; thus, the polarization direction varies within the transverse cross-section.

\begin{figure*}
	\centering
	\includegraphics[width=1\linewidth]{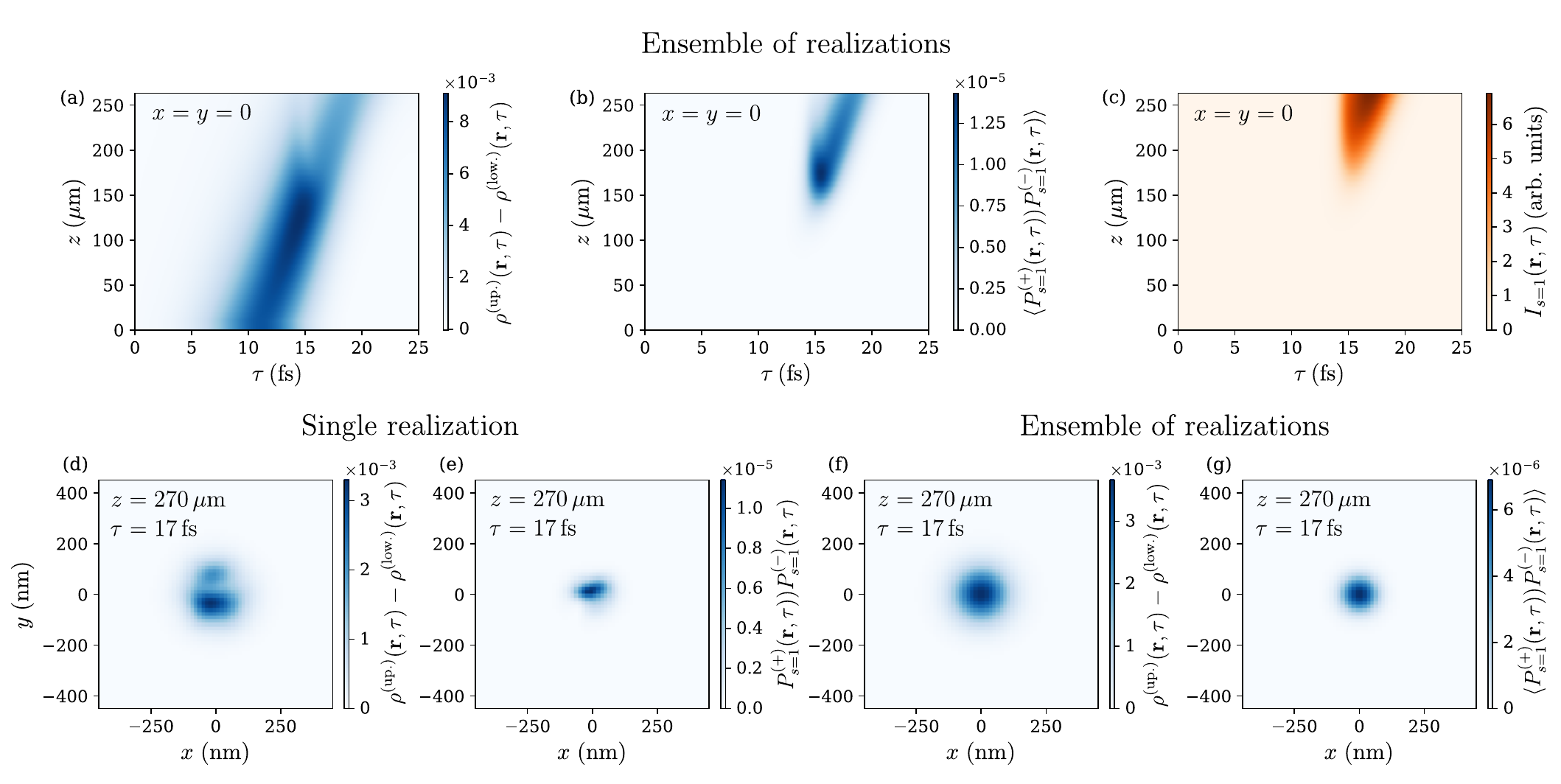}
	\caption[.]{Spatio-temporal evolution on the axis ($x=y=0$) of (a)~effective population inversion, (b)~two-particle correlation between atomic coherences, and (c)~emitted radiation intensity based on averaging over 1,300 numerical realizations; transverse profiles for a single realization at propagation distance $z=270$ \textmu m and corresponding $\tau_{\mathrm{peak}}=$ 17 fs of (d)~effective population inversion and (e) two-particle correlation between atomic coherences; (f), (g) -- same quantities as (d), (e) averaged over 1,300 numerical realizations.}
	\label{fig: medium polarization properties}
\end{figure*}

The speckles of colors observed in Figs.~\ref{fig: field polarization} (c), (e), and (g) suggest that the field polarization properties between neighboring points in the transverse cross-section are correlated. To quantify this kind of correlation, let us consider the fourth-order moment of the field taken in two points in space:
\begin{align}
\label{eq: G4 C}
C(\textbf{r}_{\perp},z,\tau) 
    =\int d\textbf{r}_{\perp}' \frac
        { \langle \textbf{S}(\textbf{r}_\bot',z,\tau)\cdot\textbf{S}(\textbf{r}_\bot+\textbf{r}_\bot',z,\tau)\rangle}
        {\langle S(\textbf{r}_\bot',z,\tau)\rangle\langle S(\textbf{r}_\bot+\textbf{r}_\bot',z,\tau)\rangle
        }.
\end{align}
For a classical light field, the expression under the integral gives a scalar product of two vectors on the Poincar\'e sphere; each of the vectors describes the polarization state of light at points $\textbf{r}_{\perp}'$ and $\textbf{r}_{\perp}'+\textbf{r}_{\perp}$. In this way, the quantity $C(\textbf{r}_{\perp},z,\tau)$ describes the proximity of the polarization state at two points separated by a distance $\textbf{r}_{\perp}$ in the transverse direction. The quantity $C(\textbf{r}_{\perp},z,\tau)$ is shown in Figs. \ref{fig: field polarization} (d), (f), (h). As expected for a quantity averaged over the ensemble of realizations, $C(\textbf{r}_{\perp},z,\tau)$ is azimuthally symmetric. By construction, the width in the transverse direction $\textbf{r}_{\perp}$ of the quantity $C(\textbf{r}_{\perp},z,\tau)$ reflects the average size of the coherent region (speckle) within which the polarization properties of the radiation are close. As Fig.~\ref{fig: field polarization} shows, the extent of $C(\textbf{r}_{\perp},z,\tau)$ indeed reflects the size of the speckle observed in single realizations. Similarly to the dynamics observed in Fig.~\ref{fig: transverse field properties}, the size of the coherent region starts from a single pixel for the SE regime and grows upon propagation, as the comparison of Figs. \ref{fig: field polarization} (d), (f), (h) shows.

\subsection{Population inversion and polarization fields of the gain medium}

Previously, we discussed the properties of the emitted field that can be observed in experiments. In addition to the field variables, numerical modeling gives access to atomic properties that cannot be directly measured.

In accordance with Eq.~(\ref{eq: gain}), the effective population inversion $\rho_{s}^{\text{(up.)}}(\textbf{r},\tau)-\rho_{s}^{\text{(low.)}}(\textbf{r},\tau)$ directly determines the amplification dynamics. Fig.~\ref{fig: medium polarization properties}\,(a) displays its spatio-temporal evolution on the axis. During the amplification stage, the population inversion is primarily conditioned by the pump. Starting from $z \approx 150$~{\textmu}m, the population inversion exhibits faint Rabi oscillations, signifying the transition to the saturation regime, where the dynamics of the atomic variables are noticeably influenced by the emitted radiation. This influence is confirmed by the analysis of Fig.~\ref{fig: medium polarization properties}\,(d), displaying the population inversion in the cross-section for a single stochastic realization. Indeed, the non-uniform distribution in the cross-section cannot be caused by the axially-symmetrical pump, supporting the conclusion that it is influenced by the emitted stochastic field. Note that this non-uniformity can be discerned only at the level of a single realization. As shown in Fig.~\ref{fig: medium polarization properties}\,(f), when the population inversion is averaged over many statistical realizations, the axial symmetry is recovered.

In two-level atomic systems with a single pair of upper and lower states $\ket{u}$ and $\ket{l}$, Rabi oscillations can be conveniently illustrated as a rotation of the Bloch vector:
\begin{equation*}
\rho_{\text{Bloch}} =\begin{bmatrix}
\rho_{ul}+\rho_{lu} \\
i\left(\rho_{ul}-\rho_{lu}\right) \\
\rho_{uu}-\rho_{ll}
\end{bmatrix}.
\end{equation*}
When decoherence is disregarded, this vector retains its length and rotates at a frequency defined by the field amplitude. Although this interpretation does not apply to multi-level systems, Figs.~\ref{fig: medium polarization properties} (d) and (e) show strong anti-correlations between the effective population inversion and polarization fields defined as follows:
\begin{subequations}
\begin{align}
\label{eq: Ppm}
\mathcal{P}^{(+)}_{s}(\textbf{r},\tau)&=\sum_{u,l}T_{lus}\rho_{ul}(\textbf{r},\tau),\\
\mathcal{P}^{(-)}_{s}(\textbf{r},\tau)&=\sum_{u,l}\rho_{lu}(\textbf{r},\tau)T_{uls},
\end{align}
\end{subequations}
where the indices $u$ and $l$ represent the upper and lower states. Since the dynamics is triggered by stochastic spontaneous emission, single stochastic realizations of the polarization field have random phases. Consequently, the mean polarization field vanishes, namely $\langle \mathcal{P}_{s}^{(\pm)}(\textbf{r},\tau)\rangle = 0$. To remove the insignificant random phases, we can look at the correlation function $\langle \mathcal{P}_{s}^{(+)}(\textbf{r},\tau)\mathcal{P}_{s}^{(-)}(\textbf{r},\tau)\rangle$ depicted in Fig.~\ref{fig: medium polarization properties} (b). According to Eq.~(\ref{eq: paraxial Omega}), the introduced polarization fields directly affect the deterministic evolution of the fields. As the saturation regime is reached, the impact of the noise terms becomes completely negligible. Consequently, starting from $z \approx 150$~{\textmu}m, the dynamics of the field is fully determined by the polarization fields, as confirmed by comparing Figs.~\ref{fig: medium polarization properties} (b) and (c). 

\section{\label{sec: Conclusions} Conclusions and outlook}

We have presented the application of stochastic quantum approach for collective light-matter interaction for the case of x-ray superfluorescence initiated by a strongly focused XFEL pump pulse. The properties of the cylindrical medium of high aspect ratio and the short pump pulse allow for a number of simplifying approximations, such as: the description based on the slowly-varying envelope functions, application of the paraxial approximation, and neglecting back-propagating wave. Under these assumptions, a system of stochastic differential equations~(\ref{eq: atomic equations}) and (\ref{eq: paraxial Omega}) has been obtained with a structure resembling Maxwell-Bloch equations augmented with appropriate noise terms for both field and atomic variables. A numerical scheme based on Eqs.~(\ref{eq:Omega-p, num}) --~(\ref{eq: Bloch equation gg, num}) has been proposed to model the resulting stochastic differential equations. The analysis presented in Sec.~\ref{subsec: Spontaneous emission} shows that the proposed noise terms are able to reproduce the temporal and spectral properties of the spontaneous emission in the forward direction---the critical quantum phenomenon that triggers ASE, and SF evolution. 

While the spontaneous emission has been accurately reproduced, verifying the dynamics in the ASE and SF regimes poses a challenge. Assuming that the workaround to avoid divergent trajectories proposed in Sec.~\ref{Gauging run-away trajectories} may impact this dynamics, a benchmark for this aspect would be highly useful. A more rigorous alternative modification is the subject of future publications.

In addition, we have extended the atomic level scheme to include the states that interact with the XFEL pump pulse. Similarly to the superfluorescence, the evolution of the XFEL pump pulse itself has been analyzed within paraxial approximation. As illustrated in Sec.~\ref{sec: Examples}, the proposed formalism allows for an extended statistical analysis of various expectation values.

The developed numerical approach is suitable for designing and analyzing x-ray SF experiments. Its general formulation allows for straightforward extension to complex level schemes, enabling the development of a quantitative theoretical model that can be compared to sXES experiments~\cite{2018'Kroll, 2020'Kroll}. Moreover, the capability to address the transverse properties of the emitted radiation can aid in interpreting complex patterns that correlate frequency and angular emission properties of x-ray SF~\cite{2022'Zhang}. As a result, it can improve the understanding and facilitates the optimization of x-ray pulse pairs production.

In the context of modeling the XLO setup~\cite{2020'Halavanau}, the developed numerical scheme provides the necessary information about spectral, angular, and polarization properties needed to model the in coupling of the x-ray SF radiation burst into the Bragg-crystal cavity. For the radiation that has made one trip within the cavity, the ability of the numerical scheme to describe the crossover from spontaneous to stimulated emission will enable the determination of the lower threshold needed for the circulated radiation to overcome the spontaneous emission. Investigating the parameter space of the geometry of the gain medium and the pump-pulse properties, optimal conditions for obtaining bright and coherent x-ray pulses can be determined and will be the subject of a future study.

\begin{acknowledgments}
A.H. acknowledges the support of the U.S. Department of Energy Contract No. DE-AC02-76SF00515 and the resources provided by the National Energy Research Scientific Computing Center (NERSC), a U.S. Department of Energy Office of Science User Facility located at Lawrence Berkeley National Laboratory, operated under Contract No. DE-AC02-05CH11231 using NERSC award ERCAP0020725. Š.K. acknowledges the research programme No. P1-0112, the research projects No. J1-8134 and PR-08965 of the Slovenian Research Agency, and the SLING consortium for providing computing resources of the HPC system NSC at the Jožef Stefan Institute. S.C. acknowledges the financial support of Grant-No.~HIDSS-0002 DASHH (Data Science in Hamburg-Helmholtz Graduate School for the Structure of the the Matter).
\end{acknowledgments}

\appendix

\section{\label{sec: Appendix: pump and decay} Dipole moment. Incoherent processes}

\subsection{Dipole moment calculations}
\label{app: dipole moment calculations}

Following~\cite{Brink1994}, the dipole matrix element for transition from state $|n'l's'j'm'\rangle$ to $|nlsjm\rangle$, where $l(l')$ and $s(s')$ denote the orbital angular momentum and spin, $j$ $(j')$ and $m$ ${(m')}$ the total angular momentum and its projection, and $n$ $(n')$ any additional quantum numbers, can be calculated as
$
\langle nlsjm|\hat{D}_{q}| n'l's'j'm'\rangle  = (-1)^{j-m}\sqrt{2j+1}
\begin{pmatrix}
j & 1 & j'\\
-m & q & m'
\end{pmatrix}
\langle nlsj||D||n'l's'j'\rangle,
$
where $\hat{D}_q$ is a component of the electric dipole operator, and $\langle nlsj||D||n'l's'j'\rangle$ is the reduced dipole matrix element, which in notation from Eq.~(\ref{eq: d via T}) corresponds to $d_0/\sqrt{3}$ due to normalization.

\subsection{Isotropic fluorescence\label{sec: Isotropic fluorescence}}

Apart from the spontaneous emission that travels along the $z$ axis and is resonant with the upper and lower states (as presented in Eq.~(\ref{eq: e and g explicit for Cu})), there is also spontaneous emission occurring in all other directions. However, since this emission does not strongly interact with the medium, it is unnecessary to consider it explicitly at the level of the fields. Nevertheless, we still need to account for the population change caused by this isotropic emission. For this reason we define isotropic fluorescence decay widths, defined as follows:
\begin{equation}
\gamma_{ij} = \frac{\omega_0^3}{3\pi\hbar\varepsilon_0c_0^3}\sum_{q=-1}^{1}\left|\langle i | \hat{D}_{q}|j\rangle \right|^2.
\end{equation}
These coefficients enter Eq.~(\ref{eq: pump decay decoherence for rho_ij}) in two forms: the dimensionless coefficients $G_{ij}^{\text{(rad.)}}$ and the inverse lifetime $\Gamma_{i}(\textbf{r},t)$. The dimensionless coefficients $G_{ij}^{\text{(rad.)}}$ characterize the increase in the populations of the lower states and are defined as the normalized spontaneous isotropic fluorescence decay widths $\gamma_{ij}/\Gamma_{\text{rad.}}$. The explicit form of $G_{ij}^{\text{(rad.)}}$ is given by:
\begin{equation}
\label{cu-G}
 \begin{tabular}{c}\\
 $\Big\{G_{lu}^{(\text{rad.})}\Big\}\,\,=\,$ 
 \end{tabular}\quad \begin{blockarray}{ccc}
 \color{darkgray}{1s_{\frac{1}{2},\,m=-\frac{1}{2}}} & \color{darkgray}{1s_{\frac{1}{2},\,m=\frac{1}{2}}}  \\
\begin{block}{(cc)c}
  1/3 & 0  & \quad \color{darkgray}{2p_{\frac{3}{2},\,m=-\frac{3}{2}}} \\
  2/9 &1/9 & \quad \color{darkgray}{2p_{\frac{3}{2},\,m=-\frac{1}{2}}}\\
  1/9 & 2/9 & \quad \color{darkgray}{2p_{\frac{3}{2},\,m=\frac{1}{2}}}\\
  0 & 1/3 & \quad \color{darkgray}{2p_{\frac{3}{2},\,m=\frac{3}{2}}}\\
\end{block}
\end{blockarray}
\end{equation}
Since the coefficients $\Gamma_{i}(\textbf{r},t)$ are influenced by other processes, we discuss them in more detail later in Appendix \ref{sub: Lifetime of the states}.

\subsection{Photoionization}

To incorporate incoherent pumping and photoionization into Eq.~(\ref{eq: pump decay decoherence for rho_ij}), we have introduced additional terms represented by the coefficients $p_{i}^{(\text{pump})}(\textbf{r},t)$. These coefficients are defined as follows:
\begin{equation}
\label{cu-rho_pump}
p_{i}^{(\mathrm{pump})}(\textbf{r},\tau) = \sum_{\mathcal{F}} J_{\mathcal{F}}(\textbf{r},\tau) S_{\mathcal{F},i}^{(\mathrm{ground})}.
\end{equation}
Here, $J_{\mathcal{F}}(\textbf{r},\tau)$ represents the flux of the electromagnetic field, and the coefficients $S_{\mathcal{F},i}^{(\mathrm{ground})}$ are the cross-sections for ionization that promotes an atom from the ground state to one of the excited states $\ket{i}$  (refer to Appendix~\ref{sec: Appendix: Photoionization cross-sections}). The process of photoionization is frequency-dependent, which requires the separate treatment of different frequency components of the electromagnetic field. This is why we introduced the index $\mathcal{F}$ to distinguish between various frequency components. Specifically, $\mathcal{F}= \mathcal{P}$ if the photoionization is induced by the pump, and $\mathcal{F}= \Omega_s$ if the photoionization is caused by the emitted SF field, where the index $s$ represents its polarization. As defined in Eq.~(\ref{eq: Nsp-via-J}), the flux $J_{\Omega_s}(\textbf{r},\tau)$ of the SF radiation is related to the Rabi frequencies $\Omega_s(\textbf{r},\tau)$ in the following way
\begin{equation}
\label{cu-J_sigma}
J_{\Omega_{s}}(\textbf{r},\tau)=
\frac{\Omega_{s}^{(+)}(\textbf{r},\tau)\Omega_{s}^{(-)}(\textbf{r},\tau)}{\frac{3}{8\pi}\lambda^2_0 \Gamma_{\text{rad.}}}.
\end{equation}
The expression for the flux $J_{\mathcal{P}}(\textbf{r},\tau)$ of the pump field can be found in Appendix \ref{app: pump dynamics}.

Apart from the flux of the field, the change in the populations in Eq.~(\ref{eq: pump decay decoherence for rho_ij}) depend on the population of the neutral ground state $\rho^{(\mathrm{ground})}(\textbf{r},\tau)$. Its evolution is governed by the following equation:
\begin{equation}
\label{cu-rho_ground}
\frac{\partial}{\partial \tau}\rho^{(\mathrm{ground})}(\textbf{r},\tau) = -\sum_{\mathcal{F},i}S_{\mathcal{F},i}^{(\mathrm{ground})}J_{\mathcal{F}}(\textbf{r},\tau)\rho^{(\mathrm{ground})}(\textbf{r},\tau).
\end{equation}

\subsection{Lifetime of the states}
\label{sub: Lifetime of the states}

{The lifetime of the states $\Gamma_i(\textbf{r},\tau)$ in Eq.~(\ref{eq: pump decay decoherence for rho_ij}) is defined primarily by three processes: Auger-Meitner effect, fluorescence and photoionization. The first two processes can be ensembled into the so-called natural decay width $\Gamma_i^{\text{(natural)}}$. The degenerate states share the same natural decay widths:
\begin{subequations}
    \label{cu-rho_decay}
    \begin{align}  
    \Gamma_u^{\text{(natural)}}=&\,\Gamma_{\text{K}}=2.24 \text{~fs}^{-1},\\ \Gamma_l^{\text{(natural)}}=&\,\Gamma_{\text{L}_3}=0.92 \text{~fs}^{-1}.
    \end{align}
\end{subequations}}
Here, the indices $u$ and $l$ represent the upper and lower states. The decay caused by subsequent photoionization of the ionized atoms can be represented in the following manner:
\begin{equation}
\label{cu-rho_ion}
\Gamma^{\mathrm{(ion.)}}_{i}(\textbf{r},\tau) = \sum_{\mathcal{F}}S_{\mathcal{F},i}^{(\mathrm{ion.})}J_{\mathcal{F}}(\textbf{r},\tau),
\end{equation}
where the cross-sections $S_{\mathcal{F},i}^{(\mathrm{ion.})}$ are defined in Appendix~\ref{sec: Appendix: Photoionization cross-sections}.

\subsection{Pump propagation}
\label{app: pump dynamics}

The propagation of the pump radiation through the medium results in absorption, and change of its temporal~\cite{2021'Inoue} and spatial profile. This change has a significant effect on the evolution of the x-ray superfluorescence; hence, modeling of the pump-pulse propagation is essential. To account for the propagation effects, we analyze the pump at the level of its electric field $\mathcal{P}(\textbf{r},t)$. Similarly to the SF field, we introduce the concept of retarded time $\tau=t-z/c$ and slowly varying amplitude $\mathcal{P}^{(+)}(\textbf{r}, \tau)$, and apply paraxial approximation: 
\begin{equation}
\mathcal{P}(\textbf{r},t)=\mathcal{P}^{(+)}(\textbf{r}, t-z/c)\mathrm{e}^{\mathrm{i}\left(k_{\mathcal{P}}z-\omega_{\mathcal{P}} t\right)}\hat{\textbf{e}}_{y}+\text{H.c.}.
\end{equation}
Here, $\omega_{\mathcal{P}}=k_{\mathcal{P}}c$ is the pump carrier frequency. Additionally, we have assume that the FEL radiation is linearly polarized along the $y$-axis. The flux of the pump field can be defined in the following way:
\begin{equation}
\label{cu-J_pump}
J_{\mathcal{P}}(\textbf{r},\tau)=\frac{2\varepsilon_0c}{\hbar\omega_{\mathcal{P}}}|\mathcal{P}^{(+)}(\textbf{r},\tau)|^2.
\end{equation}

Since we suppose that the pump field interacts with the atoms via nonresonant photoabsorption, it is sufficient to describe response of the atoms in terms of the absorption coefficient $\mu_{\mathcal{P}}(\textbf{r},\tau)$:
\begin{multline}
\label{cu-pump}
\left[\frac{\partial}{\partial z}\mp\frac{i}{2 k_{\mathcal{P}}}\left(\frac{\partial^2}{\partial x^2}+\frac{\partial^2}{\partial y^2}\right)\right]\mathcal{P}^{(\pm)}(\textbf{r},\tau)=\\-\frac{1}{2}\mu_{\mathcal{P}}(\textbf{r},\tau)\mathcal{P}(\textbf{r},\tau).
\end{multline}
The coefficient $\mu_{\mathcal{P}}(\textbf{r},\tau)$ is assumed to be real, ensuring that $\mathcal{P}^{(\pm)}(\textbf{r},\tau)=\mathcal{P}^{(\mp)*}(\textbf{r},\tau)$ and thus maintaining the electric field $\mathcal{P}(\textbf{r},t)$ as a real quantity, in contrast to the SF fields.

\subsection{Absorption of the fields}
\label{sub: Absorption of the fields}
The absorption coefficients for the pump field $\mathcal{P}(\textbf{r},\tau)$ and the SF fields $\Omega_s(\textbf{r},\tau)$ can be generally represented by the following expression:

\begin{multline}
\label{cu-kappaP}
\mu_{\mathcal{F}}(\textbf{r},\tau) = n(\textbf{r}) \sum_{i} \text{Re}\Big( \rho^{(\mathrm{ground})}(\textbf{r},\tau) S_{\mathcal{F},i}^{(\mathrm{ground})}\\ + \rho^{\mathrm{(aux.)}}(\textbf{r},\tau) S^{\mathrm{(aux.)}}{\mathcal{F}} \\+ \rho_{ii}(\textbf{r},\tau) S_{\mathcal{F},i}^{(\mathrm{ion.})} + \sigma^{(\mathrm{compound})}_{\mathcal{F}} \Big).
\end{multline}
In this expression, the index $\mathcal{F}$ distinguishes different field components, as explained after Eq.~(\ref{cu-rho_pump}). $\sigma^{(\mathrm{compound})}_{\mathcal{F}}$ represents the photoionization cross-sections for elements in the medium other than copper (for more details, refer to Appendix~\ref{sec: Appendix: Photoionization cross-sections}). It is important to note that we take the real part of the expression within the brackets, which removes the imaginary contributions from elements of the density matrix that can have arbitrary complex values. The non-linear dependence of the absorption coefficient cannot be included in the stochastic formalism without additional approximations. Given that absorption is a classical effect and the model with constant absorption coefficients is exact, our proposed absorption model must be sufficiently accurate for our purposes.

The absorption of the fields is also influenced by the cumulative population of auxiliary singly-ionized states $\rho^{(\mathrm{aux.})}(\textbf{r},\tau)$. Its evolution is governed by the following equation: 

\begin{multline}
\label{cu-rho_other}
\frac{\partial}{\partial \tau}\rho^{(\mathrm{aux.})}(\textbf{r},\tau) =- \sum_{\mathcal{F}}\rho^{(\mathrm{aux.})}(\textbf{r},\tau)S_{\mathcal{F}}^{(\mathrm{aux.})}J_{\mathcal{F}}(\textbf{r},\tau)\\+ \sum_{\mathcal{F}}\rho^{(\mathrm{ground})}(\textbf{r},\tau)S_{\mathcal{F},\,\text{aux.}}^{(\mathrm{ground})}J_{\mathcal{F}}(\textbf{r},\tau).
\end{multline}
To describe ionization that promotes atoms from the ground state to one of the auxiliary states, we introduce ionization cross-sections denoted as $S_{\mathcal{F}, \text{aux.}}^{(\mathrm{ground})}$. Their numerical values can be found in the last column of the cross-section matrix in Eq.~(\ref{eq: ground state photoionization table}). Subsequent ionization of atoms in the auxiliary states is accounted for by the cross-sections $S_{\mathcal{F}}^{(\mathrm{aux.})}$ given in Eq.~(\ref{eq: ion. state photoionization table}).

\section{Photoionization cross-sections}
\label{sec: Appendix: Photoionization cross-sections}
Photoionization cross-sections included in the Cu-K${\alpha_1}$ superfluorescence model are calculated using the GRASP~\cite{Parpia1996} and RATIP~\cite{Frietzsche2001} atomic codes. In our case, additional calculations are required to determine the partial cross-sections corresponding to individual magnetic sublevels. In accordance with the 6-level model of the Cu-K${\alpha_1}$ system presented in Eq.~(\ref{eq: e and g explicit for Cu}), the valence $4s$ electron is omitted. The initial ground state configuration of the copper atom in this approximation is [Ar]$\,3d^{10}\,^1S_0$, and the ionic states of interest are [Ar]$\,3d^{10}\, 1s^{-1}\, ^2S_{1/2}$ and [Ar]$\,3d^{10}\, 2p^{-1}\, ^2P_{3/2}$.

\begin{widetext}

\subsection{Partial photoionization cross-sections for magnetic sublevels}

Photoionization of the ground state, which causes the population inversion, is described in terms of partial ionization cross-sections encompassed in the following matrix:
\begin{equation}
\label{eq: ground state photoionization table}
 \begin{tabular}{c}\\
 $\Big\{S_{\mathcal{F},i}^{(\mathrm{ground})}\Big\}\,\,=\,\,\,\,\,$ 
 \end{tabular} \begin{blockarray}{cccccccc}
 \color{darkgray}{2p_{\frac{3}{2},m=-\frac{3}{2}}} & \color{darkgray}{2p_{\frac{3}{2},m=-\frac{1}{2}}} & \color{darkgray}{2p_{\frac{3}{2},m=\frac{1}{2}}} & \color{darkgray}{2p_{\frac{3}{2},m=\frac{3}{2}}} & \color{darkgray}{1s_{\frac{1}{2},m=-\frac{1}{2}}} & \color{darkgray}{1s_{\frac{1}{2},m=\frac{1}{2}}} &  \color{darkgray}{\text{aux.}}&\color{darkgray}{\nicefrac{\raisebox{-0.5ex}{$i\,$}}{\raisebox{-0.8ex}{$\mathcal{F}$}}}  \\
\begin{block}{(ccccccc)c}
  0.27\,\sigma^{(\mathrm{g})}_{\mathcal{P},2p} & 0.23\,\sigma^{(\mathrm{g})}_{\mathcal{P},2p} & 0.23\,\sigma^{(\mathrm{g})}_{\mathcal{P},2p} & 0.27\,\sigma^{(\mathrm{g})}_{\mathcal{P},2p} & 0.5\, \sigma^{(\mathrm{g})}_{\mathcal{P},1s} & 0.5\, \sigma^{(\mathrm{g})}_{\mathcal{P},1s} & \sigma^{(\mathrm{g})}_{\mathcal{P},a}  & \quad \color{darkgray}{\!\!\!\mathcal{P}} \\
  0.12\, \sigma^{(\mathrm{g})}_{\Omega,2p} & 0.18\, \sigma^{(\mathrm{g})}_{\Omega,2p} & 0.28\, \sigma^{(\mathrm{g})}_{\Omega,2p} & 0.42\, \sigma^{(\mathrm{g})}_{\Omega,2p} & 0 & 0 & \sigma^{(\mathrm{g})}_{\Omega,a} & \quad \color{darkgray}{\Omega_{-1}}\\
  0.42\, \sigma^{(\mathrm{g})}_{\Omega,2p} & 0.28\, \sigma^{(\mathrm{g})}_{\Omega_{+1},2p} & 0.18\, \sigma^{(\mathrm{g})}_{\Omega,2p} & 0.12\, \sigma^{(\mathrm{g})}_{\Omega,2p} & 0 & 0 & \sigma^{(\mathrm{g})}_{\Omega,a} & \quad \color{darkgray}{\Omega_{+1}}\\
\end{block}
\end{blockarray}
\end{equation}
In this matrix,  $\mathcal{F} = \mathcal{P}$ if the photoionization is caused by the pump field, and $\mathcal{F}= \Omega_{s}$ if the photoionization is induced by the emitted SF field. $\sigma^{(\mathrm{g})}_{\mathcal{P},j}$ and $\sigma^{(\mathrm{g})}_{\Omega,j}$ represent the cross-sections for photoionization. The index $i$ defines the orbital of the electron that was removed from the atom due to ionization. When $i$ is denoted as "aux.," it refers to additional states that do not directly participate in the formation of SF emission and are accounted for as a single auxiliary state.

The second photoionization of the ionized atoms is described by the following partial cross-sections:
\begin{equation}
\label{eq: ion. state photoionization table}
 \begin{tabular}{c}\\
 $\Big\{S_{\mathcal{F},i}^{(\mathrm{ion.})}\Big\}\,\,=\,\,\,\,\,\,$ 
 \end{tabular}\begin{blockarray}{ccccccc}
 \color{darkgray}{2p_{\frac{3}{2},m=-\frac{3}{2}}} & \color{darkgray}{2p_{\frac{3}{2},m=-\frac{1}{2}}} & \color{darkgray}{2p_{\frac{3}{2},m=\frac{1}{2}}} & \color{darkgray}{2p_{\frac{3}{2},m=\frac{3}{2}}} & \color{darkgray}{1s_{\frac{1}{2},m=-\frac{1}{2}}} & \color{darkgray}{1s_{\frac{1}{2},m=\frac{1}{2}}} &\color{darkgray}{\nicefrac{\raisebox{-0.5ex}{$i\,$}}{\raisebox{-0.8ex}{$\mathcal{F}$}}}  \\
\begin{block}{(cccccc)c}
  1.05\, \sigma^{(\mathrm{i})}_{\mathcal{P},2p} & 0.95\, \sigma^{(\mathrm{i})}_{\mathcal{P},2p} & 0.95\, \sigma^{(\mathrm{i})}_{\mathcal{P},2p} & 1.05\, \sigma^{(\mathrm{i})}_{\mathcal{P},2p} & \sigma^{(\mathrm{i})}_{\mathcal{P},1s} & \sigma^{(\mathrm{i})}_{\mathcal{P},1s} & \quad \color{darkgray}{\!\!\!\mathcal{P}} \\
  0.70\, \sigma^{(\mathrm{i})}_{\Omega,2p} & 0.83\, \sigma^{(\mathrm{i})}_{\Omega,2p} & 1.06\, \sigma^{(\mathrm{i})}_{\Omega,2p} & 1.41\, \sigma^{(\mathrm{i})}_{\Omega,2p} & 0.75\, \sigma^{(\mathrm{i})}_{\Omega,1s} & 1.25\, \sigma^{(\mathrm{i})}_{\Omega,1s}  & \quad \color{darkgray}{\Omega_{-1}}\\
  1.41\, \sigma^{(\mathrm{i})}_{\Omega,2p} & 1.06\, \sigma^{(\mathrm{i})}_{\Omega,2p} & 0.83\, \sigma^{(\mathrm{i})}_{\Omega,2p} & 0.70\, \sigma^{(\mathrm{i})}_{\Omega,2p} & 1.25\, \sigma^{(\mathrm{i})}_{\Omega,1s} & 0.75\, \sigma^{(\mathrm{i})}_{\Omega,1s} & \quad \color{darkgray}{\Omega_{+1}}\\
\end{block}
\end{blockarray}
\end{equation}
Here, $\sigma^{(\mathrm{i})}_{\mathcal{P}, j}$ and $\sigma^{(\mathrm{i})}_{\Omega, j}$ denote the total photoionization cross-section of an atom in the level $j$.

\end{widetext}

Photoionization of the atom in the states represented by the effective auxiliary state is accounted for via cross-sections $S_{\mathcal{F}}^{\mathrm{(aux.)}}$ given in the following table:
\begin{equation}
\label{eq: aux. state photoionization table}
 \begin{tabular}{c}\\
 $\Big\{
S_{\mathcal{F}}^{\mathrm{(aux.)}}\Big\}\,\,=\,\,\,\,\,\,$ 
 \end{tabular}\begin{blockarray}{cc}
 \color{darkgray}{\text{aux.}} &\color{darkgray}{\nicefrac{\raisebox{-0.5ex}{$i\,$}}{\raisebox{-0.8ex}{$\mathcal{F}$}}}  \\
\begin{block}{(c)c}
  \sigma_{\mathcal{P}}^{\mathrm{(a)}} & \quad \color{darkgray}{\!\!\!\mathcal{P}} \\
  \sigma_{\Omega}^{\mathrm{(a)}}  & \quad \color{darkgray}{\Omega_{-1}}\\
 \sigma_{\Omega}^{\mathrm{(a)}}& \quad \color{darkgray}{\Omega_{+1}}\\
\end{block}
\end{blockarray}
\end{equation}

Since the target medium consists of a copper nitrate solution, we must consider the photoionization of atoms other than copper. It leads to significant absorption of both the pump and emitted fields, necessitating its inclusion in the model. However, the cross-sections for ionization of additional elements in the solution are smaller than those for copper. As such, we assume that only a small fraction of these atoms becomes ionized, maintaining a population of 1 throughout the target at all times.

The effective photoionization cross-section due to the compound elements other than copper can be expressed as follows:
\begin{equation}
\sigma_{\mathcal{F}}^{(\mathrm{compound})} = \sum_{\mathrm{el.}}N_{\mathrm{el.}}\sigma_{\mathcal{F}, \mathrm{el.}}.
\end{equation}
In this equation, $\sum_{\mathrm{el.}}$ represents the sum over these additional elements, $N_{\mathrm{el.}}$ denotes the number of atoms of a given element per one copper atom, and $\sigma_{\mathcal{F}, \mathrm{el.}}$ is the corresponding cross-section for ionization with field mode $\mathcal{F}$. The elements present in the compound are hydrogen, oxygen (denoted as $\mathrm{el.}=\mathrm{O}$), and nitrogen (denoted as $\mathrm{el.}=\mathrm{N}$). However, since the photon energy of the pump and emitted fields is approximately 8-9 keV, nearly three orders of magnitude above the ionization threshold of hydrogen, the corresponding cross-sections for hydrogen can be considered negligible.

The numerical values of photoionization cross-sections for all the processes included in the simulation of the Cu-K${{{\alpha}_1}}$ system are provided in Table~\ref{csCuTab}. In the case of an 8-molar solution of copper nitrate, the number of atoms in the compound per one copper atom is $N_{\mathrm{O}}=13$ and $N_{\mathrm{N}}=2$. Further details of the calculations are outlined below.

\begin{table}[h]

	\begin{center}
	\begin{tabular}{c c | c c}
		\hline
		\hline
		Parameter & Value [nm$^2$] & Parameter & Value [nm$^2$] \\
		\hline
		$\sigma^{(\mathrm{g})}_{\mathcal{P},1s}$ & $2.53\times 10^{-6}$ & $\sigma^{(\mathrm{i})}_{\mathcal{P},1s}$ & $4.75\times 10^{-7}$ \\
		$\sigma^{(\mathrm{g})}_{\mathcal{P},2p}$ & $1.04\times 10^{-7}$ & $\sigma^{(\mathrm{i})}_{\mathcal{P},2p}$ & $3.02\times 10^{-7}$ \\
		$\sigma^{(\mathrm{g})}_{\mathcal{P},a}$ & $3.23\times 10^{-7}$ & $\sigma^{(\mathrm{i})}_{\Omega,1s}$ & $6.53\times 10^{-7}$ \\
		$\sigma^{(\mathrm{g})}_{\Omega,2p}$ & $1.52\times 10^{-7}$ & $\sigma^{(\mathrm{i})}_{\Omega,2p}$ & $4.15\times 10^{-7}$ \\
		$\sigma^{(\mathrm{g})}_{\Omega,a}$ & $4.34\times 10^{-7}$ & $\sigma_{\mathcal{P}}^{\mathrm{(a)}}$ & $3.27\times 10^{-7}$ \\
		 &  & $\sigma_{\Omega}^{\mathrm{(a)}}$ & $4.58\times 10^{-7}$ \\
		 \hline
		 $\sigma_{\mathcal{P}, \mathrm{O}}$ & $2.00\times10^{-8}$ & $\sigma_{\mathcal{P}, \mathrm{N}}$ & $1.11\times10^{-8}$ \\ 
		 $\sigma_{\Omega, \mathrm{O}}$ & $2.75\times10^{-8}$ & $\sigma_{\Omega, \mathrm{N}}$ & $1.55\times10^{-8}$ 
	\end{tabular}
	\end{center}
	\caption{\label{tab: table of cross-sections}Values of photoionization cross-sections included in the model of the Cu-K$_{\alpha_1}$ system.}
	\label{csCuTab}
\end{table}

\subsection{Calculation of the partial photoionization cross-sections}
\label{cs-derivation}

The wave function of the outgoing electron can be expanded in terms of partial waves as~\cite{Joachain1975}
\begin{equation}
\begin{split}
|\psi_c\rangle&=\sum_{l=0}^{\infty}\sum_{m_l=-l}^l|l m_l s m_s\rangle Y_{l m_l}^*(\vartheta_k, \varphi_k)\\&=\sum_{l=0}^{\infty}\sum_{m_l=-l}^l\sum_{j_f=|l-s|}^{l+s}\sum_{m_f=-j_f}^{j_f}\langle j_f m_f|l m_l s m_s\rangle \\&\times |(ls)j_f m_f\rangle Y_{l m_l}^*(\vartheta_k, \varphi_k),
\end{split}
\end{equation}
where $s=1/2$ is the electron spin and $m_s$ its projection, $l$ orbital angular momentum and $m_l$ its projection, and $\vartheta_k$ and $\varphi_k$ are the polar and azimuthal angles associated with the wave vector of the electron. The final state product function combining the ion and electron can then be expanded in terms of the total angular momentum of the system as~\cite{Brink1994}
\begin{equation}
\begin{split}
|\psi_c\rangle|J_f M_f\rangle =& \sum_{l,m_l}\sum_{j_f,m_f}\langle j_f m_f|l m_l s m_s\rangle\\&\times \sum_{J=|j_f-J_f|}^{j_f+J_f}\sum_{M=-J}^J \langle JM|j_f m_f J_f M_f\rangle\\ &\times |\left[(ls)j_f J_f\right]JM\rangle Y_{l m_l}^*(\vartheta_k, \varphi_k),
\end{split}
\end{equation}
where $J_f$ is the total angular momentum of the ion and $M_f$ its projection, $j_f$ and $m_f$ are the total angular momentum of the electron and its projection, and $J$ and $M$ correspond to the combined system of the ion and electron. The photoionization cross-section is proportional to the absolute square of photoionization amplitudes $\langle J_f M_f|\langle \psi_c|\hat{D}_q|J_i M_i\rangle$, where $J_i$ is the total angular momentum of the initial state, $M_i$ its projection, and $\hat{D}_q$ is a component of the electric dipole operator. 

In the case of photoionization from the initial neutral state to a selected state of Cu$^+$ with given $J_f$ and $M_f$ by field with polarization mode $q$, $J_i=M_i=0$ and the cross-sections of interest are
\begin{equation}
\label{cs_partial_ground}
\begin{split}
\sigma(J_f, M_f, q)&=\xi \sum_{m_s}\int \mathrm{d}\Omega_k |\langle J_f M_f|\langle \psi_c|\hat{D}_q|00\rangle|^2 \\&= \xi\sum_l \sum_{j_f} \langle 1 q|j_f (q-M_f) J_f M_f\rangle^2 \\&\times|\langle \left[(ls)j_f J_f\right]1||D||0\rangle|^2
\end{split},
\end{equation}
where $\mathrm{d}\Omega_k=\sin\vartheta_k\mathrm{d}\vartheta_k\mathrm{d}\varphi_k$, and $\xi$ is a constant factor that depends on the specific form of the dipole transition operator and unit system used in the calculation (conventions differ between different references and atomic codes). This expression determines the individual cross-sections in the last two rows of $S_{\mathcal{F}i}^{(\mathrm{ground})}$ corresponding to ionization with the circularly polarized modes of the emitted field. In the derivation of Eq.~(\ref{cs_partial_ground}) the following properties of spherical harmonics and Clebsch-Gordan coefficients were used~\cite{Brink1994}:
\begin{widetext}
\begin{subequations}
\begin{align}
\int \mathrm{d}\Omega_k Y_{lm_l}(\vartheta_k, \varphi_k) Y^*_{l' m_l'}(\vartheta_k, \varphi_k)=\delta_{l,l'}\delta_{m_l,m_l'},\\
\sum_{m_1,m_2}\langle JM|j_1 m_1 j_2 m_2\rangle \langle j_1 m_1 j_2 m_2|J'M'\rangle=\delta_{J, J'}\delta_{M, M'},\\
\langle j_1 m_1 j_2 m_2|JM\rangle \neq 0\ \Leftrightarrow \ m_1+m_2=M.
\end{align}
\end{subequations}
In the chosen coordinate system, the pump pulse is linearly polarized along the $y$ axis. cross-sections for ionization with the pump field can be expressed as
\begin{equation}
\begin{split}
\sigma(J_f, M_f, y)&=\xi \sum_{m_s}\int \mathrm{d}\Omega_k \big|\langle J_f M_f|\langle \psi_c|\,{\big(\hat{D}_{-1}+\hat{D}_{+1}\big)}/{\sqrt{2}}\,|00\rangle\big|^2\\&=\frac{1}{2}\left[\sigma(J_f, M_f, q=-1)+\sigma(J_f, M_f, q=+1)\right],
\end{split}
\end{equation}
and correspond to the individual cross-sections in the first row of $S_{\mathcal{F}i}^{(\mathrm{ground})}$. 

Reduced dipole matrix elements $\langle \left[(ls)j_f J_f\right]1||D||0\rangle$ (also called photoionization amplitudes) are part of the output of the RATIP code~\cite{Frietzsche2001}, and can be used to calculate the prefactors in matrix $S_{\mathcal{F}i}^{(\mathrm{ground})}$. The code also outputs cross-sections, which are calculated for unpolarized light and are averaged over initial states and summed over final states. Unpolarized light can be treated as a linear combination of two incoherent linearly polarized beams of equal intensity~\cite{Manson1982}. The calculated cross-sections, which correspond to the total cross-sections $\sigma_{\mathcal{F},i}^{(\mathrm{g})}$, can in our notation be written as
\begin{equation}
\sigma(J_f)=\frac{1}{2}\sum_{q=-1, 1}\sum_{M_f}\sigma(J_f, M_f, q)=\xi \sum_l\sum_{j_f}|\langle \left[(ls)j_f J_f\right]1||D||0\rangle|^2.
\end{equation}

In the case of Cu$^+$ to Cu$^{++}$ ionization, the cross-sections of interest are
\begin{equation}
\begin{split}
\sigma(J_i,M_i,q)&=\zeta\sum_{J_f}\sum_{M_f}\sum_{m_s}\int \mathrm{d}\Omega_k |\langle J_f M_f|\langle \psi_c|\hat{D}_q|J_i M_i\rangle|^2 \\&
= \zeta \sum_{J_f}\sum_l\sum_{j_f}\sum_J \langle J (M_i+q)|J_i M_i 1 q\rangle^2 |\langle\left[(ls)j_f J_f\right]J||D||J_i\rangle|^2,
\end{split}
\end{equation}
where $J_i$ is the total angular momentum of the initial ionic state of Cu$^+$ and $M_i$ its projection. These cross-sections correspond to the individual cross-sections in the last two rows of $S_{\mathcal{F}i}^{(\mathrm{ion.})}$. The cross-section for ionization with the pump pulse can similarly as above be expressed as
\begin{equation}
\sigma(J_i, M_i, y)=\frac{1}{2}\left[\sigma(J_i, M_i, q=-1)+\sigma(J_i, M_i, q=+1)\right],
\end{equation}
and corresponds to the individual cross-sections in the first row of $S_{\mathcal{F}i}^{(\mathrm{ion.})}$. Again, the photoionization cross-sections and amplitudes are calculated with the RATIP code. In our notation these cross-sections can be expressed as
\begin{equation}
\begin{split}
\sigma(J_i)&=\frac{1}{2}\sum_{q=-1, 1}\frac{1}{2 J_i+1}\sum_{M_i}\sigma(J_i, M_i, q)\\&=\zeta\frac{1}{3(2J_i+1)}\sum_{J_f}\sum_l\sum_{j_f}\sum_J (2J+1) |\langle\left[(ls)j_f J_f\right]J||D||J_i\rangle|^2,
\end{split}
\end{equation}
in the derivation of which the following symmetry relation was used~\cite{Brink1994}:
\begin{equation}
\langle j_1 m_1 j_2 m_2|JM\rangle = (-1)^{j_2+m_2}\sqrt{\frac{2J+1}{2j_1+1}}\langle j_2 (-m_2)JM|j_1 m_1\rangle.
\end{equation}
These cross-sections also correspond to the total cross-sections $\sigma_{\mathcal{F},i}^{(\mathrm{i})}$. Because of the averaging over the initial states, the relation between the partial and total cross-sections is $\sum_{M_i}\sigma(J_i,M_i,q)=(2J_i+1)\sigma(J_i)$.

\end{widetext}

\section{\label{sec: Appendix: Equivalence of Ito and Stratonovich forms}Stochastic Differential Equations in the Ito form}

Consider a system of stochastic differential equations for a vectorial stochastic variable $\textbf{x}(t)$ in the Ito form: 
\begin{align}
\label{eq: Ito form, general f}
\frac{dx_i(t)}{dt} = A_i(\textbf{x},t) + \sum_j B_{ij}(\textbf{x},t) f_i(t).
\end{align}
Here $f_i(t)$ are normalized Gaussian white noise terms
\begin{equation}
\label{eq: correlator appendix}
    \langle f_i(t)f_j(t')\rangle=\delta_{ij}\delta(t-t').
\end{equation}
The stochastic Ito equation can be related to the following finite difference scheme
\begin{align}
\label{eq: Ito form, general}
\Delta x_i^{(t)} = A_i^{(t)} \Delta t + \sum_j B_{ij}^{(t)} \varepsilon^{(t)}_j\sqrt{\Delta t},
\end{align}
where $\varepsilon^{(t)}_j$ are normalized Gaussian random numbers. Note, that the equations in Sec. \ref{sec: Stochastic equations} involve complex noise terms in contrast to the examples given in this section. The complex noise term $f_{\text{com.}}(t)$ can be expressed through two real noise terms
\begin{equation*}
    f_{\text{com.}}(t)=\frac{1}{\sqrt{2}}(f_{1}(t)+i f_{2}(t)).
\end{equation*}

The convenience of the Ito form lies in the ease of the numerical implementation. According to the Ito interpretation (\ref{eq: Ito form, general}), the stochastic integration requires a new independent random contribution every time increment and involves dynamic variables from the previous time step.

{\section{\label{sec: derivations} Derivation of the stochastic differential equations}

\subsection{Bloch equations}

Before deriving the stochastic equations used to simulate x-ray superfluorescence, we begin by formulating the deterministic Maxwell–Bloch equations. These equations can be derived by employing a fully factorized ansatz for the density matrix of the system
\begin{equation}
\label{eq: classical decomposition}
\rho(t) = \prod_a \hat{\rho}_a(t)\prod_{\mathbf{k},s} \hat{\Lambda}(\alpha_{\mathbf{k},s}(t), \alpha^\dagger_{\mathbf{k},s}(t)).
\end{equation}
 In this context, each atom is characterized by an independent one-particle density matrix denoted as $\hat{\rho}_a(t) = \sum_{p,q} \rho_{a,pq}(t) \hat{\sigma}_{a,pq}$, while the field modes are described using projectors $\hat{\Lambda}(\alpha_{\mathbf{k},s}(t), \alpha^\dagger_{\mathbf{k},s}(t))$ as defined in Eq.~(\ref{eq: stochastic ansatz for the density matrix}). Consequently, the state of the system is defined the by variables $\rho_{a,pq}(t)$ representing the elements of the atomic one-particle density matrices, and $\alpha_{\mathbf{k},s}(t)$ as well as $\alpha^\dagger_{\mathbf{k},s}(t)$ play the role of the field amplitudes. To derive the equations of motion for these variables, we employ the decomposition presented in Eq.~(\ref{eq: classical decomposition}) and substitute it into the following master equation
\begin{equation}
\label{eq: full master equation}
\begin{split}
    \dot{\hat{\rho}}(t)=\mathcal{L}[\hat{\rho}(t)]=&\frac{i}{\hbar}\left[\hat{\rho}(t),\hat{H}_f+\sum_a\hat{H}_a+\hat{V}\right]\\&+\hat{\mathcal{L}}_{\text{incoh.}}[\hat{\rho}(t)]+\hat{\mathcal{L}}_{\text{absorp.}}[\hat{\rho}(t)].
    \end{split}
\end{equation}
This allows us to construct equations for the expectation values as follows:
\begin{equation*}
\begin{split}
   \dot{\alpha}_{\textbf{k},s}(t)&=\text{Tr}(\hat{a}_{\textbf{k},s}\dot{\hat{\rho}}(t))=\text{Tr}(\hat{a}_{\textbf{k},s}\mathcal{L}[\hat{\rho}(t)]), \\\dot{\alpha}_{\textbf{k},s}^{\dag}(t)&=\text{Tr}(\hat{a}_{\textbf{k},s}^{\dag}\dot{\hat{\rho}}(t))=\text{Tr}(\hat{a}_{\textbf{k},s}^{\dag}\mathcal{L}[\hat{\rho}(t)]), \\\dot{\rho}_{a,pq}(t)&=\text{Tr}(\hat{\sigma}_{a,pq}\dot{\hat{\rho}}(t))=\text{Tr}(\hat{\sigma}_{a,pq}\mathcal{L}[\hat{\rho}(t)]).
\end{split}
\end{equation*}
To create a closed system of equations, it is important to note that the ansatz in Eq.~(\ref{eq: classical decomposition}) factorizes second-order correlators as follows:
\begin{subequations}
\label{eq: truncation rules}
\begin{align}
    \text{Tr}(\hat{\sigma}_{a,pq}\hat{a}_{\textbf{k},s}\hat{\rho}(t))&=\rho_{a,qp}(t)\alpha_{\textbf{k},s}(t),\\
    \text{Tr}(\hat{\sigma}_{a,pq}\hat{a}_{\textbf{k},s}^{\dag}\hat{\rho}(t))&=\rho_{a,qp}(t)\alpha_{\textbf{k},s}^{\dag}(t).
\end{align}
\end{subequations}
\begin{widetext}
The equations for the field variables $\alpha_{\textbf{k},s}(t)$ and $\alpha^\dag_{\textbf{k},s}(t)$ can be divided into two parts: one arising from unitary evolution and the other from absorption. The unitary evolution is described as follows:

\begin{subequations}
   \label{eq: equations for the classical field modes app} 
\begin{align}
    \dot{\alpha}_{\textbf{k},s}(t)\Big|_{\text{unitary}}=&-i\omega_{k}\alpha_{\textbf{k},s}(t)+d_0g_0\sum_{u,l}T_{lu,s}\rho_{a,ul}(t) e^{-i\textbf{k}\cdot\textbf{r}_a},\\
\dot{\alpha}_{\textbf{k},s}^\dag(t)\Big|_{\text{unitary}}=&\,\,i\omega_{k}\alpha_{\textbf{k},s}^\dag(t)+d_0g_0\sum_{u,l}T_{ul,s}\rho_{a,lu}(t) e^{i\textbf{k}\cdot\textbf{r}_a},
\end{align}
\end{subequations}
where the indices $u$ and $l$ represent the upper and lower states. Apart from that, we have employed Eq.~(\ref{eq: d via T}).  To describe absorption, we define the electric fields $\mathcal{E}^{(\pm)}_{s}(\textbf{r},t)$:
\begin{subequations}
\label{eq: Electric fields}
\begin{align}
    \mathcal{E}^{(+)}_{s}(\textbf{r},t)&=\,\,i\hbar\sum_\textbf{k} g_0\alpha_{\textbf{k},s}(t)e^{i\textbf{k}\cdot\textbf{r}},\\
    \mathcal{E}^{(-)}_{s}(\textbf{r},t)&=-i\hbar\sum_\textbf{k} g_0\alpha_{\textbf{k},s}^{\dag}(t)e^{-i\textbf{k}\cdot\textbf{r}}.
\end{align}  
\end{subequations}
Then, the absorption enters the equations of motion in the following way:
\begin{equation}
    \frac{\partial}{\partial t}\mathcal{E}^{(\pm)}_{s}(\textbf{r},t)\Big|_{\text{absorp.}}=-\frac{\mu_s(\textbf{r},t)}{2}\mathcal{E}^{(\pm)}_{s}(\textbf{r},t).
\end{equation}
For the sake of convenience, we have also divided the equations pertaining to the atomic variables into two parts. The part responsible for incoherent processes can be expressed as:
\begin{equation}
\label{eq: incoh equations appendix}
\begin{split}
   \dot\rho_{a,pq}(t)\Big|_{\text{incoh.}}=&
    -(\Gamma_{p}(\textbf{r},t)+\Gamma_{q}(\textbf{r}_a,t))\rho_{pq}(\textbf{r}_a,t)/2
    +\delta_{pq}\left(p_{p}^{(\text{pump})}(\textbf{r}_a,t)\rho^{(\mathrm{ground})}_{a}(t)+\Gamma_{\text{rad.}}\sum_{k}G_{pk}^{(\text{rad.})}\rho_{a,kk}(t)\right).
    \end{split}
\end{equation}
This equation is essentially a discrete version of Eq.~(\ref{eq: pump decay decoherence for rho_ij}) in original time $t$. The unitary evolution is described by the following part:
\begin{equation}
\label{eq: bloch equations appendix}
\begin{aligned}
 \dot{\rho}_{a,pq}\!\left(t\right)\Big|_{\text{unitary}}=-i\omega_{pq}{\rho}_{a,pq}\!\left(t\right)&+\frac{id_0}{\hbar}\sum_{r,s}\Bigg[\mathcal{E}_s^{(+)}\!\left(\textbf{r}_a,t\right)\Big(T_{p>r,s}{\rho}_{a,rq}\!\left(t\right)-{\rho}_{a,pr}\!\left(t\right)T_{r>q,s}\Big)\\&+\mathcal{E}^{(-)}_s\!\left(\textbf{r}_a,t\right)\sum_{r}\Big(T_{p<r,s}{\rho}_{a,rq}\!\left(t\right)-{\rho}_{a,pr}\!\left(t\right)T_{r<q,s}\Big)\Bigg],
\end{aligned}
\end{equation}
where $p>q$ means that index $p$ corresponds to the subset of upper states $\{\ket{e}\}$ whereas index $q$ --- to the subset of ground states $\{\ket{g}\}$. In the equations, we have employed the electric fields $\mathcal{E}^{(\pm)}_{s}(\textbf{r},t)$ that conveniently assemble the field amplitudes $\alpha_{\textbf{k},s}(t)$ and $\alpha_{\textbf{k},s}^\dag(t)$.

Truncation of the second-order correlators presented in Eq.~(\ref{eq: truncation rules}) and used in the derivations of Eqs.~(\ref{eq: equations for the classical field modes app})~-- (\ref{eq: bloch equations appendix}) shows that the resulting equations are valid only for the systems with strong classical behavior. Let us find the neglected terms in the master equation and analyze their structure. If we insert the decomposition from Eq.~(\ref{eq: classical decomposition}) in the master equation (\ref{eq: full master equation}) and apply Eqs.~(\ref{eq: equations for the classical field modes app})~-- (\ref{eq: bloch equations appendix}), we notice that the right-hand side $\mathcal{L}[\hat{\rho}(t)]$ of Eq.~(\ref{eq: full master equation}) is restored only partially
\begin{equation}
\label{eq: missing terms}
    \mathcal{L}[\hat{\rho}(t)]-\dot{\hat{\rho}}(t)=\sum_{b,\,\textbf{k},\,s}\hat{\chi}_{b;\textbf{k},s}(t)\prod_{a\neq b}\hat{\rho}_a(t)\prod_{\substack{\textbf{k}'\neq \textbf{k},\\ s' \neq s}} \hat{\Lambda}(\alpha_{\mathbf{k}',s'}(t), \alpha^\dagger_{\mathbf{k}',s'}(t)).
\end{equation} 
The time derivative of Eq.~(\ref{eq: classical decomposition}) can give rise to terms where either a single atomic one-particle density matrix $\hat{\rho}_a$ or a single field projector $\hat{\Lambda}(\alpha_{\mathbf{k},s}(t), \alpha^\dagger_{\mathbf{k},s}(t))$ is modified. Consequently, the remaining terms in Eq.~(\ref{eq: missing terms}) intertwine the atomic and field degrees of freedom. These terms can be expressed as:
\begin{subequations}
\label{eq: chi-terms}
\begin{equation}
    \hat{\chi}_{a;\textbf{k},s}(t)=\sum_{p,\,q}\left[\chi_{a,pq;\textbf{k},s}^{(+)}(t)\, \frac{\partial}{\partial \alpha_{\textbf{k},s}(t)}+\chi_{a,pq;\textbf{k},s}^{(-)}(t)\, \frac{\partial}{\partial \alpha_{\textbf{k},s}^\dag(t)}\right]\hat{\sigma}_{a,pq}\,\hat{\Lambda}(\alpha_{\mathbf{k},s}(t), \alpha^\dagger_{\mathbf{k},s}(t)),
\end{equation}
where
\begin{align}
   \chi_{a,pq;\textbf{k},s}^{(+)}(t)=&\,d_0g_0\bigg(\sum_{r}T_{p<r,s}{\rho}_{rq}(t)-{\rho}_{pq}(t)\sum_{u,l}T_{lu,s}{\rho}_{ul}(t)\bigg)e^{-i\textbf{k}\cdot\textbf{r}_a}, \\
   \chi_{a,pq;\textbf{k},s}^{(-)}(t)=&\,d_0g_0\bigg(\sum_{r}{\rho}_{pr}(t)T_{r>q,s}-{\rho}_{pq}(t)\sum_{u,l}T_{ul,s}{\rho}_{lu}(t)\bigg)e^{i\textbf{k}\cdot\textbf{r}_a}.
\end{align}
\end{subequations}
Here, the indices $u$ and $l$ represent the upper and lower states. 

\subsection{Noise terms}

While~$\hat{\chi}_{a;\textbf{k},s}(t)$ may initially seem complex, the uncompensated right-hand side of Eq.~(\ref{eq: missing terms})  essentially entangles individual atoms and single field modes and does not introduce additional intricate higher-order correlations. As demonstrated in Ref.~\cite{chuchurka2023stochastic}, the terms in Eq.~(\ref{eq: chi-terms}) can be correctly recaptured by introducing suitable stochastic terms into Eqs.~(\ref{eq: equations for the classical field modes app})~-- (\ref{eq: bloch equations appendix}) in the following manner:
\begin{subequations}
\label{eq: additional terms to the Bloch equations}
\begin{align}
    \dot{\alpha}_{\textbf{k},s}(t)\big |_{\text{noise}}=\xi_{\textbf{k},s}(t),\quad  &\,\dot{\alpha}_{\textbf{k},s}^\dag(t)\big |_{\text{noise}}=\xi_{\textbf{k},s}^\dag(t),\\
 \dot{\rho}_{a,pq}\!\left(t\right)\big |_{\text{noise}}&\,=\eta_{a,pq}(t).
\end{align}
\end{subequations}
Here, we introduce a set of Gaussian white noise terms, namely~$\xi_{\textbf{k},s}(t)$, $\xi_{\textbf{k},s}^\dag(t)$, and $\eta_{a,pq}(t)$, with the following correlation properties:
\begin{subequations}
\label{eq: correlator def}
\begin{align}
    \big\langle \xi_{\textbf{k},s}(t) \eta_{a,pq}(t')\big\rangle & = \kappa_{a,pq;\textbf{k},s}^{(+)}(t)\delta(t-t'),\\
    \big\langle \xi_{\textbf{k},s}^\dag(t) \kappa_{a,pq}(t')\big\rangle & = \kappa_{a,pq;\textbf{k},s}^{(-)}(t)\delta(t-t'),
\end{align}
\end{subequations}
which we will determine later. We assume that the correlators of the remaining pairs of noise terms are zero. Two non-zero correlators in Eq.~(\ref{eq: correlator def}) prove sufficient for reproducing the missing terms in Eq.~(\ref{eq: missing terms}). Additionally, we treat the noise terms as integrated in the Ito sense. See Appendix~\ref{sec: Appendix: Equivalence of Ito and Stratonovich forms} for more details. 

Typically, stochastic equations are solved using the Monte Carlo approach. The proper statistics of the dynamic variables~$\rho_{a,pq}\left(t\right)$, $\alpha_{\textbf{k},s}(t)$, and $\alpha_{\textbf{k},s}^\dag(t)$ are reconstructed by solving multiple equations with a randomly sampled stochastic contribution in accordance with their statistical properties. Since the equations do not couple the variables from different realizations, their integration can be parallelized, offering significant performance advantages compared to methods based on the direct decomposition of the wave function into some basis set. To sample the density matrix, we insert each realization of the dynamic variables $\rho_{a,pq}\left(t\right)$, $\alpha_{\textbf{k},s}(t)$, and $\alpha_{\textbf{k},s}^\dag(t)$ into the decomposition in Eq.~(\ref{eq: classical decomposition}) and combine the resulting factorized density matrices into a normalized linear combination:
\begin{equation}
\label{eq: app stochastic density matrix}
    \rho(t)=\Big\langle\prod_a \hat{\rho}_a(t)\prod_{\mathbf{k},s} \hat{\Lambda}(\alpha_{\mathbf{k},s}(t), \alpha^\dagger_{\mathbf{k},s}(t))\Big\rangle,
\end{equation}
resulting in a non-factorizable density matrix. Similar to the approach in Ref.~\cite{chuchurka2023stochastic}, this linear combination serves to restore the missing entangled terms in Eq.~(\ref{eq: missing terms}). Although the stochastic ansatz does not alter the expression for the first term~$\mathcal{L}[\hat{\rho}(t)]$, it does modify the derivative~$\dot{\hat{\rho}}(t)$ by introducing additional terms proportional to~$\kappa_{a,pq;\textbf{k},s}^{(+)}(t)$ and $\kappa_{a,pq;\textbf{k},s}^{(-)}(t)$, a~concept known as Ito's lemma.

Consider an arbitrary function $S$ that depends on the stochastic variables~${\rho}_{a,pq}$, $\alpha_{\textbf{k},s}$, and $\alpha_{\textbf{k},s}^\dag$. Ito's lemma reads:
\begin{equation}
\label{app: ito}
\begin{split}
    \frac{dS}{dt}=&\sum_{a,p,q}\frac{\partial S}{\partial {\rho}_{a,pq}} \frac{d {\rho}_{a,pq}}{d t}+\sum_{\textbf{k},s}\frac{\partial S}{\partial {\alpha}_{\textbf{k},s}} \frac{d {\alpha}_{\textbf{k},s}}{d t}+\sum_{\textbf{k},s}\frac{\partial S}{\partial {\alpha}^\dag_{\textbf{k},s}} \frac{d {\alpha}^\dag_{\textbf{k},s}}{d t}+\sum_{\mathclap{\substack{a,p,q\\\textbf{k},s}}}\left[\frac{\partial^2 S}{\partial {\rho}_{a,pq}\partial{\alpha}_{\textbf{k},s}} \kappa_{a,pq;\textbf{k},s}^{(+)}+\frac{\partial^2 S}{\partial {\rho}_{a,pq}\partial{\alpha}_{\textbf{k},s}^\dag} \kappa_{a,pq;\textbf{k},s}^{(-)}\right].
\end{split}
\end{equation}
As a result, the complete derivative of the density matrix presented in Eq.~(\ref{eq: app stochastic density matrix}) gains the following additional components:
\begin{align*}
    \dfrac{d \hat{\rho}(t)}{dt}\bigg|_{\text{noise}}=\bigg\langle&\sum_{\substack{a,p,q\\\textbf{k},s }}\bigg[\kappa_{a,pq;\textbf{k},s}^{(+)}(t)\, \frac{\partial}{\partial \alpha_{\textbf{k},s}(t)}+\kappa_{a,pq;\textbf{k},s}^{(-)}(t)\, \frac{\partial}{\partial \alpha_{\textbf{k},s}^\dag(t)}\bigg]\hat{\sigma}_{a,pq} \prod_{b\neq a}\hat{\rho}_{b}(t)\prod_{\textbf{k},s}\hat{\Lambda}(\alpha_{\mathbf{k},s}(t), \alpha^\dagger_{\mathbf{k},s}(t))\bigg\rangle,
\end{align*} 
that entangle pairs of atoms and have exactly the same form as the right-hand side of Eq.~(\ref{eq: missing terms}). Consequently, if the correlators of the noise terms have the following form
\begin{equation*}
    \kappa_{a,pq;\textbf{k},s}^{(+)}(t) = \chi_{a,pq;\textbf{k},s}^{(+)}(t), \quad \kappa_{a,pq;\textbf{k},s}^{(-)}(t) = \chi_{a,pq;\textbf{k},s}^{(-)}(t).
\end{equation*}
Eqs.~(\ref{eq: equations for the classical field modes app}) -- (\ref{eq: bloch equations appendix}) accompanied by the noise terms in Eq.~(\ref{eq: additional terms to the Bloch equations}) fully satisfy master equation~(\ref{eq: full master equation}).

To simulate~$\xi_{\textbf{k},s}(t)$, $\xi_{\textbf{k},s}^\dag(t)$, and $\eta_{a,pq}(t)$, we have to decompose them in terms of independent noise terms. There is no unique decomposition, however, the structure of Eq.~(\ref{eq: chi-terms}) suggests the most compact one
\begin{subequations}
\label{eq: noise terms represented}
\begin{align}
    \dot{\alpha}_{\textbf{k},s}(t)\big |_{\text{noise}}&=\xi_{\textbf{k},s}(t)=d_0 g_0\sum_{a}f_{a,s}(t) e^{-i\textbf{k}\cdot\textbf{r}_a},\\
    \dot{\alpha}_{\textbf{k},s}^\dag(t)\big |_{\text{noise}}&=\xi_{\textbf{k},s}^\dag(t)=d_0 g_0\sum_{a}g_{a,s}(t) e^{i\textbf{k}\cdot\textbf{r}_a},
\end{align}
\begin{equation}
\begin{aligned}
    \dot{\rho}_{a,pq}(t)\Big|_{\text{noise}}=\,\,\eta_{a,pq}(t)=&\,\sum_s\bigg(\sum_{r}{\rho}_{a,pr}(t)T_{r>q,s}-{\rho}_{a,pq}(t)\sum_{u,l}T_{ul,s}{\rho}_{a,lu}(t)\bigg)g^\dag_{a,s}(t)\\&+\sum_s\bigg(\sum_{r}T_{p<r,s}{\rho}_{a,rq}(t)-{\rho}_{a,pq}(t)\sum_{u,l}T_{lu,s}{\rho}_{a,ul}(t)\bigg) f^\dag_{a,s}(t),
\end{aligned}
\end{equation} 
\end{subequations}
where $f_{a,s}(t)$, $f_{a,s}^\dag(t)$, $g_{a,s}(t)$, and $g_{a,s}^\dag(t)$ are Gaussian white noise terms independent of the variables $\alpha_{\textbf{k},s}(t)$, $\alpha_{\textbf{k},s}^\dag(t)$, and $\rho_{a,pq}(t)$. The noise terms $f_{a,s}(t)$ and $f_{a,s}^\dag(t)$ are statistically independent of $g_{a,s}(t)$ and $g_{a,s}^\dag(t)$. These elementary noise terms have the following correlation properties
\begin{subequations}
\label{eq: elementary noise terms}
\begin{align}
    \langle f_{a,s}(t)f_{a',s'}(t')\rangle=&\,\langle f_{a,s}^\dag(t)f_{a',s'}^\dag(t')\rangle=0,\\
    \langle f_{a,s}(t)f_{a',s'}^\dag(t')\rangle &\,= \delta_{ss'}\delta_{aa'}\delta(t-t'),
\end{align}
\end{subequations}
that can only be sampled by complex-valued Gaussian white noise terms. Similar stochastic properties hold for $g_{a,s}(t)$ and $g^\dag_{a,s}(t)$.

\subsection{Fields in the coordinate space}

The deterministic unitary evolution of the atomic variables is characterized by Eq.~(\ref{eq: bloch equations appendix}) which incorporates the field variables $\alpha_{\textbf{k},s}(t)$ and $\alpha_{\textbf{k},s}^\dag(t)$ combined into electric fields $\mathcal{E}_s^{(\pm)}(\textbf{r},t)$ as indicated in Eq.~(\ref{eq: Electric fields}). Instead of explicitly tracking the dynamics of the variables $\alpha_{\textbf{k},s}(t)$ and $\alpha_{\textbf{k},s}^\dag(t)$, we derive the equations for $\mathcal{E}_s^{(\pm)}(\textbf{r},t)$.

In the paraxial approximation, the field modes propagate nearly parallel to the $z$ axis. Furthermore, we assume that only the field traveling along with the pump pulse significantly contributes to the dynamics of the atomic variables. Consequently, we make the assumption that $\omega_k=kc \approx \left[k_z+\frac{\textbf{k}_\bot^2}{2k_0}\right]c$. We apply this approximation to Eq.~(\ref{eq: equations for the classical field modes app}), along with the noise terms presented in Eq.~(\ref{eq: noise terms represented}). The resulting equations are then summed over the paraxial wave vectors, yielding the following expression:
\begin{equation}
\label{eq: field equations}
\begin{split}
    \left[\frac{\partial}{c\partial t}+\frac{\partial}{\partial z}\mp\frac{i}{2 k_0}\left(\frac{\partial^2}{\partial x^2}+\frac{\partial^2}{\partial y^2}\right)+\frac{\mu_s(\textbf{r},t)}{2}\right]\mathcal{E}_{s}^{(\pm)}(\textbf{r},t)&=\pm i\frac{k_0 d_0}{2 \varepsilon_0 }P^{(\pm)}_s(\textbf{r},t),
\end{split}
\end{equation}
where we have introduced polarization fields containing both the deterministic atomic variables $\rho_{a,pq}(t)$ and noise terms $g_{a,s}(t)$ and $f_{a,s}(t)$:
\begin{subequations}
\begin{align*}
    P^{(+)}_s(\textbf{r},t)=& \sum_a\left(\sum_{u,l}T_{lu,s}\rho_{a,ul}(t)+f_{a,s}(t)\right) e^{ik_0(z-z_a)}\delta_\varepsilon(\textbf{r}-\textbf{r}_a) , \\ P^{(-)}_s(\textbf{r},t)=&\sum_a\left(\sum_{u,l}T_{ul,s}\rho_{a,lu}(t)+g_{a,s}(t)\right) e^{-ik_0(z-z_a)}\delta_\varepsilon(\textbf{r}-\textbf{r}_a).
    \end{align*}
\end{subequations}
Here, $\delta_\varepsilon(\textbf{r}-\textbf{r}_a)$ represents the sum $\sum_{\textbf{k}}e^{i(\textbf{k}-\textbf{k}_0)\cdot\left(\textbf{r}-\textbf{r}_a\right)}/V$. The summation is carried out over a relatively large set of paraxial wave vectors $\textbf{k}\approx \textbf{k}_0$ included in the electric fields. The resulting function is localized, resembling the functionality of a delta-function. Further, we assume that the transverse part of $\delta_\varepsilon(\textbf{r}-\textbf{r}_a)$ is infinitely small behaving as a true delta-function for the transverse coordinates. In exchange, we introduce damping to the Laplace operator $\nicefrac{\partial^2}{\partial x^2}+\nicefrac{\partial^2}{\partial y^2}$ for non-paraxial modes. Consequently, we replace $\delta_\varepsilon(\Delta\mathbf{r})$ into a product of two components as follows:
\begin{equation*}
\delta_\varepsilon(\Delta\textbf{r})\quad \to \quad \delta(\Delta\textbf{r}_\bot)\delta_\varepsilon(\Delta z).
\end{equation*}

The number of longitudinal modes included in the electric field defines the spatial and temporal scale at which the fields' envelopes remain constant. The corresponding spatial scale defines the width of the delta function $\delta_\varepsilon(\Delta z)$.

\section{Field variables in retarded time}
\label{app: retarded time}

As motivated in Sec.~\ref{sec: Stochastic equations}, it is convenient to replace the original time variable $t$ with the retarded time $\tau = t - z/c$, which effectively incorporates the propagation effects. However, it slightly modifies the correlation properties of the noise terms. To provide a more detailed demonstration, let us formally integrate Eq.~(\ref{eq: field equations}), which yields the following expression for the field $\mathcal{E}^{(+)}_s(\textbf{r},t)$:
\begin{equation}
\label{eq: the field produced by atoms 0}
\begin{aligned}
    \mathcal{E}_{s}^{(+)}(\textbf{r},t) &= \pm i\frac{k_0 d_0}{2 \varepsilon_0 }\int_{z'<z}G_s(\textbf{r},\textbf{r}')P^{(+)}_s(\textbf{r}',t-(z-z')/c)d\textbf{r}'.
\end{aligned}
\end{equation}
Here, we express the solution in terms of the Green functions $G_s(\textbf{r},\textbf{r}')$ corresponding to the following equation:
\begin{equation}
\label{eq: field equations green}
\begin{aligned}
    \left[\frac{\partial}{\partial z}-\frac{i}{2 k_0}\left(\frac{\partial^2}{\partial x^2}+\frac{\partial^2}{\partial y^2}\right)+\frac{\mu_s(\textbf{r})}{2}\right]G_{s}(\textbf{r},\textbf{r}') &= \delta(\textbf{r}-\textbf{r}'),
\end{aligned}
\end{equation}
where $\delta(\textbf{r}-\textbf{r}')$ is the proper delta-function. In this section, we assume a stationary absorption coefficient $\mu_s(\mathbf{r})$, which simplifies the derivations without affecting the final result. Apart from the omitted time derivative and time-dependence of the absorption coefficient $\mu_s(\mathbf{r})$, the left-hand side of this equation retains the same form as the original Eq.~(\ref{eq: field equations}).

Let us examine the field $\mathcal{E}^{(+)}_s(\textbf{r},t)$ generated by an individual atom positioned at point~$\textbf{r}_a$. For $z<z_a$, there is no field since we have neglected back propagation. However, for $z>z_a$, the field can be described as follows:

\begin{equation}
\label{eq: the field produced by a single atom 1}
\begin{aligned}
\mathcal{E}^{(+)}_s(\textbf{r},t) =&\, i\frac{k_0 d_0}{2 \varepsilon_0 } \int_{-\infty}^{\infty} dz' G_s(\textbf{r},\textbf{r}')\bigg(\sum_{u,l}T_{lu,s}\rho_{a,ul}(t-z/c+z'/c)+f_{a,s}(t-z/c+z'/c)\bigg) e^{ik_0(z'-z_a)} \delta_\varepsilon(z'-z_a)\big|_{\textbf{r}'_\bot=\textbf{r}_{\bot, a}}
\end{aligned}
\end{equation}
Here, we have replaced $z$ with infinity in the upper integration limit. This is justified by the paraxial approximation, which reproduces the field only at a sufficient distance from the atom, where $\delta_\varepsilon(z'-z_a)$ is negligibly small. On the other hand, within the paraxial approximation, the self-interaction of a single atom mediated by the field—a cause of spontaneous decay—cannot be accurately reproduced. It must be treated separately at the level of lifetimes. See Sec.~\ref{sec: Inclusion of the pump and decay processes} and Appendix~\ref{sec: Isotropic fluorescence} for more details.

The deterministic part of Eq.~(\ref{eq: the field produced by a single atom 1}) can be significantly simplified. Since the atomic coherence $\rho_{a,ul}$ is driven by the field with the carrier frequency $\omega_0$, multiplying it by $e^{ik_0(z'-z_a)}$ results in a slowly varying function. Consequently, for a sufficiently small width of $\delta_\varepsilon(z'-z_a)$, the integration sign together with the longitudinal delta-function in the deterministic part can be easily omitted:
\begin{equation}
\label{eq: the field produced by a single atom 2}
\begin{aligned}
\mathcal{E}^{(+)}_s(\textbf{r},t)\Big|_{\text{det.}} &= i\frac{k_0 d_0}{2 \varepsilon_0 } G_s(\textbf{r},\textbf{r}_{a})\sum_{u,l}T_{lu,s}\rho_{a,ul}(t-(z-z_a)/c).
\end{aligned}
\end{equation}

More care is required when integrating the noise terms. There is no timescale on which the noise terms change slowly, making it impossible to approximate the integral. For the sake of brevity, we can define the following "smoothed" noise terms:
\begin{equation*}
\begin{aligned}
\tilde{f}_{a,s}(t) & =\int_{-\infty}^{\infty} f_{a,s}(t-t') e^{-i\omega_0 t'} \delta_\varepsilon( t') dt',\\
\tilde{g}_{a,s}(t) & =\int_{-\infty}^{\infty} g_{a,s}(t-t') e^{i\omega_0 t'} \delta_\varepsilon( t') dt',
\end{aligned}
\end{equation*}
where the effective delta-function $\delta_\varepsilon(t')$ has the width of $\delta_\varepsilon(\Delta z)$ divided by the speed of light. The new function $\delta_\varepsilon(\Delta t)$ is consequently involved in the following correlation functions:
\begin{subequations}
\label{eq: modified elementary noise terms app}
\begin{align}
\langle \tilde{f}_{a,s}(t)\tilde{f}_{a',s'}(t')\rangle & =\langle f_{a,s}^\dag(t)f_{a',s'}^\dag(t')\rangle=0,\\
\langle \tilde{f}_{a,s}(t)f_{a',s'}^\dag(t')\rangle & = \delta_{ss'}\delta_{aa'}\delta_\varepsilon(t-t')e^{-i\omega_0 (t-t')}.
\end{align}
\end{subequations}
The stochastic properties of $\tilde{g}_{a,s}(t)$ and $g^\dag_{a,s}(t)$ are similar, with the only difference being a change in the sign of $i$. As a result, the noise contribution in Eq.~(\ref{eq: the field produced by a single atom 1}) gets the following form:
\begin{equation}
\label{eq: the field produced by a single atom 3}
\mathcal{E}^{(+)}_s(\textbf{r},t)\Big|_{\text{noise}} = i\frac{k_0 d_0}{2 \varepsilon_0 } G_s(\textbf{r},\textbf{r}_{a})\tilde{f}_{a,s}(t-(z-z_a)/c).
\end{equation}
In contrast to the correlation properties in Eq.~(\ref{eq: elementary noise terms}) that require a specific Ito's interpretation of the time integration (see Appendix \ref{sec: Appendix: Equivalence of Ito and Stratonovich forms}), the noise terms that conform to the correlation properties in Eq.~(\ref{eq: modified elementary noise terms app}) can be sampled using smooth functions and possess a simpler physical interpretation.

Note that both the deterministic part in Eq.~(\ref{eq: the field produced by a single atom 2}) and the noise part in Eq.~(\ref{eq: the field produced by a single atom 3}) depend on the retarded time $t-z/c$. We can explicitly imprint it into the field variables with the following redefinition:
\begin{equation}
\label{eq: D via Omega app}
\Omega^{(\pm)}_{s}(\textbf{r},\tau)=\frac{d_0}{\hbar}\mathcal{E}^{(\pm)}_{s}(\textbf{r},\tau+z/c)e^{\pm i\omega_0\tau}.
\end{equation}
Here, we express the field variables in terms of the Rabi frequency to simplify the equations and figure out the characteristic parameters of the problem. Besides, we compensate frequently oscillating multipliers $e^{\pm i\omega_0\tau}$, so the fields $\Omega^{(\pm)}_{s}(\textbf{r},\tau)$ represent the slowly varying envelopes.

To account for more then one atom, we add a summation over index $a$ to Eqs.~(\ref{eq: the field produced by a single atom 2}) and~(\ref{eq: the field produced by a single atom 3}). Using the definition of the Rabi frequencies, we arrive at the following expression
\begin{equation}
\label{eq: full field expressions app}
 \Omega^{(+)}_{s}(\textbf{r},\tau)=i\frac{3}{8\pi}\lambda^2_0\Gamma_{\text{rad.}} \sum_{a:\, z_a<z}G_s(\textbf{r},\textbf{r}_a)\bigg(\sum_{u,l}T_{lu,s}\rho_{a,ul}(\tau+z_a/c)+\tilde{f}_{a,s}(\tau+z_a/c)\bigg)e^{ i\omega_0\tau},
\end{equation}  
where $\lambda_0$ is the carrier wavelength and $\Gamma_{\text{rad.}}$ is the spontaneous emission rate calculated based on $d_0$ and given by $\Gamma_{\text{rad.}}=\omega_0^3 d_0^2/[3 \pi \varepsilon_0 \hbar c^3]$. The resulting fields, including $\Omega^{(-)}_{s}(\textbf{r},\tau)$, can be interpreted as solutions of the following equations:
\begin{equation}
\label{eq: full field equations app}
    \left[\frac{\partial}{\partial z}\mp\frac{i}{2 k_0}\left(\frac{\partial^2}{\partial x^2}+\frac{\partial^2}{\partial y^2}\right)+\frac{\mu_s(\textbf{r},\tau)}{2}\right]\Omega_{s}^{(\pm)}(\textbf{r},\tau)=i\frac{3}{8\pi}\lambda^2_0\Gamma_{\text{rad.}}P_s^{(\pm)}(\textbf{r},\tau),
\end{equation}
where the polarization fields $P_s^{(\pm)}(\textbf{r},\tau)$ read as follows:
\begin{subequations}
\label{eq: continuous dipole moments}
\begin{align}
    P^{(+)}_s(\textbf{r},\tau)=& \sum_a\left(\sum_{u,l}T_{lu,s}\rho_{a,ul}(\tau+z_a/c)+\tilde{f}_{a,s}(\tau+z_a/c)\right) e^{ i\omega_0\tau}\delta(\textbf{r}-\textbf{r}_a) , \\ P^{(-)}_s(\textbf{r},\tau)=&\sum_a\left(\sum_{u,l}T_{ul,s}\rho_{a,lu}(\tau+z_a/c)+\tilde{g}_{a,s}(\tau+z_a/c)\right)e^{- i\omega_0\tau}\delta(\textbf{r}-\textbf{r}_a).
    \end{align}
\end{subequations}
Note, that we have reintroduced the time dependence of the absorption coefficient $\mu_s(\textbf{r},\tau)$ in Eq.~(\ref{eq: full field equations app}). By following the rules outlined in Eq.~(\ref{eq: retarded time for the aux. variables}),  this absorption coefficient is now explicitly dependent on the retarded time $\tau$.

\end{widetext}

\section{Collective and continuous variables}
\label{app: collective variables}

\subsection{Wave equations}

Since the Green function $G_s(\textbf{r},\textbf{r}')$ in Eq.~(\ref{eq: full field expressions app}) exhibits slow change from one atom to another, it becomes possible to group closely situated atoms into collective variables. We can divide the entire medium into small regions $\{R_i\}$, each with a volume $\Delta V$, and containing $\Delta N$ atoms. Equation~(\ref{eq: full field expressions app}) suggests the following definition of the collective coherences for each region~$R_i$:
\begin{subequations}
    \label{eq: app collective coherences}
\begin{align}
     \rho_{ul}^{(i)}(\tau) &= e^{i\omega_0 \tau}\sum_{a\in\Delta V} \rho_{a,ul}(\tau+z_a/c)/\Delta N,\\
 \rho_{lu}^{(i)}(\tau) &= e^{-i\omega_0 \tau}\sum_{a\in\Delta V} \rho_{a,lu}(\tau+z_a/c)/\Delta N.
\end{align}
Indices $u$ and $l$ represent upper and lower excited states. As mentioned before, the atomic coherences $\rho_{a,ul}$ are driven by the field with the carrier frequency $\omega_0$, so this carrier frequency is also imprinted in the coherences. To counteract these frequent oscillations and obtain the slowly varying amplitudes, we have multiplied the coherences by $e^{\pm i\omega_0 \tau}$. Similarly to Eq.~(\ref{eq: app collective coherences}), we define collective atomic variables for the remaining density matrix elements:
\begin{align}
    \rho_{u_1u_2}^{(i)}(\tau) = \frac{1}{\Delta N}\sum_{a \in \Delta V} \rho_{a,u_1u_2}(\tau+z_a/c),\\
 \rho_{l_1l_2}^{(i)}(\tau) = \frac{1}{\Delta N}\sum_{a \in \Delta V} \rho_{a,l_1l_2}(\tau+z_a/c).
\end{align}
\end{subequations}
Here, the indices $u_i$ and $l_i$ represent the upper and lower states.
 By analogy, we define similar collective noise terms for each region $R_i$:
\begin{subequations}
    \label{eq: app collective noise terms everywhere}
    \begin{align}
     f_s^{(i)}(\tau) &= e^{i\omega_0 \tau}\sum_{a\in\Delta V} \tilde{f}_{a,s}(\tau+z_a/c)/\Delta V,\\
 g_s^{(i)}(\tau) &= e^{-i\omega_0 \tau}\sum_{a\in \Delta V} \tilde{g}_{a,s}(\tau+z_a/c)/\Delta V.
\end{align}
\end{subequations}
Incorporating the collective variables into Eq.~(\ref{eq: full field expressions app}) yields the following expression:
\begin{multline}
\label{eq: full field expressions collective app}
 \Omega^{(+)}_{s}(\textbf{r}_i,\tau)\\=i\frac{3}{8\pi}\lambda^2_0\Gamma_{\text{rad.}} \sum_{j:\, z_j<z_i}G_s(\textbf{r}_i,\textbf{r}_j)\bigg(n(\textbf{r}_j)\sum_{u,l}T_{lu,s}\rho_{ul}^{(j)}(\tau)\\+f_{s}^{(j)}(\tau)\bigg)\Delta V,
\end{multline}  
where $z_j$ and $z_i$ represent the location of the regions $R_j$ and $R_i$. The condition $z_j < z_i$ implies that the atoms within the region $R_j$ do not interact with each other but are solely influenced by the external fields generated by the atoms from the other regions. 

In the limit of infinitesimally small $\Delta V$, we can introduce the continuous variables as follows:
\begin{equation*}
\begin{matrix}
\rho_{pq}^{(i)}(\tau) &  & \rho_{pq}(\textbf{r},\tau),\\
f_s^{(i)}(\tau) & \quad\quad\to\quad\quad & f_{s}(\textbf{r},\tau),\\
g_s^{(i)}(\tau) &  & g_{s}(\textbf{r},\tau).
\end{matrix}    
\end{equation*}
In terms of these continuous variables, both the fields $\Omega^{(+)}_s(\textbf{r},\tau)$ in~Eq.~(\ref{eq: full field expressions collective app}) and $\Omega^{(-)}_s(\textbf{r},\tau)$ can be described by the partial differential equations outlined in Sec.~\ref{sec: Stochastic wave equations for the field amplitudes}. As mentioned before, we make the assumption that a given region is effected by external electric fields originating from atoms in other regions. In the limit of infinitesimally small $\Delta V$, this naturally leads to Ito's interpretation when performing integration along the $z$ axis. More about Ito stochastic differential equations can be found in Appendix~\ref{sec: Appendix: Equivalence of Ito and Stratonovich forms}.

\subsection{Bloch equations}

Passing from one atom to another, the electric fields experience slight perturbations due to diffraction and interactions with the atoms. It is primarily the propagation along the $z$ axis that significantly affects these fields. As a result, the slowly varying variables $\Omega^{(\pm)}_{s}(\textbf{r},\tau)$ remain uniform across individual regions, but this uniformity is exclusively observed for the fixed retarded time that conveniently accounts for the propagation effects.

If the atomic dynamics is solely determined by the electric fields, the atoms within the individual regions evolve identically. The only distinction is that the dynamics of the neighboring atoms is shifted in time due to the finite speed of light propagation. Consequently, the individual atomic variables can be approximated by the corresponding collective variables. For example, in the case of the atomic coherence $\rho_{a,ul}$, we can express it as follows:
\begin{equation}
\label{eq: transition to the collective variables}
    \rho_{a,ul}(\tau+z_a/c) \approx \rho_{ul}^{(i)}(\tau)e^{-i\omega_0\tau}.
\end{equation}
Here, $\rho_{ul}^{(i)}(\tau)$ corresponds to the region $R_i$ encompassing the atom $a$.

However, in addition to the field variables in Eq.~(\ref{eq: bloch equations appendix}), we must also consider the pump fields in Eq.~(\ref{eq: incoh equations appendix}) and, most importantly, the noise terms in Eq.~(\ref{eq: noise terms represented}). Similar to the variables $\Omega^{(\pm)}_{s}(\textbf{r},\tau)$, when examining small individual regions, the pump field is primarily influenced by propagation along the $z$ axis, a factor that is conveniently addressed by the concept of retarded time. 

The situation with the noise terms requires more attention, since they completely change their values from one atom to another. Note that the noise terms $f_{a,s}(t)$ and $g_{a,s}(t)$ do not operate independently but always appear in groups as defined in Eq.~(\ref{eq: app collective noise terms everywhere}). Consequently, there is no need for independent noise terms $f^\dag_{a,s}(t)$ and $g^\dag_{a,s}(t)$ for each atom. Within the region $R_i$, the following correlation properties apply to each atom:
\begin{align}
\label{eq: noise_terms_correlation}
\langle f_s^{(i)}(\tau) f^\dag_{a,s}(t') \rangle &= \delta_{ss'}\delta_\varepsilon(t'-z_a/c-\tau)e^{i\omega_0(t'-z_a/c)}/\Delta V,\nonumber\\
\langle g_s^{(i)}(\tau) g^\dag_{a,s}(t') \rangle &= \delta_{ss'}\delta_\varepsilon(t'-z_a/c-\tau)e^{-i\omega_0(t'-z_a/c)}/\Delta V.
\end{align}
These correlation properties can be simultaneously restored for each atom by a single pair of independent noise terms $f^{(i)\dag}_{s}(\tau)$ and $g^{(i)\dag}_{s}(\tau)$ defined for the entire region $R_i$:
\begin{subequations}
\label{eq: noise_terms_restoration}
\begin{align}
f_{a,s}^\dag(\tau+z_a/c) &= f^{(i)\dag}_s(\tau)e^{i\omega_0 \tau},\\
g_{a,s}^\dag(\tau+z_a/c) &= g^{(i)\dag}_s(\tau)e^{-i\omega_0 \tau}.
\end{align}
\end{subequations}
The collective noise terms $f^{(i)}_{s}(\tau)$ and $f^{(i)\dag}_{s}(\tau)$ must exhibit the following correlation properties:
\begin{subequations}
\label{eq: collective_noise_terms_correlation}
\begin{align}
\langle f_{s}^{(i)}(\tau)f_{s'}^{(i)}(\tau')\rangle = \langle f_{a,s}^{(i)\dag}(\tau)f_{a',s'}^{(i)\dag}(\tau')\rangle = 0,\\
\langle f_{s}^{(i)}(\tau)f_{s'}^{(i')\dag}(\tau')\rangle = \delta_{ss'}\delta_{ii'}\delta_\varepsilon(\tau-\tau')/\Delta V.
\end{align}
\end{subequations}
The collective noise terms $g_{s}^{(i)}(\tau)$ and $g_{s}^{(i)\dag}(\tau)$ possess similar stochastic properties.

As the noise terms are found to be identical for atoms within the individual regions, we can directly substitute the discrete variables $\rho_{a,pq}(t)$ in Eqs.~(\ref{eq: incoh equations appendix}), (\ref{eq: bloch equations appendix}), and (\ref{eq: noise terms represented}) with the corresponding collective variables $\rho_{ul}^{(i)}(\tau)$, as indicated by Eq.~(\ref{eq: transition to the collective variables}). Furthermore, in the limit of infinitesimal $\Delta V$, we introduce the following continuous variables:
\begin{equation*}
\begin{matrix}
\rho_{pq}^{(i)}(\tau) &  & \rho_{pq}(\textbf{r},\tau),\\
f_s^{(i)\dag}(\tau) & \quad\quad\to\quad\quad& f_{s}^\dag(\textbf{r},\tau),\\
g_s^{(i)\dag}(\tau) &  & g_{s}^\dag(\textbf{r},\tau).
\end{matrix}    
\end{equation*}
The final equations can be found in Section~\ref{sec: Stochastic Bloch equations}. 

Expressing the correlation properties in Eq.~(\ref{eq: collective_noise_terms_correlation}) with respect to the continuous noise terms, we have:
\begin{subequations}
\label{eq: total correlator app}
\begin{align}
\langle f_{s}(\textbf{r},\tau)f_{s'}(\textbf{r}',\tau')\rangle &= \langle f_{s}^{\dag}(\textbf{r},\tau)f_{s'}^{\dag}(\textbf{r}',\tau')\rangle = 0,\\
\langle f_{s}(\textbf{r},\tau)f_{s'}^{\dag}(\textbf{r}',\tau')\rangle = \delta_{ss'}&\delta(z-z')\delta_\varepsilon(\tau-\tau')\delta_\varepsilon(\textbf{r}_\bot-\textbf{r}_\bot').
\end{align}
\end{subequations}
The continuous noise terms $g_{s}(\textbf{r},\tau)$ and $g_{s}^{\dag}(\textbf{r},\tau)$ exhibit similar stochastic properties. The delta-function $\delta(z-z')$ simply reflects the Ito's interpretation of the integration along the~$z$ axis. As previously mentioned, $\delta_\varepsilon(\tau-\tau')$ is a localized function that serves a purpose similar to that of a delta-function. Its width is determined by the number of longitudinal modes required for an accurate representation of the field. For further details, refer to Appendix~\ref{app: retarded time}. However, some clarifications are needed regarding the transverse correlations represented by~$\delta_\varepsilon(\textbf{r}_\bot-\textbf{r}_\bot')$. 

Originally, in Eq.~(\ref{eq: collective_noise_terms_correlation}), the transverse correlation properties are determined by the transverse dimensions of the individual regions into which we have divided the medium. As the size of these regions approaches zero, the transverse part of the total correlator in Eq.~(\ref{eq: total correlator app}) is expected to become infinitely narrow. Nevertheless, note that the noise terms are part of the wave equations for the fields, the solution of which is expected to be regularized by damping non-paraxial modes. This consideration allows us to "smear out" the correlation properties and define the width of the transverse correlator based on the span of relevant transverse modes required for an accurate representation of the paraxial fields.

\section{\label{sec: Illustr}Noise terms reproducing the spontaneous emission}

In this section, we provide a simplified illustration of how the interplay of the noise terms, present in both the equations for the fields and the atomic variables, reproduces the spontaneous emission that is subsequently amplified. Similarly to Appendix~\ref{app: collective variables}, we split the medium into small regions where atoms are assumed to evolve identically. For simplicity, we analyze how one of these regions participates in the collective dynamics. For this reason, we ignore the presence of all the other regions and consider a small, localized collection of two-level atoms occupying a volume $\Delta V$. All the comprising atoms are characterized by the atomic variables $\rho_{pq}(\textbf{r},\tau) \to \rho_{pq}^{(1)}(\tau)$, which are assumed to be identical for each atom. Similarly, we have the same noise terms for the whole region, $g^{(\dag)}(\textbf{r},\tau) \to g^{(\dag)}_1(\tau)/\sqrt{\Delta V}$ and $f^{(\dag)}_1(\tau) \to f^{(\dag)}_1(\tau)/\sqrt{\Delta V}$. Additionally, the fields $\Omega^{(\pm)}(\textbf{r},\tau)$ have only one polarization.

Initially, the atomic coherences are zero. If there is no external field resonant with the transition, the dynamics of the comprising atoms is defined by the incoherent processes and the noise terms. Integrating Eq.~(\ref{eq: atomic equations}), the coherences $\rho_{ul}^{(1)}(\tau)$ and $\rho_{lu}^{(1)}(\tau)$ take the following form:
\begin{equation}
\label{eq: integrated atomic equations}
\begin{aligned}
\rho_{ul}^{(1)}(\tau)= T_{{ul}}\,\int _0^\tau d\tau' \rho_{uu}(\tau')e^{-(\Gamma_u+\Gamma_l)(\tau-\tau')/2}g^{\dag}_1(\tau')/\sqrt{\Delta V},\\
\rho_{lu}^{(1)}(\tau)=T_{{lu}}\,\int _0^\tau d\tau' \rho_{uu}(\tau')e^{-(\Gamma_u+\Gamma_l)(\tau-\tau')/2}f^{\dag}_1(\tau')/\sqrt{\Delta V},
\end{aligned}
\end{equation}
where we only consider one polarization and omit the index $s$. As mentioned in Sec. \ref{sec: Stochastic Bloch equations} after Eq.~(\ref{eq: noise decomposition}), we have omitted the noise contributions that exhibit a quadratic dependence on the atomic variables ${\rho}_{pq}(\textbf{r},\tau)$. Since the noise terms $g^{\dag}_1(\tau)$ and $f^{\dag}_1(\tau)$ are uncorrelated, there is no macroscopic dipole moment:
\begin{equation*}
    \langle \rho_{lu}^{(1)}(\tau)\rangle= \langle \rho_{ul}^{(1)}(\tau)\rangle = \langle \rho_{ul}^{(1)}(\tau)\rho_{lu}^{(1)}(\tau')\rangle =0,
\end{equation*}
which is absolutely coherent with the assumptions that the neighboring atoms are independent and do not experience any external influence. The change in the mean populations of the atomic levels is solely caused by the finite lifetime and the pump, as the noise terms average out to zero.

As we can see, the noise terms do not directly affect the dynamics of the atoms. Their primary purpose is to facilitate the generation of spontaneous fields. To demonstrate this, we integrate Eq.~(\ref{eq: paraxial Omega}), which leads to the following expressions for the fields $\Omega^{(\pm)}(\tau)$ generated by the atoms in the small region:
\begin{widetext}
\begin{equation}
\begin{pmatrix}
\Omega_{\text{det.}}^{(+)}(\tau)\\
\Omega_{\text{noise}}^{(+)}(\tau)
\end{pmatrix}=i\frac{3}{8\pi}\lambda^2_0\Gamma_{\text{rad.}}\Delta z\begin{pmatrix}T_{lu}n\rho_{ul}^{(1)}(\tau)\\f_1(\tau)/\sqrt{\Delta V}\end{pmatrix},\quad
\begin{pmatrix}
\Omega_{\text{det.}}^{(-)}(\tau)\\
\Omega_{\text{noise}}^{(-)}(\tau)
\end{pmatrix}=-i\frac{3}{8\pi}\lambda^2_0\Gamma_{\text{rad.}}\Delta z\begin{pmatrix}T_{ul}n\rho_{lu}^{(1)}(\tau)\\g_1(\tau)/\sqrt{\Delta V}\end{pmatrix},
\end{equation}
where $n$ is the concentration. We have neglected the diffraction effects to simplify the expressions. Substituting the coherences from Eq.~(\ref{eq: integrated atomic equations}), we get the following expressions:

\begin{subequations}
\label{eq: spontaneous Omegas}
\begin{equation}
\begin{pmatrix}
\Omega_{\text{det.}}^{(+)}(\tau)\\
\Omega_{\text{noise}}^{(+)}(\tau)
\end{pmatrix}=i\frac{3}{8\pi}\frac{\lambda^2_0\Gamma_{\text{rad.}}\Delta z}{\sqrt{\Delta V}}\begin{pmatrix}|T_{lu}|^2n\,\int _0^\tau d\tau' \rho_{uu}^{(1)}(\tau')e^{-(\Gamma_u+\Gamma_l)(\tau-\tau')/2}g^{\dag}_1(\tau')\\f_1(\tau)\end{pmatrix},
\end{equation}
\begin{equation}
\begin{pmatrix}
\Omega_{\text{det.}}^{(-)}(\tau)\\
\Omega_{\text{noise}}^{(-)}(\tau)
\end{pmatrix}=-i\frac{3}{8\pi}\frac{\lambda^2_0\Gamma_{\text{rad.}}\Delta z}{\sqrt{\Delta V}}\begin{pmatrix}|T_{ul}|^2n\,\int _0^\tau d\tau' \rho_{uu}^{(1)}(\tau')e^{-(\Gamma_u+\Gamma_l)(\tau-\tau')/2}f^{\dag}_1(\tau')\\g_1(\tau)\end{pmatrix}.
\end{equation}
\end{subequations}

\end{widetext}
We characterize the field by the first order correlation functions $J_s(\textbf{r},\tau,\tau')$ defined in 
Eq.~(\ref{eq: Jsp-t1-t2_def}). Adopting the notation from this section and omitting the polarization index, we write:
\begin{equation}
\label{eq: app field correlator def}
J(\tau,\tau')=\frac{\langle\Omega^{(+)}(\tau)\Omega^{(-)}(\tau')\rangle}{\frac{3}{8\pi}\lambda^2_0 \Gamma_{\text{rad.}}}.
\end{equation}
Analyzing Eq.~(\ref{eq: spontaneous Omegas}), we notice that $\Omega_{\text{det.}}^{(-)}$ and $\Omega_{\text{det.}}^{(+)}$ are not correlated. Same with $\Omega_{\text{noise}}^{(-)}$ and $\Omega_{\text{noise}}^{(+)}$. $J(\tau,\tau')$ reads then as follows:
\begin{multline}
\label{eq: interplay}
J(\tau,\tau')=\left[\frac{3}{8\pi}\lambda^2_0 \Gamma_{\text{rad.}}\right]^{-1}\Big(\left\langle\Omega_{\text{det.}}^{(-)}(\tau)\Omega_{\text{noise}}^{(+)}(\tau')\right\rangle\\+\left\langle\Omega_{\text{noise}}^{(-)}(\tau)\Omega_{\text{det.}}^{(+)}(\tau')\right\rangle\Big).
\end{multline}
Eq.~(\ref{eq: interplay}) shows, how the noise terms from the atomic and field equations can finally meet and give non-zero correlations: $\Omega_{\text{det.}}^{(\pm)}(\tau)$ include the integrated noise terms from the equations for the atomic variables $f_1^\dag(\tau)$ and $g_1^\dag(\tau)$ and $\Omega_{\text{noise.}}^{(\pm)}(\tau)$  contain $f_1(\tau)$ and $g_1(\tau)$. Incorporating Eq.(\ref{eq: spontaneous Omegas}) into  Eq.~(\ref{eq: interplay}) yields the following correlator:
\begin{multline}
    \label{eq: relevant observable}
   J(\tau,\tau')= \frac{3}{8\pi}\frac{\lambda^2_0}{\Delta x\Delta y} \Gamma_{\text{rad.}}
n\Delta z  |T_{{ge}}|^2\\\times
    \rho_{uu}^{(1)}(\min(\tau,\tau'))e^{-(\Gamma_u+\Gamma_l)|\tau-\tau'|/2},
\end{multline}
which is the Lorentzian spectrum of the spontaneous emission. Summing up the spontaneous emission from the other regions yields Eq.~(\ref{eq: Jsp-t1-t2_num}). Here, $\lambda^2_0/[\Delta x\Delta y]$ represents the solid angle over which the spontaneous emission propagates.

The spontaneous emission interacts with the atoms in neighboring regions, stimulating them to decay faster, which results in increased emission. This leads to the phenomenon of amplified spontaneous emission.}

\section{\label{sec: Numerical} Numerical realization}

\subsection{Noise terms, atomic and field variables on a grid}

For the current implementation, we use a uniform rectangular grid with step size $\Delta x, \Delta y, \Delta z, \Delta \tau$ and follow the atomic and field variables at grid nodes denoted by a four-dimensional index $\texttt{xyz\texttau}$:

\begin{equation}
\rho_{{ij}}(\textbf{r},\tau)\,\to\,\rho_{{ij},\texttt{xyz\texttau}}, \quad \Omega_s^{(\pm)}(\textbf{r},\tau)\,\to\,\Omega_{s,\texttt{xyz\texttau}}^{(\pm)}.
\end{equation}

 The noise contributions are modeled with the help of Gaussian random numbers with the following correlation properties  
\begin{subequations}
    \label{eq: xi deiscrete correlation properties, num}
\begin{align}
\langle 
\xi_{s,\texttt{xyz}\texttt{\texttau}}^{(\pm)}
\xi_{s',\texttt{x}'\texttt{y}'\texttt{z}'\texttt{\texttau}'}^{(\pm)*}
\rangle  
&=
\delta_{ss'}\,\delta_{\texttt{x}\texttt{x}'}\,\delta_{\texttt{y}\texttt{y}'}\,\delta_{\texttt{z}\texttt{z}'}\,\delta_{\texttt{\texttau\texttau}'},\\
\langle 
\xi^{(\pm)}_{s,\texttt{xyz}\texttt{\texttau}}
\xi_{s',\texttt{x}'\texttt{y}'\texttt{z}'\texttt{\texttau}'}^{(\pm)}
\rangle  
&=\langle\xi^{(\mp)}_{s,\texttt{xyz}\texttt{\texttau}}
\xi_{s',\texttt{x}'\texttt{y}'\texttt{z}'\texttt{\texttau}'}^{(\pm)}
\rangle = 0,
\end{align}
\end{subequations}
that can be directly used to discretize the noise terms
\begin{align}
\label{eq: original discretization}
\begin{bmatrix}
           {f}_s\!\left(\textbf{r},\tau\right) \\
           {f}_s^{\dag}\!\left(\textbf{r},\tau\right) \\
           {g}_s\!\left(\textbf{r},\tau\right) \\
           g_s^{\dag}\!\left(\textbf{r},\tau\right)
         \end{bmatrix}
\,\to\,\begin{bmatrix}
           \xi_{s,\texttt{xyz}\texttt{\texttau}}^{(+)} \\
           \xi_{s,\texttt{xyz}\texttt{\texttau}}^{(+)*} \\
           \xi_{s,\texttt{xyz}\texttt{\texttau}}^{(-)} \\
           \xi_{s,\texttt{xyz}\texttt{\texttau}}^{(-)*}
         \end{bmatrix}/{\sqrt{\Delta z \Delta x\Delta y \Delta \tau}}.
\end{align}

\subsection{Diffusion gauges}
\label{app: diffusion gauge}
In section \ref{Gauging run-away trajectories}, we proposed to use the drift gauges for removing run-away realizations from the stochastic differential equations. The rigorous application of the drift gauges requires re-weighting trajectories from the final statistical sample. We aim to skip the re-weighting procedure. To mitigate the effect of this approximation, we attempt to minimize the need of the substitution in Eq.~(\ref{eq: gauge transformation}). We achieve this by reducing the difference between atomic coherences $\rho_{ul}(\textbf{r},\tau)$ and $\rho_{lu}^*(\textbf{r},\tau)$ through the use of the diffusion gauge discussed in Sec. \ref{Gauging run-away trajectories}. Throughout this appendix, we use indices $u$ and $u'$ to denote the upper excited states, while $l$ and $l'$ represent the lower excited states.

Note that the correlation properties in Eq.~(\ref{eq: xi deiscrete correlation properties, num}) do not change under the following transformation:
\begin{align}
\label{eq: gauged random numbers}
\xi_{s,\texttt{xyz}\texttt{\texttau}}^{(\pm)}&\to \xi_{s,\texttt{xyz}\texttt{\texttau}}^{(\pm)}R_{s,\texttt{xyz}\texttt{\texttau}},\\\nonumber
\xi_{s,\texttt{xyz}\texttt{\texttau}}^{(\pm)*}&\to\xi_{s,\texttt{xyz}\texttt{\texttau}}^{(\pm)*}/R_{s,\texttt{xyz}\texttt{\texttau}},
\end{align}
where $R_{s,\texttt{xyz}\texttt{\texttau}}$ must be statistically independent from the the noise terms $\xi_{s,\texttt{x}'\texttt{y}'\texttt{z}'\texttt{\texttau}'}^{(\pm)}$ for $\texttt{z}'\geq\texttt{z}$. 

Suitable gauging coefficients $R_{s,\texttt{xyz}\texttt{\texttau}}$ must minimize the following expression for each polarization $s$, time $t$ and coordinates $\textbf{r}$:
\begin{equation}
\Big\langle\Big|\sum_{eg}T_{ges}\left(\rho_{eg}(\textbf{r},\tau)- \rho_{ge}^*(\textbf{r},\tau)\right)\Big|^2\Big\rangle.
\end{equation}
Assuming coherences $\rho_{uu'}(\textbf{r},\tau)$ and $\rho_{ll'}(\textbf{r},\tau)$ to be small, coefficients $R_{s,\texttt{xyz}\texttt{\texttau}}$ can be expressed in the following way:
\begin{equation*}
R_{s,\texttt{xyz}\texttt{\texttau}}=\sqrt{\frac{16\pi g_{s,\texttt{xyz}\texttt{\texttau}}}{3\lambda^2\Gamma_{\text{rad.}}\Delta z}}
\end{equation*}
where $g_{s,\texttt{xyz}}$ depends on the discretized version of $\rho_{s}^{(\text{up.})}(\textbf{r},\tau)$ and $\rho_{s}^{(\text{low.})}(\textbf{r},\tau)$ from equation (\ref{eq: E-and-G, num}):
\begin{equation}
\label{eq: choice of g}
g_{s,\texttt{xyz}\texttt{\texttau}}=\frac{\rho_{s,\texttt{xyz\texttau}}^{(\text{up.})}}{\rho_{s,\texttt{xyz\texttau}}^{(\text{up.})}-\rho_{s,\texttt{xyz\texttau}}^{(\text{low.})}}.
\end{equation}
Inter-level coherences $\rho_{uu'}(\textbf{r},\tau)$ and $\rho_{ll'}(\textbf{r},\tau)$ are not created during the pump stage; they develop during the interaction with the SF field. Since the noise terms are significant during spontaneous emission and ASE stages when a strong SF field has not yet developed, the full consideration of the inter-level coherences for noise-term calculations would be a small correction compared to the populations of the levels at the stages under consideration.

\subsection{Numerical scheme for the field variables}

Given the atomic variables at grid nodes $\rho_{pq,\texttt{xyz\texttau}}$, we can propagate the field variables using multislicing approach~\cite{2017'Li_multislice, 2009'Voelz}. To simplify the differentiation along the $x$- and $y$-axis and achieve spectral accuracy, we make use of Fourier transform in the $xy$-plane
\begin{equation*}
\begin{pmatrix}\Omega_{s,\,\text{det.}}^{(\pm)}\\
\Omega_{s,\,\text{noise}}^{(\pm)}
\end{pmatrix}_{\!\!\texttt{k}_x\texttt{k}_y\texttt{z\texttau}}\!\!\!\!=\,\,\sum_{\textbf{x},\textbf{y}}\mathcal{F}^{\,\texttt{xy}}_{\texttt{k}_x\texttt{k}_y}\,\begin{pmatrix}\Omega_{s,\,\text{det.}}^{(\pm)}\\
\Omega_{s,\,\text{noise}}^{(\pm)}
\end{pmatrix}_{\!\!\texttt{xyz\texttau}},
\end{equation*}
where $\mathcal{F}^{\,\texttt{xy}}_{\texttt{k}_x\texttt{k}_y}$ denotes the components of the Fourier transfrom, and $\texttt{k}_x$ and $\texttt{k}_y$ indicate the Fourier components of the fields. To denote the inverse Fourier transform, we swap the indices, namely $\left(\mathcal{F}^{-1}\right)^{\texttt{xy}}_{\texttt{k}_x\texttt{k}_y}=\mathcal{F}_{\,\texttt{xy}}^{\texttt{k}_x\texttt{k}_y}$. Assuming the introduced notation, the integrating scheme takes the form:  
\begin{multline}
\label{eq:Omega-p, num}
\begin{pmatrix}
\Omega_{s,\,\text{det.}}^{(+)}\\
\Omega_{s,\,\text{noise}}^{(+)}
\end{pmatrix}_{\!\!\texttt{xy(z+1)\texttau}}\!\!\!\!=\mathcal{G}_{\texttt{xyz\texttau}}\sum_{\texttt{k}_x,\texttt{k}_y}\mathcal{F}_{\,\texttt{xy}}^{\texttt{k}_x\texttt{k}_y}\,\mathcal{K}_{\texttt{k}_x\texttt{k}_y}\begin{pmatrix}
\Omega_{s,\,\text{det.}}^{(+)}\\
\Omega_{s,\,\text{noise}}^{(+)}
\end{pmatrix}_{\!\!\texttt{k}_x\texttt{k}_y\texttt{z\texttau}}\\+i\begin{pmatrix}\gamma
n_\texttt{xyz}\Delta V\!\displaystyle\sum_{u,\,l}  T_{{lu}s} 
    \rho_{ul,\texttt{xyz\texttau}}\\2\sqrt{\frac{\gamma  g_{s,\texttt{xyz\texttau}}}{2\Delta\tau}}
    \xi_{s,\texttt{xyz\texttau}}^{(+)}\end{pmatrix},
\end{multline}
\begin{multline}
\label{eq:Omega-m, num}
\begin{pmatrix}
\Omega_{s,\,\text{det.}}^{(-)}\\
\Omega_{s,\,\text{noise}}^{(-)}
\end{pmatrix}_{\!\!\texttt{xy(z+1)\texttau}}\!\!\!\!=\mathcal{G}_{\texttt{xyz\texttau}}^{*}\sum_{\texttt{k}_x,\texttt{k}_y}\mathcal{F}_{\,\texttt{xy}}^{\texttt{k}_x\texttt{k}_y}\,\mathcal{K}_{\texttt{k}_x\texttt{k}_y}^*\begin{pmatrix}\Omega_{s,\,\text{det.}}^{(-)}\\
\Omega_{s,\,\text{noise}}^{(-)}
\end{pmatrix}_{\!\!\texttt{k}_x\texttt{k}_y\texttt{z\texttau}}\\-i\begin{pmatrix}\gamma
n_\texttt{xyz}\Delta V\!\displaystyle\sum_{u,\,l}  T_{{ul}s} 
    \rho_{lu,\texttt{xyz\texttau}}\\2\sqrt{\frac{\gamma  g_{s,\texttt{xyz\texttau}}}{2\Delta\tau}}
    \,\xi_{s,\texttt{xyz\texttau}}^{(-)}\!\end{pmatrix}.
\end{multline}
Here, we have already performed the transformation (\ref{eq: gauged random numbers}) and introduced the gauging function $g_{s,\texttt{xyz\texttau}}$ from equation (\ref{eq: choice of g}).  For brevity, we have also introduced an effective radiative decay rate
\begin{equation}
\label{eq: gamma}
    \gamma=\frac{3}{8\pi}\times \frac{\lambda^2}{\Delta x \Delta y}\times\Gamma_{\text{rad.}},
\end{equation}
where $\lambda^2/\Delta x \Delta y$ represent the solid angle over which the paraxial modes propagate. The ratio $\gamma/\Gamma_{\text{rad.}}$ defines the proportion of the spontaneous emission participating in the amplification process. This coefficient turns out to be a universal constant further appearing in the equations for the discrete atomic variables. Apart from that, we have also introduced the elementary volume $\Delta V=\Delta x\Delta y\Delta z$, the atomic density $n_{\texttt{xyz}}$ defined on the grid, and two additional matrices describing absorption and diffraction upon propagation along the medium
\begin{align} 
\label{eq: kernels}
&\mathcal{G}_{\texttt{xyz\texttau}}=\exp\!\left[\left(\frac{\mu_{\texttt{xyz\texttau}}}{2}\mp i \delta_{\texttt{xyz\texttau}}k_0\right)\Delta z\right],\nonumber\\&\mathcal{K}_{\texttt{k}_x\texttt{k}_y}=\exp\!\left[-i\frac{k_x^2+k_y^2}{2k_0}\Delta z\right],
\end{align}
where $\mu_{\texttt{xyz\texttau}}$ and $\delta_{\texttt{xyz\texttau}}$ represent $\mu(\textbf{r},\tau)$ and $\delta(\textbf{r},\tau)$ on a grid.

To remove run-away trajectories, we have to adopt the strategy from Sec.~\ref{Gauging run-away trajectories} to Eqs.~(\ref{eq:Omega-p, num}) and~(\ref{eq:Omega-m, num}). At each time step, we find $\texttt{xyz}$-points satisfying the condition in Eq.~(\ref{eq: condition for the correction}) and transform Eqs.~(\ref{eq:Omega-p, num}) and~(\ref{eq:Omega-m, num}) in accordance with Eq.~(\ref{eq: gauge transformation}).

With the proposed scheme, we can achieve high stability and first-order accuracy for the integration of the deterministic part along the $z$ axis. In addition, Eqs.~(\ref{eq:Omega-p, num}) and~(\ref{eq:Omega-m, num}) can also be used for integrating pump fields.

According to Section \ref{sec: Stochastic equations}, the noise terms should exhibit finite correlations in time and along the transverse directions. However, the proposed discretized noise terms associated with distinct nodes on the grid are completely independent. This lack of correlation in the transverse direction is not a problem, as the propagation along the $z$ axis will introduce these correlations by cutting the non-paraxial modes. To restore finite correlations in time, the variables used to construct observables must be averaged over neighboring time nodes.  

\subsection{Numerical scheme for the atomic variables}

We use an approach similar to split-step method, and treat the increment of the regular part of the equations for atomic variables with  suitable explicit high order algorithm, while for the noise part we use an explicit Euler-Murayama scheme
\begin{equation*}
\rho_{pq,\texttt{xyz(\texttau}+1)}=\rho_{pq,\texttt{xyz\texttau}}+\Delta\rho_{pq,\texttt{xyz\texttau}}|_{\text{det.}}+\Delta\rho_{pq,\texttt{xyz\texttau}}|_{\text{noise}}.
\end{equation*}

In this article, the time integration of the regular part of the atomic variables is performed separately for each $\texttt{xyz}$-point with Runge-Kutta fourth-order algorithm 
\begin{widetext}
    
\begin{align}
\label{eq: Bloch equation ee, num}
\frac{\Delta\rho_{uu',\,\texttt{xyz\texttau}}}{\Delta\tau}\Big|_{\text{det.}}&=
   \left[-i\Delta \omega_{uu'}\rho_{uu'}
   +i\sum_{l,\, s} \left( 
   \Omega_{s,\,\text{det.}}^{(+)}T_{{ul}s}\rho_{{lu'}}
   -\Omega_{s,\,\text{det.}}^{(-)}\rho_{{ul}}T_{{lu'}s}
   \right)\right]_{\text{RK},\,\texttt{xyz\texttau}} \\  
   \label{eq: Bloch equation eg, num}
\frac{\Delta\rho_{ul,\,\texttt{xyz\texttau}}}{\Delta\tau}\Big|_{\text{det.}}&=
   \left[-i\Delta \omega_{{ul}}\rho_{{ul}}
   -i\sum_{s} \Omega_{s,\,\text{det.}}^{(+)}
   \left( 
   \sum_{u'}\rho_{{uu'}} T_{{u'l}s} - \sum_{l'}T_{{ul'}s} \rho_{{l'l}}
   \right)\right]_{\text{RK},\,\texttt{xyz\texttau}}
   \\
   \label{eq: Bloch equation ge, num}
\frac{\Delta\rho_{lu,\,\texttt{xyz\texttau}}}{\Delta\tau}\Big|_{\text{det.}}&=\left[ i \Delta \omega_{{ul}}\rho_{{lu}}
   +i\sum_{s}\Omega_{s,\,\text{det.}}^{(-)}
   \left( 
   \sum_{u'}T_{{lu'}s}\rho_{{u'u}}  - \sum_{l'}\rho_{{ll'}}T_{{l'u}s} \right)\right]_{\text{RK},\,\texttt{xyz\texttau}} \\
   \label{eq: Bloch equation gg, num}\frac{\Delta\rho_{ll',\,\texttt{xyz\texttau}}}{\Delta\tau}\Big|_{\text{det.}}&=
   \left[-i \Delta \omega_{{ll'}}\rho_{{ll'}}
   +i\sum_{u,\, s}\left( 
   \Omega_{s,\,\text{det.}}^{(-)}T_{{lu}s}\rho_{{ul'}}
   -\Omega_{s,\,\text{det.}}^{(+)}\rho_{{lu}}T_{{ul'}s}
   \right)\right]_{\text{RK},\,\texttt{xyz\texttau}}
\end{align}
Note that the equations only contain the deterministic fields $\Omega_{s,\,\text{det.}}^{(\pm)}(\textbf{r},\tau)$. The noise parts of the fields $\Omega_{s,\,\text{noise}}^{(\pm)}(\textbf{r},\tau)$ must be taken into account together with the other noise terms at the level of Euler-Murayama scheme
\begin{align}
\label{eq: Bloch equation ee, num, noise}
\frac{\Delta\rho_{uu',\texttt{xyz\texttau}}}{\Delta\tau}\Big|_{\text{noise}}=&i\sum_{l,\, s} \left( 
   \Omega_{s,\,\text{noise}}^{(+)}T_{{ul}s}\rho_{{lu'}}
   -\Omega_{s,\,\text{noise}}^{(-)}\rho_{{ul}}T_{{lu'}s}
   \right)_{\texttt{xyz\texttau}}\\
\label{eq: Bloch equation eg, num, noise}
\frac{\Delta\rho_{{ul},\texttt{xyz\texttau}}}{\Delta\tau}\Big|_{\text{noise}}=&\sum_{s} \Bigg[i\Omega_{s,\,\text{noise}}^{(+)}
   \left( 
   \sum_{l'} T_{{ul'}s} \rho_{{l'l}}-\sum_{u'} \rho_{{uu'}} T_{{u'l}s}
   \right) +\sqrt{\frac{\gamma
            g_{s}^{-1}}{2\Delta\tau}}\xi_{s}^{(-)*}\sum_{u'}\rho_{{uu'}} T_{{u'l}s}\Bigg]_{\texttt{xyz\texttau}}\\
\label{eq: Bloch equation ge, num, noise}
\frac{\Delta\rho_{lu,\texttt{xyz\texttau}}}{\Delta\tau}\Big|_{\text{noise}}=&\sum_{s}\Bigg[i\Omega_{s,\,\text{noise}}^{(-)}
   \left( \sum_{u'}
   T_{{lu'}s}\rho_{{u'u}}  - \sum_{l'}\rho_{{ll'}} T_{{l'u}s} \right) +\sqrt{\frac{\gamma
            g_{s}^{-1}}{2\Delta\tau}}\, \xi_{s}^{(+)*}\sum_{u'}T_{{lu'}s}\rho_{{u'u}}\Bigg]_{\texttt{xyz\texttau}} \\  
\label{eq: Bloch equation gg, num, noise} 
\frac{\Delta\rho_{ll',\texttt{xyz\texttau}}}{\Delta\tau}\Big|_{\text{noise}}=&
   \sum_{u,\, s}\Bigg[i\left( 
   \Omega_{s,\,\text{noise}}^{(-)}T_{{lu}s}\rho_{{ul'}} 
   -\Omega_{s,\,\text{noise}}^{(+)}\rho_{{lu}}T_{{ul'}s}
   \right)
 + \sqrt{\frac{\gamma
            g_{s}^{-1}}{2\Delta\tau}}\left(\xi_s^{(+)*} T_{{lu}s} \rho_{{ul'}}
   +\xi_s^{(-)*} \rho_{{lu}}T_{{ul'}s}\right)\Bigg]_{\texttt{xyz\texttau}}
\end{align}

As mentioned in Sec. \ref{sec: Stochastic Bloch equations} after Eq.~(\ref{eq: noise decomposition}), we have omitted the noise contributions that exhibit a quadratic dependence on the atomic variables ${\rho}_{pq}(\textbf{r},\tau)$.

\end{widetext}

\subsection{Qualitative analysis of the noise terms}

The proportionality of the noise-term increments to $\sqrt{\Delta t}$ in equations for atomic variables~(\ref{eq: Bloch equation ee, num, noise}) --~(\ref{eq: Bloch equation gg, num, noise}) is inherent for stochastic differential equations. 

In contrast to the deterministic source, the noise source in equations~(\ref{eq:Omega-p, num}) --~(\ref{eq:Omega-m, num}) is not proportional to grid size $\Delta z$. Consequently, the ratio between the noise and deterministic contribution is inversely proportional to $\Delta z$. This inverse proportionality can be understood following the arguments presented for superradiance of distributed systems~\cite{1982'Gross}: the smaller the grid size is, the larger the quantum fluctuations of atomic coherence are due to the finite number of the emitters within the grid voxel. In our case, the higher spatial resolution we would like to achieve, the larger the noise-term values reflecting the larger relative role of quantum effects would be, and the larger amount of realizations we would need to run in order to achieve smooth profiles for observables of interest.

\bibliography{bib1.bib}
\end{document}